\def\draftversion{false}
\RequirePackage{ifthen}
\ifthenelse{\equal{\draftversion}{true}}{
  \documentclass[floatfix,galley,prb]{revtex4}
}{
  \documentclass[floatfix, twocolumn, prb]{revtex4}
}

\usepackage{filecontents,pdfpages}
\usepackage{color}
\usepackage{multirow}
\usepackage{epsfig}
\usepackage{graphicx}
\usepackage{amsmath}
\usepackage{amssymb}
\usepackage{bm}
\usepackage{dcolumn}
\usepackage{braket}
\usepackage{bbold}
\usepackage{sidecap}
\usepackage{float}
\usepackage{hyperref}
\usepackage{pdfpages}
\hypersetup{hidelinks}
\usepackage{pdfpages}
\usepackage{pdfpages}

\begin{document}

\title{High throughput screening for spin-gapless semiconductors in quaternary Heusler compounds}
\author{Qiang~Gao}
\author{Ingo~Opahle}
\author{Hongbin~Zhang}
\email[corresp.\ author: ]{hzhang@tmm.tu-darmstadt.de}
\affiliation{Institute of Materials Science, Technische Universit$\ddot{a}$t Darmstadt, 64287 Darmstadt, Germany}

\date{\today}

\begin{abstract}
Based on high throughput density functional theory calculations, we performed systematic screening for spin-gapless semiconductors (SGSs)
in quaternary Heusler alloys XX$^\prime$YZ (X, X$^\prime$, and Y are transition metal elements without Tc, and Z is one of B, Al, Ga, In, Si, Ge, Sn, Pb, P, As, Sb, and Bi). 
Following the empirical rule, we focused on compounds with 21, 26, or 28 valence electrons, resulting in 12, 000
possible chemical compositions.
After systematically evaluating the thermodynamic, mechanical, and dynamical stabilities, we successfully identified 70 stable SGSs, confirmed by
explicit electronic structure calculations with proper magnetic ground states.
It is demonstrated that all four types of SGSs can be realized, defined based on the spin characters of the bands around the Fermi energy,
and the type-II SGSs show promising transport properties for spintronic applications.
The effect of spin-orbit coupling is investigated, resulting in large anisotropic magnetoresistance and anomalous Nernst effects.

\end{abstract}

\maketitle

\def\scr{\scriptsize}
\ifthenelse{\equal{\draftversion}{true}}{
  \marginparwidth 2.7in
  \marginparsep 0.5in
  \newcounter{comm} 
  \def\commnext{\stepcounter{comm}}
  \def\commtext{{\bf\color{blue}[\arabic{comm}]}}
  \def\commmar{{\bf\color{blue}[\arabic{comm}]}}
  \def\hzm#1{\commnext\marginpar{\small HZ\commmar: #1}\commtext}
  \def\qgm#1{\commnext\marginpar{\small QG\commmar: #1}\commtext}
}{
  \def\hzm#1{}
  \def\qgm#1{}
}

\def\Red#1{\textcolor{red}{#1}}
\def\Blue#1{\textcolor{blue}{[#1]}}
\def\Magenta#1{\textcolor{magenta}{#1}}

\section{Introduction}
\label{intro}
In recent years, spin-gapless semiconductors (SGSs) have drawn intensive attention to the spintronics community. 
Basically, SGSs are half metals with the majority spin channel being semimetallic, {\it i.e.}, the gap is zero, while there
is a finite band gap in the minority spin channel. 
Based on how the spin characters of bands touching the Fermi level, there are four types of SGSs as sketched in Fig.~\ref{fig:mixfigure}(a).~\cite{wang-2008-prl}
In the type-I SGSs, the valence band maximum (VBM) and conduction band minimum (CBM) are in the same spin channel 
while there is a gap in the opposite spin channel. This is the conventional SGS as mentioned.
Moreover, the CBM and VBM can hold different spin characters, hereafter dubbed as type-II SGS.
Additionally, if the VBM (CBM) is of one spin character while the CBMs (VBMs) are originated from both spin channels, 
type-III (type-IV) SGSs will be defined.
In principle, the VBM and CBM can touch each other at the same or different $k$-points,
corresponding to the direct or indirect zero band gap.
In comparison to the usual half metals, 
the 100\% spin polarized carriers can be excited from the valence to conduction bands with no energy cost, 
leading to new functionalities and potential applications in logic gates. 
For instance, the spin polarized transport properties of SGSs can be tuned by shifting the Fermi energy with finite gate voltages,~\cite{wang-2008-prl, c-felsher-2013-prl} 
which are promising for future spintronic applications.


Based on first principles calculations, it was originally proposed that Co-doped PbPdO$_2$ can host the SGS state.~\cite{wang-2008-prl}  
However, its Curie temperature (T$_C$) is just about 180K,~\cite{pbpdo2apl} well below the room temperature. 
The first above room temperature SGS was experimentally observed in the inverse Heusler Mn$_2$CoAl (T$_C$=720 K). ~\cite{c-felsher-2013-prl}
Later on, the Heusler compounds have been considered as outstanding candidates for SGSs.
For example, ternary Heusler Ti$_2$MnAl, quaternary Heusler CoFeMnSi, and DO$_3$ type Heusler V$_3$Al have been predicted theoretically 
to be SGSs~\cite{skaftouros-2013-apl, xu-2013-epl, gao-2013-apl} 
and also confirmed by experimental measurements.~\cite{feng-2015-pssppl, bainsla-2015-prb, jamer-2015-prb} 
Interestingly, during the explorations of SGSs in the Heusler compounds, an empirical rule has been discovered. 
That is, the Heusler compounds with 18, 21, 26, or 28 valence electrons are more probable to realize the SGS phase.~\cite{skaftouros-2013-apl, qiang_2015, sg9} 
However, there has been no systematic study to design novel SGS Heusler systems.
Particularly, there are still a few questions about SGSs to be understood. 
For instance, all four types of SGSs should in principle exist but most experimentally studied 
systems are of type-I and type-II.~\cite{skaftouros-2013-apl, xu-2013-epl, qiang_2015}         
A particular intriguing question is the effect of spin-orbit coupling (SOC) on the transport properties of SGSs, {\it i.e.}, whether a band gap can be opened 
with nontrivial topological properties. 
Wang has proposed recently that SGSs are promising for massless and dissipationless spintronics and quantum anomalous Hall effects.~\cite{nsr-wang}  
In this regard, SGSs with direct band touching will be very interesting, since they may host nontrivial topological properties after considering SOC.


On the other hand, high throughput (HTP) screening based on density functional theory (DFT) calculations 
has been proven to be an efficient way to search for materials with desired properties.~\cite{Stefano-Curtarolo-2012-comp-mat-sci, Ingo-Opahle-2013-njp}
Using the AFLOWLIB database, Carrete  {\it et al.} have done HTP calculations on approximately 79, 000 half-Heusler compounds 
and found 75 systems which are thermodynamically stable, where the thermal conductivities and thermoelectric performance have also been evaluated.~\cite{j-Carrete-2014-prx} 
The Heusler compounds with ten valence electrons (X$_2$YZ , X = Ca, Sr, and Ba; Y= Au; Z = Sn, Pb, As, Sb, and Bi) are demonstrated 
to have ultra-low lattice thermal conductivities according to He's HTP calculations.~\cite{jiangang-he-2016-prl}
Whereas, in a more recent HTP study, He {\it et al.} have identified 99 new nonmagnetic semiconductors following the 18 valence electron rule with promising thermoelectric properties.~\cite{he-18}
Furthermore, for spintronics applications, Ma {\it et al.} have performed a systemic HTP study on 405 inverse Heusler alloys resulting in 14 stable semiconductors and 10 half metals.~\cite{ma-arxive} 
Focusing on the magnetic properties, Sanvito {\it et al.} did HTP calculations on 36, 540 Heusler alloys, leading to 248 thermodynamically stable compounds with 20 magnetic cases.~\cite{Sanvito}
Moreover, 21 antiferromagnetic Heusler compounds with high N$\acute{e}$el temperature have been proposed for spintronic applications.~\cite{Jan-2017}
Last but not least, among 286 Heusler compounds, HTP screening calculations suggest 62\% have a tetragonal structure due to the peak-and-valley character in the density of states.~\cite{Parkin-phyrev-appl} 

In this work, we have carried out a systematic HTP screening for SGSs in Heusler compounds (including DO$_3$ binary, ternary, and quaternary Heusler systems). 
Based on the empirical rule, we considered 12, 000 systems with 21, 26, or 28 
valence electrons and identified 80 
novel SGSs, which are thermodynamically stable based on the formation energies. 
Among them, 70 
are both mechanically and dynamically stable.  
It is noted that the Heusler alloys with 18 valence electrons are also promising to realize SGSs, which will be investigated in the future.
We have identified all four types of SGSs in the quaternary Heusler compounds, together with one case showing direct band touching at the Fermi energy.
The longitudinal and transversal transport properties were also evaluated based on the semi-classical transport theory, revealing that 
SGSs are promising materials for spintronic applications.
It is demonstrated that the magnetization direction can be used to tailor the electronic structure and hence the physical properties for SGSs with heavy elements,
due to the anisotropy caused by SOC.

\section{Computational Details}

\begin{figure}[htp]

\begin{center}
\includegraphics[width=8.5cm]{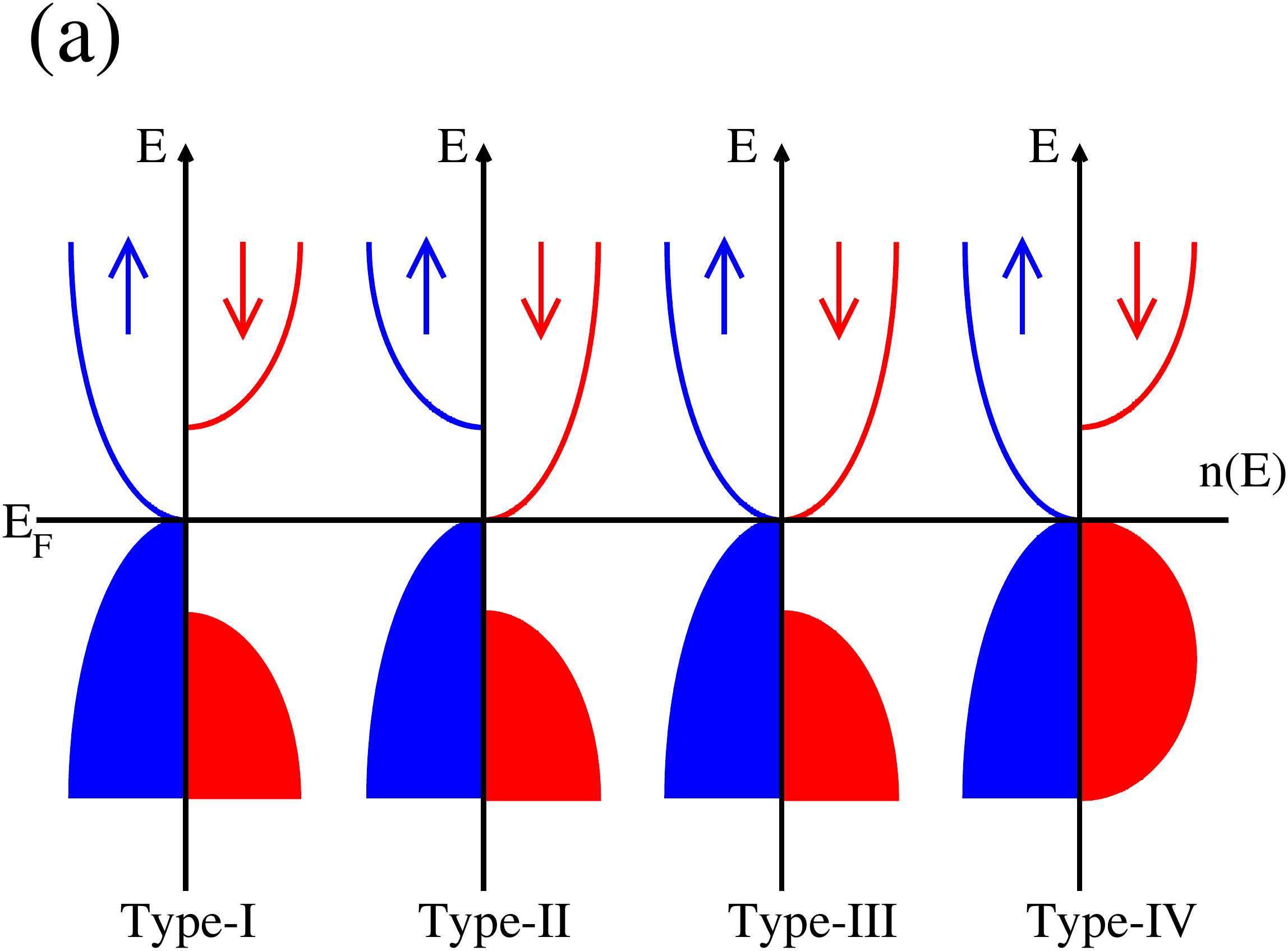}
\end{center}

\vspace{0.15cm}

\includegraphics[width=9cm]{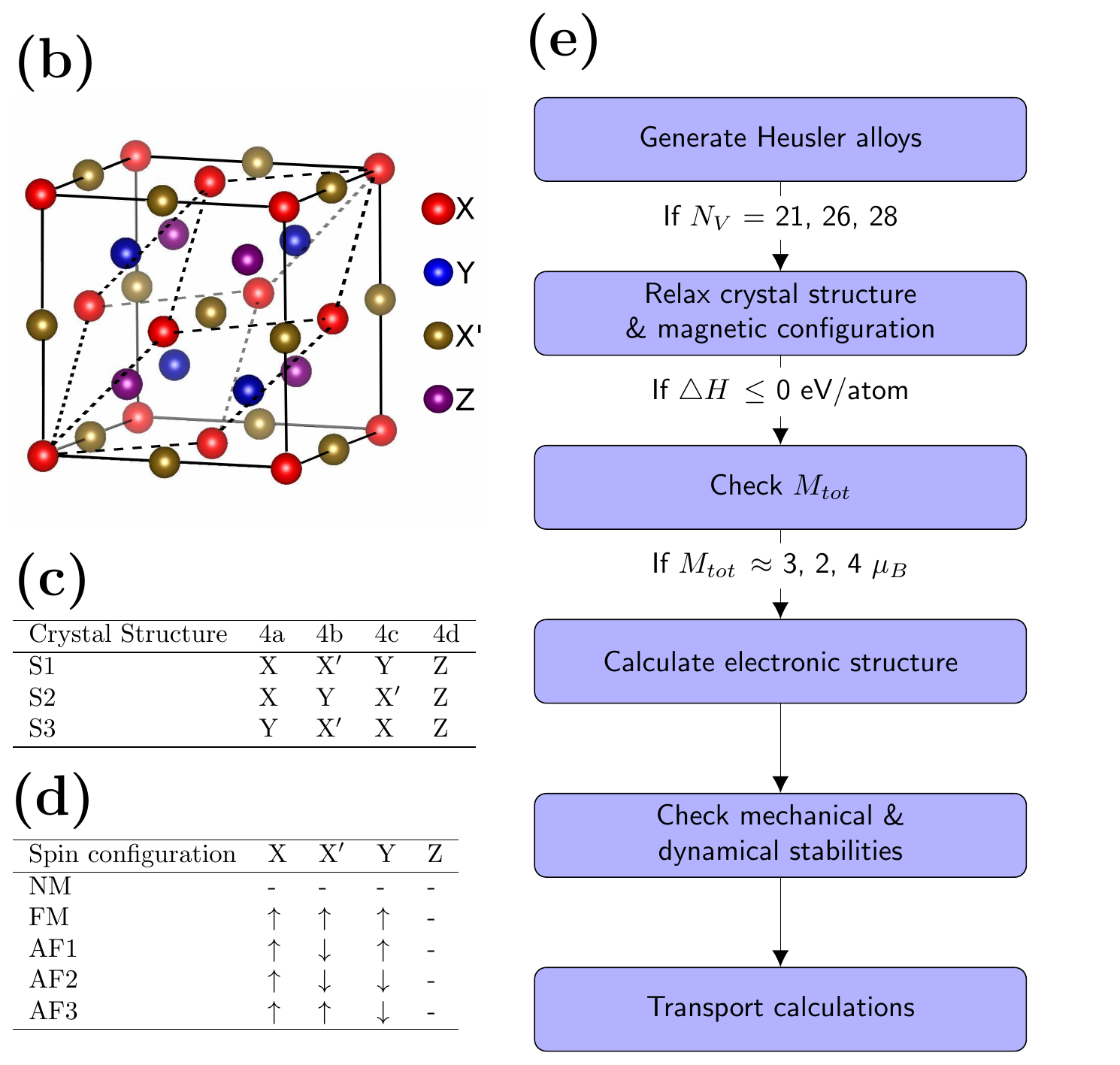}

\caption{(Color online)  (a) Sketches of the density of states for the four types of SGSs, defined based on the touching schemes of
the majority (marked in blue) and minority (marked in red) bands.  
(b) The crystal structure of quaternary Heusler XX$^\prime$YZ, where the solid (dashed) lines indicate the conventional cubic (primitive rhombohedral) cell. 
(c) The three possible site occupations for a quaternary Heusler alloy with a specific chemical composition. 
(d) The possible spin configurations within the primitive cell.
(e) The work flow for the present HTP screening.} 

\label{fig:mixfigure}
\end{figure}

We considered quaternary Heusler compounds with a general chemical formula XX$^\prime$YZ, where X, X$^\prime$, and Y are the transition metal elements except for the radioactive Tc,
and Z is one of the main group elements among B, Al, Ga, In, Si, Ge, Sn, Pb, P, As, Sb, and Bi. 
For convenience, the ternary and binary (DO$_3$-type) Heusler systems are considered as quaternary Heusler by allowing X, X$^\prime$, or Y to be the same element.
As shown in Fig.~\ref{fig:mixfigure}(b), 
quaternary Heusler XX$^\prime$YZ has the so-called LiMgPdSn-type structure with space group F$\bar{4}$3m (space group 216),
consisting of 4 Wyckoff positions 4a(0,0,0), 4c($\frac{1}{4}$,$\frac{1}{4}$,$\frac{1}{4}$), 4b($\frac{1}{2}$,$\frac{1}{2}$,$\frac{1}{2}$), and 4d($\frac{3}{4}$,$\frac{3}{4}$,$\frac{3}{4}$).\cite{heusler-1, heusler-2}  
According to the empirical rule for the number of valence electrons (N$_\text{V}$), all the possible chemical composition with 21, 26, and 28 valence electrons
are generated, leading to about 12,000 possible compounds.  Moreover, three site occupations are considered for each chemical coposition, as shown in Fig.~\ref{fig:mixfigure}(c).~\cite{heusler-stru} Lastly,
we consider that all the transition metal elements (X, X$^\prime$, and Y) are magnetic while the main group element (Z) is non-magnetic (NM).
For each chemical composition in each site-occupation, we 
consider five spin configurations, namely, the NM, FM, AF1, AF2, and AF3 phases (Fig.~\ref{fig:mixfigure}(d)).  

The HTP screening has been carried out in an automated way following the work flow shown in Fig.~\ref{fig:mixfigure}(e), managed with 
our in-house developed high-throughput environment (HTE).~\cite{Ingo-Opahle-2013-njp, ingo-highthrough} 
The DFT calculations are performed using the Vienna ab initio Simulation Package (VASP).~\cite{vasp1, vasp2} 
For each composition-occupation case, the structural relaxation is done in a two-step manner to save computational time. 
In the first step, ultrasoft pseudopotentials (US-PP)~\cite{{chaoyan}} are used in combination with the PW91~\cite{vasp3} exchange correlation functional, where 
the cutoff energy for the plane wave basis is set to 250 eV and  and a k-space density of 40 \AA$^{-1}$.
The follow-up finer relaxation is done using the projector augmented plane wave (PAW) method with the exchange-correlation functional under the generalized gradient approximation (GGA) parameterized by Perdew, Burke, and Ernzerhof (PBE).~\cite{vasp4} 
The cutoff energy for the plane wave expansion is increased to 350 eV and the $k$-mesh density is increased to 50 \AA$^{-1}$ to achieve good convergence.
The structural relaxations are done for each magnetic configuration mentioned above.


After obtaining the magnetic ground state together with the optimized crystalline structures, 
the formation energy ($\bigtriangleup H$) is evaluated to verify the thermodynamic stability, {\it i.e.}, the stability 
with respect to decomposing into constituting elements. 
For a general quaternary Heusler XX$^\prime$YZ, the formation energy is expressed as
\begin{equation}
 \bigtriangleup H^{XX'YZ}=\frac{1}{4}[E^{XX'YZ}-(E^{X}+E^{X'}+E^{Y}+E^{Z})],  
\end{equation}
where  E$^\text{X}$, E$^\text{X'}$, E$^\text{Y}$ and E$^\text{Z}$ are the energies of elements X, X$^\prime$, Y, and Z in their bulk forms, while E$^{XX^\prime YZ}$ is the ground state energy of XX$^\prime$YZ. 
The electronic structure together with the magnetic moments of the compounds with negative formation energies are calculated with a denser $k$-mesh of $21\times 21 \times 21$ using the full-potential local-orbital minimum-basis band structure scheme (FPLO).~\cite{fplo1, fplo2} 
The SGS phase can be identified by examining the value of magnetic moments ({\it i.e.}, being an integer following the Slater-Pauling rule as discussed below) and the 
band structure directly.

For the candidate SGSs, we further checked the mechanical and dynamical stability. The mechanical stability describes the stability of the crystal against deformations or distortions in terms of strain, which
can be obtained based on the elastic constants (C$_{ij}$). Basically, the elastic constants are associated with the second-order change of the internal energy for a crystal under an arbitrary deformation of strain as 
 \begin{equation}
 C_{ij}=\frac{1}{V_0}(\frac{\partial ^2E}{\partial \varepsilon_i\partial \varepsilon_j}),
 \end{equation}
 where E is the internal energy, V$_0$ is the equilibrium volume of the crystal, and $\varepsilon_i$ or $\varepsilon_j$ denote applied strains. 
 For a cubic crystal system (such as Heusler compounds in this work), the elastic constant matrix has only three independent elements as
\begin{equation}
C_{cubic=}{
\left[ \begin{array}{cccccc}
C_{11} & C_{12} & C_{12}&0&0&0\\
 C_{12} &C_{11} & C_{12}&0&0&0\\
C_{12} & C_{12}&C_{11}&0&0&0\\
0&0&0&C_{44}&0&0\\
0&0&0&0&C_{44}&0\\
0&0&0&0&0&C_{44}
\end{array} 
\right ]}.
\end{equation}
Correspondingly, the Born stability conditions~\cite{elstic-huang} suggest
 \begin{equation}
 C_{11}+2C_{12}>0,  C_{11}-C_{12}>0,  C_{44}>0,
 \end{equation}
which are related the bulk, tetragonal, and shear moduli, respectively. 

On the other hand, the dynamical stability describes the change of the total energy with respect to the internal degrees of freedom, {\it i.e.}, the atomic displacements. 
In the harmonic approximation, the total energy of a crystal can be expressed as in terms of displacements $D_{\mathbf{R}\sigma}$
 \begin{equation}
E=E_0+\frac{1}{2}\sum_{\mathbf{R},\sigma}\sum_{\mathbf{R}',\sigma'}D_{\mathbf{R}\sigma}\Phi^{\sigma\sigma'}_{\mathbf{R}\mathbf{R'}}D_{\mathbf{R}'\sigma'},
 \end{equation}
 \hzm{The caption of Figure 1 should be modified}
where $\mathbf{R}$ is the position, $\sigma$ is the Cartesian index, and $\Phi^{\sigma\sigma'}_{\mathbf{R}\mathbf{R'}}$ is the interatomic force constant matrix. 
The dynamical stability is determined by the dynamical matrix $D(\mathbf{q})$, which can be obtained from Fourier transformation of $\Phi(\mathbf{R})$ as following
 \begin{equation}
D(\mathbf{q})=\frac{1}{M}\sum_{R}\Phi(\mathbf{R})e^{-i\mathbf{qR}}
 \end{equation}
where $\mathbf{q}$ is the wave vector of phonon. Dynamical stability indicates that $D(\mathbf{q})$ is positive-definite, meaning all the phonons have real and positive frequencies $\omega(\mathbf{q})$. 
The phonon dispersion calculations are carried out using the Phonopy~\cite{{phonopy}} package with force constants obtained from VASP.

Finally, the transport properties are studied for a few representative SGS candidates, including the anomalous Hall conductivity (AHC)  and the longitudinal conductivity. 
The AHC is calculated by integrating the Berry curvature ($\Omega(\mathbf{k})$) over the whole Brillouin zone (BZ) as $\sigma_{xy}=\frac{e^2}{\hbar}\int_{BZ}\Omega (\mathbf{k})d^3\mathbf{k}$, 
\hzm{Anomalous Hall conductivity is $\sigma_{xy}$, not $\rho_{xy}$.}
with the Berry curvature given by
 \begin{equation}
\Omega_{xy}(\mathbf{k})=2\text{Im}\sum_{n<E_F}\sum_{m \neq n}\frac{<\psi_{n{\mathbf k}}|\nu_x|\psi_{m{\mathbf k}}> <\psi_{m{\mathbf k}}|\nu_y|\psi_{n{\mathbf k}}>} {(\epsilon_{m{\bf k}}- \epsilon_{n{\mathbf k}})^2},
 \end{equation}
where $\psi_{\alpha{\mathbf k}}$ is the spinor Bloch wave function corresponding to the eigenenergy $\epsilon_{\alpha{\bf k}}$, and $\nu_i$ is the $i$-th Cartesian component of the velocity operator.
In our calculations, in order to achieve numerical convergence, the AHC is obtained using the Wannier interpolation technique based on
the Maximally Localised Wannier Functions (MLWFs).~\cite{wannier90}
Furthermore, the longitudinal conductivities at finite temperature (300 K) for SGSs are calculate based on the semiclassical theory with the BoltzTrap~\cite{boltztrap} code. 
Here the energy independent relaxation time ($\tau$) is used to approximate the distribution function as
 \begin{equation}
(\frac{\partial f}{\partial t})_s=-\frac{f-f_0}{\tau}
 \end{equation}
where $f_0$ and $f$ are the equilibrium and nonequilibrium distribution functions, respectively. 
The conductivity is expressed by 
 \begin{equation}
\sigma_{\alpha \beta}(T,\mu) = \frac{1}{V}\int \bar{\sigma}_{\alpha \beta}(\epsilon)[-\frac{\partial f_0(T, \epsilon, \mu)}{\partial \epsilon}]d\epsilon
 \end{equation}
 where $\alpha$ and $\beta$ are the Cartesian indices, $V$ and $\mu$ indicate the unit cell volume and the chemical potential, respectively.
 The transport distribution function $\bar{\sigma}_{\alpha \beta}(\epsilon)$ can be evaluated by 
  \begin{equation}
\bar{\sigma}_{\alpha \beta}(\epsilon) = \frac{e^2}{N}\sum_{i,\mathbf{k}}\tau \cdot \nu_\alpha(i,\mathbf{k}) \cdot \nu_\beta(i,\mathbf{k}) \cdot \frac{\delta (\epsilon-\epsilon_{i,\mathbf{k}})}{d\epsilon}
 \end{equation}
  \begin{equation}
\nu_\alpha (i,\mathbf{k}) = \frac{1}{\hbar}\bigtriangledown_k \epsilon_{i,\mathbf{k}}
 \end{equation}
where $\mathbf{k}$, i, and N are the wave vector, band index, and the number of the sampled $\mathbf{k}$ points.
\hzm{If you list two things, use ``A and B" --- do not use ``A, and B". If you list three or more, use ``A, B, and C". Change the last sentence and do not make such mistakes in the future.}
For bonding analysis, the crystal orbital Hamilton population (COHP) was evaluated using the LOBSTER code.~\cite{cohp}

\section{results and discussion}

\subsection{Validation}

{\tiny
\begin {table}[htp]
\begin{center}
\caption{\footnotesize Comparisons between our HTP calculations and previous reported SGSs. For the mechanical and dynamical stabilities, ``1" (``0") indicates the system is stable (unstable). ``Ref. Exp." and ``Ref. Cal." denote experimental and computational references.}

\vspace{-0.cm}
\begin{tabular}{llllllllllllllll}

\hline
\hline
	Compound & latt.  &M$_{tot}$ &  $\bigtriangleup H$ &Mechanical  &Dynamical  &\\
&  (\AA)&    ($\mu_B$)& (eV/at.)& stability&  stability &   \\

	\hline
	$N_V=21$\\
\hline

						Ti$_2$CoSi & 6.081  & 3.00  & -0.3718 &  0   &     1  &  \\
	Ref. Cal.~\cite{skaftouros-2013-apl}&  6.030 &  3.03 &  &  &   &  \\
		\hline
						MnCrTiSi & 5.855  & 3.02  & -0.4103 & 1&1   & \\
	Ref. Cal.~\cite{kog_jap_2013}& 5.860 &  2.98 &   &&  &  \\
	\hline
										MnCrVAl & 5.897  &3.00  & -0.2110 &  1&1   &\\
	Ref. Cal.~\cite{kog_jap_2013}& 5.900 &  2.99 &  &   &   & 	 \\
	\hline
										MnVTiAs & 5.978  &2.90   & -0.2353 & 1& 1  & \\
	Ref. Cal.~\cite{kog_jap_2013}& 5.990 &  2.87 &  &   &   & 	  \\
	\hline
	
	CoVTiAl & 5.978  &3.00  & -0.3248 &  1& 1  &  \\
	Ref. Cal.~\cite{kog_jap_2013}& 6.040 &  3.00 &  &   &    \\
	\hline

		FeVTiSi & 5.978  &3.02  & -0.4351 & 1&1   &  \\
	Ref. Cal.~\cite{kog_jap_2013}& 5.910&  2.99 &  &    \\
	\hline
	FeCrTiAl & 5.964  &3.02   & -0.2920 & 1&1   &  \\
	Ref. Cal.~\cite{kog_jap_2013}& 5.970 &  3.00 &  &   &    \\
	\hline
	
							CoVHfGa & 6.193  &2.95    & -0.2434 &  1&1 \\
	Ref. Cal.~\cite{yang_2016}& 6.260 &  3.00 &  &  &   & 	 \\
	\hline
												CrFeHfGa & 6.127  &3.00  & -0.1858 &  1 &1  &\\
	Ref. Cal.~\cite{yang_2016}& 6.261 &  3.02 &  &  &   & 	  \\
	\hline
	
								ZrCoVIn & 6.445  &2.97    & -0.0632 &0 &  1  &  \\
	Ref. Cal.~\cite{qiang_2015}& 6.468 &  3.00  &  -0.3500&   & 	  &  \\
	\hline
								ZrFeCrIn & 6.408  &3.02  & 0.0279 &  1&  0    &  \\
	Ref. Cal.~\cite{qiang_2015}& 6.419 &  3.00  &  -0.0325& &  & 	 \\
	\hline
									ZrFeCrGa & 6.177  &3.00   & -0.1690 &1   &  1     & \\
	Ref. Cal.~\cite{qiang_2015}& 6.184 &  3.00 &  -0.2400&   &     &	 \\
	\hline
										ZrFeVGe & 6.199  &3.06  & -0.2069 & 1&   0    &  \\
	Ref. Cal.~\cite{qiang_2015}& 6.210 &  3.00  &  -0.2500&   & 	  &\\

	\hline
$N_V=26$\\
\hline
	Mn$_2$CoAl & 5.729  &2.01  & -0.2666 &  1&1   & \\
	Ref. Exp.~\cite{c-felsher-2013-prl}& 5.798 &  2.00     &   & 	&  \\
		Ref. Cal.~\cite{xing_2008}& 5.760 &  2.00   &   &   \\
	\hline
	CoFeCrAl & 5.692 &2.00   & -0.1931    & 1&1 &\\
	Ref. Exp.~\cite{Nehra}& 5.736 &  2.00     &   & 	 &  \\
		Ref. Cal.~\cite{gao_jac}& 5.710 &  2.00  & -0.2500 & &  &  \\
	\hline
	CoFeCrGa & 5.717 &2.00  & -0.0686 &   1&1  &  \\
	Ref. Exp.~\cite{Bainsla_prb_2015_1}& 5.736 &  2.00      &   & && \\
		Ref. Cal.~\cite{gao_jac}& 5.730 &  2.00    &   & &&	  \\
	\hline	
											CoFeTiAs & 5.835  &2.00   & -0.3615 & 1&1   & \\
	Ref. Cal.~\cite{kog_jap_2013}& 5.850 &  1.99  &  &   &  &	 \\
	\hline
										CoMnCrSi & 5.669  &2.00   & -0.3280 &  1&1  & \\
	Ref. Cal.~\cite{xu-2013-epl}& 5.630 &  2.00  & -0.3750 &   & 	  &  \\
	\hline
	FeMnCrSb & 6.059  &2.00   & 0.0996 &  1&1  & \\
	Ref. Cal.~\cite{xu-2013-epl}& 5.980 &  2.00 &  &   & 	& \\
	\hline
	ZrCoFeP & 5.941  &2.00  & -0.3491   &  0&0&\\
	Ref. Cal.~\cite{qiang_2015}& 5.944 &  2.00 & -0.6500 &   &  & 	 \\
	\hline
	
	$N_V=28$\\
	
			\hline

	CoFeMnSi & 5.597  &4.00   & -0.3833 &  1&1   & \\
	Ref. Exp.~\cite{bainsla-2015-prb}& 5.658 &  4.00 &  &   &   & 	 \\
		Ref. Cal.\cite{dai_jap_2009}& 5.609 &  4.00 &  &   &   &  \\
	\hline
	Mn$_2$CuAl &5.710  &0.00   & -0.1066 &  0&0&  \\
	Ref. Cal.~\cite{luo_2016_jac}& 5.650 &  0.00 &  &  &   & 	 \\
	\hline
	Cr$_2$ZnSi &5.972  &0.00  & 0.08745 & 1 &0    & \\
	Ref. Cal.~\cite{zhang_epl}& 5.850 &  0.00 &  &  &   & 	 \\
	\hline
         Cr$_2$ZnGe &6.123 &0.00   & 0.1898 &  1&1  & \\
	Ref. Cal.~\cite{zhang_epl}& 6.140 &  0.22 &  &  &   & 	 \\
	\hline
         Cr$_2$ZnSn &6.413 &0.00  & 0.3079 &  1& 0 &\\
	Ref. Cal.~\cite{zhang_epl}& 6.530 &  0.14 &  &  &   &  \\

\hline
\hline	

\label{table:literature}
\end{tabular}

\end{center}

\end{table}

}

To validate the stability criteria implemented in the HTE, we collected all the previously reported Heusler SGSs,
and compared with our DFT results (TABLE~\ref{table:literature}).
The lattice constants, total magnetic moments (TABLE~\ref{table:literature}), and the electronic structure (not shown) 
are in good agreement with the literature. 
However, even though the formation energies for most of the reported Heusler SGSs are negative, ZrFeCrIn and Cr$_2$ZnX (X = Si, Ge, and Sn)
turn out to be thermodynamically unstable in our HTP calculations.
For ZrFeCrIn, in the previous calculations,~\cite{qiang_2015} the energies of composite elements with the fcc structure
are considered, which leads to an underestimation of the formation energy. 
This explains also the big difference for the formation energy of ZrCoVIn.
For the Cr$_2$ZnX compounds, only the inverse Heusler structure is considered in Ref.~\onlinecite{zhang_epl}. 
According to our calculations, all three compounds end up with the non-magnetic (NM) full Heusler structure, with positive formation energies.
That is, even though the electronic structure might be interesting with the hypothetical crystal structures, 
the stability should be checked before making valid predictions.
It is noted that for quaternary compounds, the thermodynamical stability with respect to other competing binary, ternary, and quaternary phases, {\it i.e.}, the distance to the convex hull
should also be evaluated. 
We note that 55 previously unknown, thermodynamically stable (low convex hull) quaternary Heusler compounds are discovered among 2,000,000 compounds by using machine learning method.~\cite{apsmeeting}

Furthermore, it is observed that the mechanical stability or the dynamical stability criteria are also critical for some previously predicted compounds. 
For instance, according to our calculations, Ti$_2$CoSi and ZrCoVIn are mechanically unstable, ZrFeVGe is dynamically unstable,
and ZrCoFeP and Mn$_2$CuAl are both mechanically and dynamically unstable.
We note that such compounds may still be synthesized experimentally using molecular beam epitaxy, which
is known to be efficient in obtaining metastable crystalline phases.
For all the systems which have been experimentally synthesized, such as CoFeCrAl, CoFeCrGa, CoFeMnSi, and Mn$_2$CoAl,
we observed that they fulfill all three stability criteria based on our calculations. 
This confirms the reliability of our theoretical framework to do HTP screening for novel SGSs.


{\small
\begin{center}
\begin{table*}
\caption{Basic information of the newly predicted SGS candidates with negative formation energies in our research. The red compounds are either mechanically or dynamically unstable. }
\begin{tabular}{|l|l|l|l|l|l|l|l|l|l|l|lllllllllllllllllllll}

\cline{1-11}
XX$^\prime$YZ& a$_{opt}$&M$_{tot}$&$\bigtriangleup H$&SGS&  &XX$^\prime$YZ& a$_{opt}$&M$_{tot}$&$\bigtriangleup H$&SGS\\
(4a,4b,4c,4d)& (\AA) &($\mu_B$)&(eV/atom) &type& &(4a,4b,4c,4d)& (\AA) &($\mu_B$)&(eV/atom) &type\\
\cline{7-11}\cline{1-5}
\multicolumn{5}{|c|}{N$_\text{V}$=21}& &
						\multicolumn{5}{c|}{N$_\text{V}$=26} \\
	\cline{7-11}\cline{1-5}					
\multicolumn{1}{|c|}{ IrVYSn }& 6.720 & 3.00 &-0.0942&SOC-I&    &\multicolumn{1}{c|}{ CoOsTiSb }& 6.255 & 2.00 &  -0.1635  & I\\
\multicolumn{1}{|c|}{ CoVYSn }& 6.620 & 3.00 &  -0.0862  &II&    &\multicolumn{1}{c|}{ CoFeHfSb }& 6.232 & 2.00 &  -0.2847  & I\\
\multicolumn{1}{|c|}{ CoVScSn }& 6.402 & 3.00 &  -0.2049  &III&    &\multicolumn{1}{c|}{ CoOsZrSb }& 6.453 & 2.00 &  -0.1075  & I\\
\multicolumn{1}{|c|}{ IrVScSn }& 6.518 & 3.00 &  -0.2488  & SOC-II&    &\multicolumn{1}{c|}{ RhFeTiSb }& 6.259 & 1.95 &  -0.3896  & I\\
\multicolumn{1}{|c|}{ RhVScSn }& 6.518 & 3.00 &  -0.2773  &I &    &\multicolumn{1}{c|}{ CoFeTiSb }& 6.074 & 2.00 &  -0.2948  & I\\
\multicolumn{1}{|c|}{ CoVYGe }& 6.377 & 3.00 &  -0.0763  &II &    &\multicolumn{1}{c|}{ IrFeTiSb }& 6.287 & 1.99 &  -0.2932  & III\\
\multicolumn{1}{|c|}{ CoVScGe }& 6.145 & 3.00 &  -0.2749  &II &    &\multicolumn{1}{c|}{ CoRuTiSb }& 6.228 & 2.00 &  -0.3261  & I\\
\cline{7-11}

\multicolumn{1}{|c|}{ IrVScGe }& 6.300 & 3.00 &  -0.3025  & II&    &\multicolumn{1}{c|}{ CoFeNbGe }& 5.961 & 2.00&  -0.2374  & I\\
\multicolumn{1}{|c|}{ RhVScGe }& 6.290 & 3.00 &  -0.3318  & II&    &\multicolumn{1}{c|}{ \color{red}CoOsNbSn }& 6.352 & 2.00 &  -0.0609  & I\\
\multicolumn{1}{|c|}{ RhVYGe }& 6.512 & 3.00 &  -0.1377  &III &    &\multicolumn{1}{c|}{\color{red} CoRuTaSn }& 6.303 & 2.00 &  -0.1268  & I\\
\multicolumn{1}{|c|}{ CoVYSi }& 6.297 & 3.00 &  -0.1077  & II&    &\multicolumn{1}{c|}{ IrFeTaSn }& 6.354 & 1.98 &  -0.1782  & I\\
\multicolumn{1}{|c|}{ CoVScSi }& 6.058 & 3.00 &  -0.355  &II &    &\multicolumn{1}{c|}{ CoOsTaGe }& 6.143 & 2.00 &  -0.0702  & I\\
\multicolumn{1}{|c|}{ IrVScSi }& 6.215 & 3.00 &  -0.4254  & SOC-II&    &\multicolumn{1}{c|}{\color{red} CoOsTaSi }& 6.064 & 1.99 &  -0.2546  &I \\
\multicolumn{1}{|c|}{ RhVScSi }& 6.210 & 3.00 &  -0.4242  &II &    &\multicolumn{1}{c|}{ CoOsTaSn }& 6.332 & 2.00 &  -0.007  & I\\
\multicolumn{1}{|c|}{ RhVYSi }& 6.438 & 3.00 &  -0.1862  & III&    &\multicolumn{1}{c|}{ CoFeTaGe }& 5.938 & 2.00 &  -0.2475  &I \\
\cline{1-5}
\multicolumn{1}{|c|}{ PtVScAl }& 6.369 & 3.00 &  -0.4431  & SOC-I&    &\multicolumn{1}{c|}{ CoFeTaSi }& 5.856 & 2.00 &  -0.4222  & I\\
\multicolumn{1}{|c|}{ PtVYAl }& 6.608 & 3.00 &  -0.2477  & I&    &\multicolumn{1}{c|}{ CoFeTaSn }& 6.154 & 2.00 &  -0.1353  & I\\

\cline{7-11}
\multicolumn{1}{|c|}{ PtVYGa }& 6.600 & 3.00 &  -0.1867  & I&    &\multicolumn{1}{c|}{ IrCoNbAl }& 6.162 & 1.99 &  -0.5563  & I\\
\cline{1-5}
\multicolumn{1}{|c|}{ FeCrHfAl }& 6.142 & 3.00 &  -0.2456  &II &    &\multicolumn{1}{c|}{ IrCoNbGa }& 6.173 & 2.00 &  -0.4043  & I\\
\multicolumn{1}{|c|}{ OsCrHfAl }& 6.299 & 3.00 &  -0.403  & II&    &\multicolumn{1}{c|}{ IrCoNbIn }& 6.360 & 2.00 &  -0.1326  & I\\
\multicolumn{1}{|c|}{ RuCrHfAl }& 6.284 & 3.00 &  -0.4544  &II &    &\multicolumn{1}{c|}{ IrCoTaAl }& 6.140 & 2.00 &  -0.5579  &I \\
\multicolumn{1}{|c|}{ FeCrTiAl }& 5.964 & 3.00 &  -0.292  & II&    &\multicolumn{1}{c|}{\color{red} IrCoTaGa }& 6.150 & 2.00 &  -0.388  & I\\
\multicolumn{1}{|c|}{ FeCrZrAl }& 6.194 & 3.00 &  -0.2156  & III&    &\multicolumn{1}{c|}{\color{red} IrCoTaIn }& 6.336 & 2.00 &  -0.1622  & I\\
\multicolumn{1}{|c|}{ OsCrZrAl }& 6.347 & 3.00 &  -0.3543  & SOC-II&    &\multicolumn{1}{c|}{\color{red} CoCoNbAl }& 5.970 & 2.00 &  -0.4312  & I\\
\multicolumn{1}{|c|}{ RuCrZrAl }& 6.335 & 3.00 &  -0.4154  & III&    &\multicolumn{1}{c|}{\color{red} CoCoNbGa }& 5.968 & 2.00 &  -0.3299  & I\\

\cline{1-5}
\multicolumn{1}{|c|}{ FeCrScSi }& 5.992 & 3.00 &  -0.279  & II&    &\multicolumn{1}{c|}{\color{red} CoCoNbIn }& 6.179 & 2.00 &  -0.0869  &I \\
\cline{7-11}
\multicolumn{1}{|c|}{ FeCrScSn }& 6.364 & 3.00 &  -0.0891  &II &    &\multicolumn{1}{c|}{ IrCoTiPb }& 6.380 & 2.00 &  -0.0571  & I
\\

\multicolumn{1}{|c|}{ FeCrYSi }& 6.236 & 3.00 &  -0.0081  & III&    &\multicolumn{1}{c|}{ IrCoTiSn }& 6.276 & 2.00 &  -0.3789  &I
\\
\multicolumn{1}{|c|}{ OsCrYSi }& 6.386 & 3.00 &  -0.0246  & SOC-III&    &\multicolumn{1}{c|}{ IrCoTiSi }& 5.965 & 2.00 &  -0.6805  &I \\
\cline{1-5}\cline{7-11}
\multicolumn{1}{|c|}{ CoVHfAl }& 6.211 & 3.00 &  -0.2896  & I&    &\multicolumn{1}{c|}{{\color{red} CoRuCrAl }}& 5.848 & 2.01 &  -0.2802  & II
 \\

\multicolumn{1}{|c|}{ IrVHfAl }& 6.346 & 3.00 &  -0.4634  & II&    &\multicolumn{1}{c|}{ NiCrMnAl }& 5.809 & 2.00 &  -0.2127  & III\\

\multicolumn{1}{|c|}{ RhVHfAl }& 6.342 & 3.00 &  -0.3855  & II&    &\multicolumn{1}{c|}{ NiReCrAl }& 5.920 & 1.97 &  -0.1633  & II\\
\multicolumn{1}{|c|}{ CoVZrAl }& 6.258 & 3.00 &  -0.2662  & I&    &\multicolumn{1}{c|}{ CoOsCrAl }& 5.866 & 2.00 &  -0.2412  &II \\
\cline{7-11}
\multicolumn{1}{|c|}{ CoVZrGa }& 6.238 & 3.00 &  -0.2317  &I &    &\multicolumn{5}{c|}{N$_\text{V}$=28}
\\
\cline{1-5}\cline{7-11}
\multicolumn{1}{|c|}{ {\color{red} IrTiZrSn }}& 6.651 & 2.98 &  -0.3335  & II &    &\multicolumn{1}{c|}{ NiFeMnAl }& 5.731 & 4.00 &  -0.2773  & IV  & \\
\cline{7-11}
 \multicolumn{1}{|c|}{ IrTiZrSi }& 6.385 & 2.96 &  -0.4232  &II &    & \multicolumn{5}{c|}{Continue with $N_V=21$}\\

\cline{1-5}
\cline{7-11}
 \multicolumn{1}{|c|}{ FeVNbAl }& 6.117 & 2.99 &  -0.2012  &II& &\multicolumn{1}{c|}{ MnCrNbAl }& 6.077 & 3.00 &  -0.1912  &II\\

\multicolumn{1}{|c|}{ FeVTaAl }& 6.097 & 2.99 &  -0.2202  & II&  &\multicolumn{1}{c|}{ MnCrTaAl }& 6.053 & 2.99 &  -0.2124  & II \\
\cline{1-5}\cline{7-11}
\multicolumn{1}{|c|}{ MnCrZrGe }& 6.157 & 2.99 &  -0.1473  & II  &  &\multicolumn{1}{c|}{ FeVHfGe }& 6.158 & 3.00 &  -0.2094  &II \\

\multicolumn{1}{|c|}{ MnCrZrSi }& 6.076 & 3.00 &  -0.2569  & II & &\multicolumn{1}{c|}{ FeVHfSi }& 6.079 & 3.00 &  -0.3187  & II\\

\multicolumn{1}{|c|}{ MnCrZrSn }& 6.393 & 3.00 &  -0.0593  & II & & \multicolumn{1}{c|}{ FeVHfSn }& 6.386 & 3.00 &  -0.129  & II\\

\cline{1-11}

\end{tabular}
\label{Table:new}
\end{table*}

\end{center}
}

\subsection{New SGS candidates}

Having confirmed the SGSs reported in the literature, HTP calculations are done following the work flow (Fig.~\ref{fig:mixfigure}(e)). 
In total, we have identified 80 new SGS candidates with negative formation energies, as summarized in TABLE~\ref{Table:new}.
Among such candidates, 70 systems are also mechanically and dynamically stable (TABLE~\ref{Table:new}).
 More detailed information about the mechanical stabilities (including elastic constants), dynamical stabilities, and, magnetic moments (including partial magnetic moments) are summarized in the big table of \textbf{Section S1} in the Supplemental Material, and the band structures shown in \textbf{Section} \textbf{S3} and \textbf{S4} of the Supplemental Material.

In the following, we will focus on the 70 cases which fulfill all the three stability criteria (TABLE~\ref{Table:new}).
It is observed that all four types of SGSs as sketched in Fig.~\ref{fig:mixfigure} (a) can be found, namely, there are 28 (32, 9, and 1) type-I (type II, type-III, and type-IV) SGSs, respectively.


According to TABLE~\ref{Table:new}, most newly predicted SGSs are quaternary Heusler compounds.
We have only found three new ternary SGSs ({\it e.g.}, Co$_2$NbX (X = Al, Ga, and In), but they are either dynamically or mechanically unstable. 
As shown in TABLE~\ref{table:literature}, there are six ternary Heusler ({\it e.g.}, Ti$_2$CoSi, Mn$_2$CoAl, Mn$_2$CuAl, Cr$_2$ZnX (X=Si, Ge, and Sn)) 
proposed to be SGSs. 
It is mentioned above that the Cr$_2$ZnX  (X=Si, Ge, and Sn) would be more stable in the full Heusler structure, in contrast to previous calculations.~\cite{zhang_epl}
The Ti$_2$CoSi and Mn$_2$CuAl are both mechanically unstable, thus dedicated experimental techniques such as molecular beam epitaxy 
should be used to synthesize such compounds.
In this regard, Mn$_2$CoAl is a special case.
Our analysis on systems with the same number of valence electrons as Mn$_2$CoAl, such as Mn$_2$FeSi, reveals that
the CBM and VBM have very strong overlap, destroying the SGS behavior. 
Generally speaking, the occurrence of the SGS phase depends significantly on detailed hybridization of the atomic orbitals, as discussed below. 
Therefore, the empirical rule on the number of electrons serves only as a qualitative guide, and explicit DFT calculations
on the electronic structure are required to identify such phases.

\subsubsection{Magnetization}

Essentially, SGSs are half metals thus the total magnetic moments are expected to be integers, and they should obey the Slater-Pauling rule.~\cite{kog_jap_2013, qiang_2015}
According to TABLE~\ref{Table:new}, it is obvious that when N$_\text{V}$ is 26 and 28, the resulting magnetic moments are 2.0 $\mu_B$ and 4.0 $\mu_B$ following $M_{tot}=(N_{V}-24)\mu_B$, 
where $M_{tot}$ and $N_{V}$ are the total magnetic moment and number of valence electrons per unit cell, respectively. 
For the cases with N$_\text{V}$ being 21, the total magnetic moments are 3.0 $\mu_B$ following $M_{tot}=(N_{V}-18)$. 
This is consistent with the expected values based on the Slater-Pauling rule.

Such behaviors of the magnetization for Heusler compounds can be understood based on the atomic models, as demonstrated in previous studies.~\cite{kog_jap_2013, qiang_2015}
Generally, the magnitude of the magnetic moments is caused by the competition between the crystal field splitting (between t$_{2g}$ and e$_g$ states) and the 
exchange splitting (between the majority and minority spin channels).~\cite{apl-material, heusler4, parking_progress}
In Ref.~\onlinecite{kog_jap_2013}, a picture with bonding and antibonding t$_{2g}$ and e$_g$ bands is applied to interpret the 
quaternary Heusler compounds with one magnetic ion, due to significant hybridization between the $d$-orbitals. 
It is realized that such a picture has to be generalized in order to understand the magnetization of the quaternary Heusler SGSs,
especially for cases with more than one types of magnetic atoms.


The $t_{2g}$-$e_g$ picture is valid for compounds with one magnetic ion.
For instance, as shown by the density of states for PtVYAl (Fig.~S1(a) in the Supplementary), 
the $t_{2g}$ states in the majority spin channels are occupied, resulting in a total magnetization of 3.0 $\mu_B$ per formula unit.
This is generally true for other cases with N$_V$=21, such as XVScSn (X = Co, Ir, and Rh), PtVYAl, and FeCrScSi.
For the N$_V$=26 cases, the t$_{2g}$ shells in both spin channels are filled, while the e$_g$ state is only occupied in the majority spin channel, leading to a total magnetization of 2.0 $\mu_B$ per formula unit,
as demonstrated by IrFeTiSb (Fig.~S1(b) in the Supplementary).
  
  \begin{figure}[htp]

\includegraphics[width=8.cm]{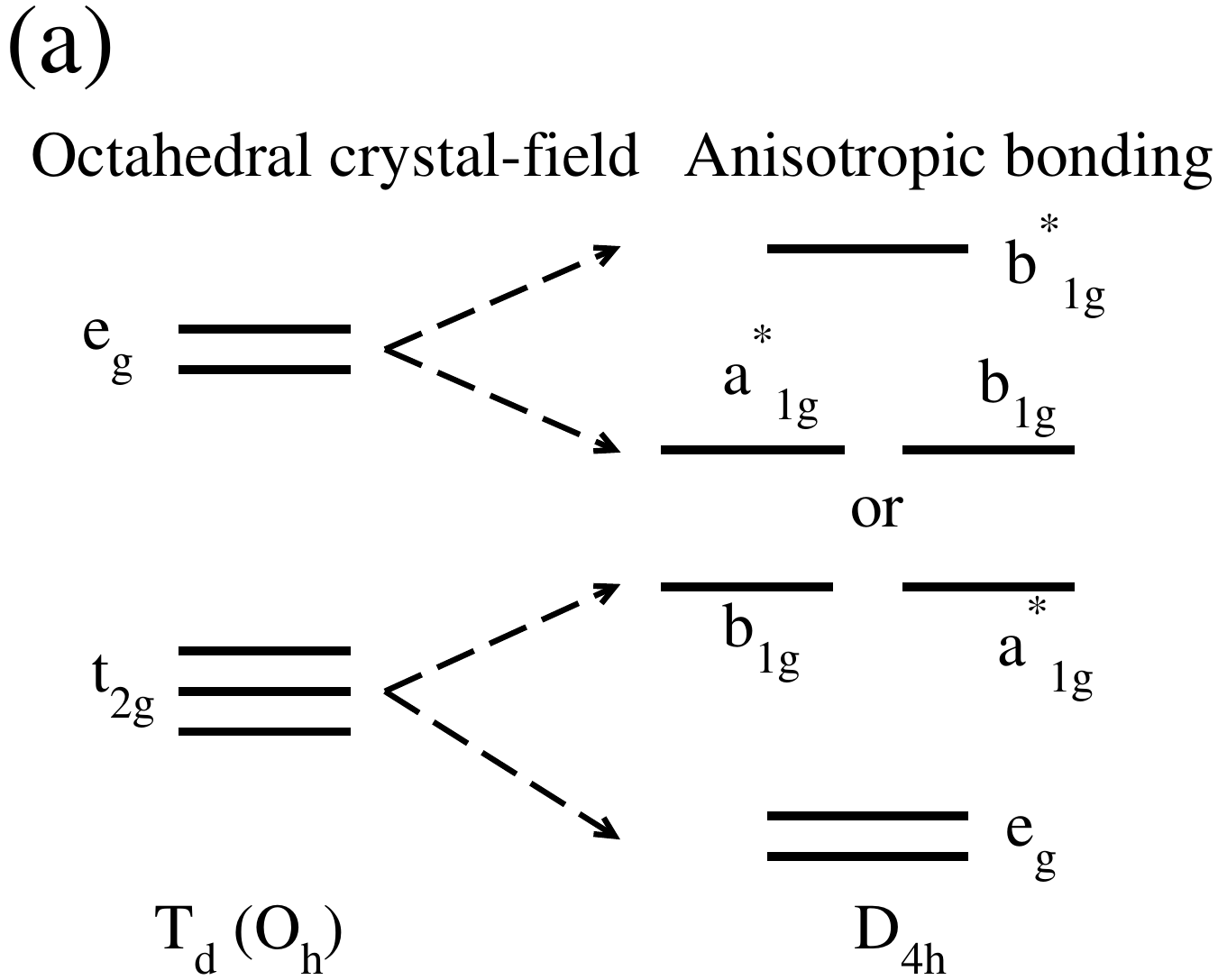}

\vspace{0.25cm}

\includegraphics[width=8.cm]{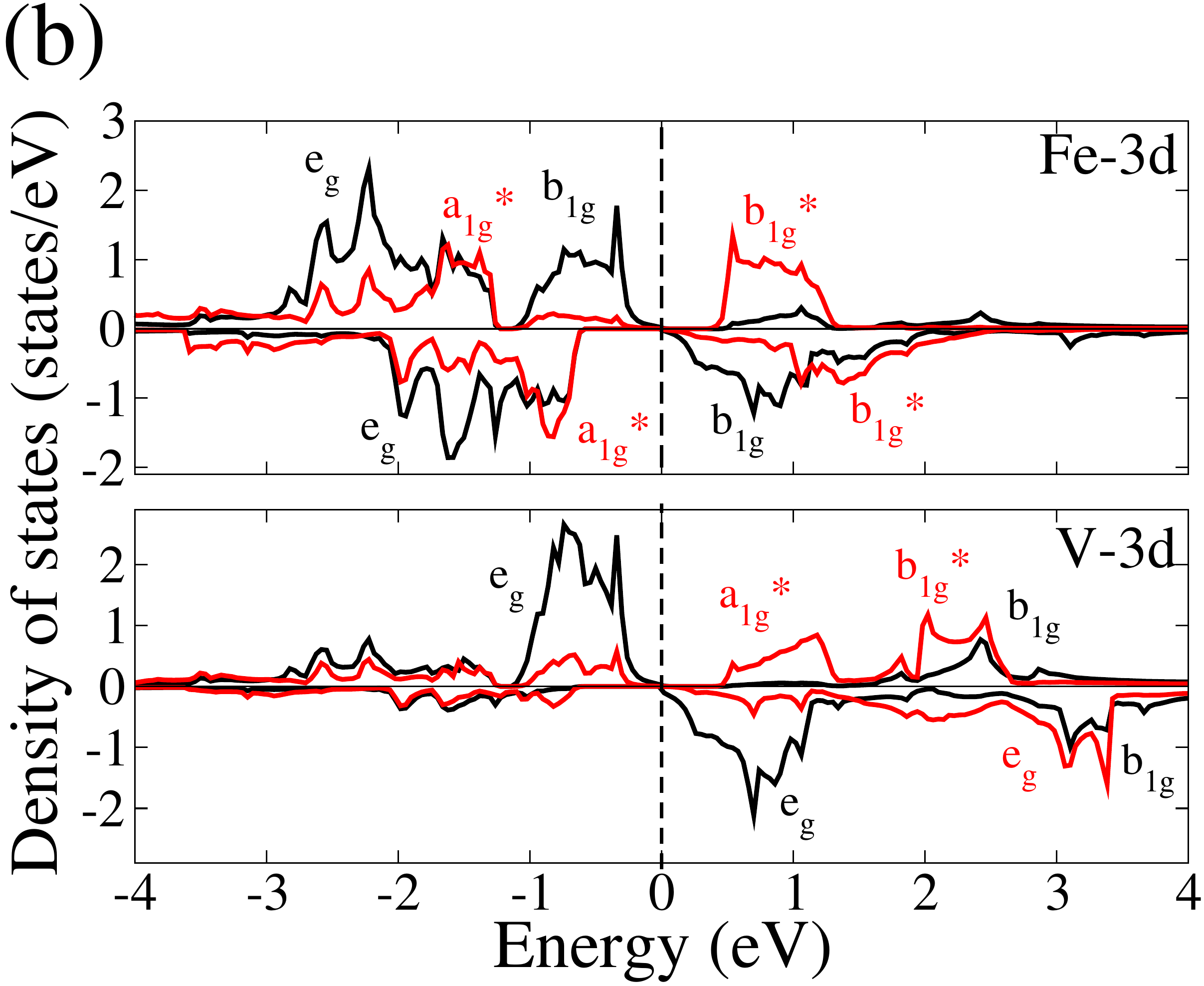}

\caption{(Color online) (a)  Orbital splitting in octahedral crystal field environment and the corresponding anisotropic bonding environment. Although the energy level of original e$_g$ is higher than that of original 
$t_{2g}$. The resulting $a_{1g}^*$ and  $b_{1g}^*$ may be not higher than the $b_{1g}$ in principle due to the energy level is shifted.
(b)  Local density of states (LDOS) for two magnetic ions in FeVNbAl. The black and red curves are the original $t_{2g}$ and $e_g$ shells. Up: LDOS for Fe with one magnetic moment. Down: LDOS for V with 2 magnetic moments.  }

\label{fig:crysalfield}
\end{figure}

The crystal field splittings changes greatly for systems with two or more magnetic atoms. 
It is well known that for full Heusler with chemical formula X$_2$YZ, the site symmetry for both Y and Z are m$\bar{3}$m (O$_h$), while that for X is $\bar{4}$3m (T$_d$).
The $d$-shell will split into $t_{2g}$ and $e_g$ subshells in both O$_h$ and T$_d$ crystallographic symmetries, where the difference is that the $t_{2g}$ is lower (higher) in
 energy for the O$_h$ (T$_d$) case, respectively.~\cite{Galanakis_2002}
For quaternary Heuslers, the site symmetry for X, X$^\prime$, Y, and Z sites is the same, {\it i.e.}, of the T$_d$ type. 
However, it is observed that the $e_g$-$t_{2g}$ picture is not applicable in Heusler alloys with two magnetic ions.
Taking FeVNbAl as an example, for both Fe and V atoms, as indicated by the partial density of states shown in Fig.~\ref{fig:crysalfield}(b), 
the crystal fields behave like the case with D$_{4h}$ symmetry, where the $t_{2g}$ shell splits into
a two-fold $e_g$ and an one-fold $b_{1g}$ subshells, and the $e_g$ shell into $a_{1g}^*$ and $b_{1g}^*$ (Fig.~\ref{fig:crysalfield}(a)).
 In the octahedral crystal field environment, it is already clear that the $t_{2g}$ orbitals are lower in energy than the $e_g$ orbitals.
 For the quaternary Heusler compounds, it is observed that the $b_{1g}$ states originated from the $t_{2g}$ shell can be either higher or lower in energy than the $e_g$-derived $a_{1g}^*$  subshells (Fig.~\ref{fig:crysalfield}(a)), as discussed in detail below.

Such a pattern of crystal field splitting can be attributed to the bonding strength of different atomic pairs.
For FeVNbAl, although the nearest neighbor V-Nb bond length (2.64~\AA) is the same as that of the nearest neighbor V-Al bonds,
the integrated COHP for the V-Nb bonds is about -4.04 eV, which is much stronger than the V-Al bonding with an integrated COHP of -1.33 eV. 
Moreover, the next nearest neighbor V-Fe bond length is about 3.05~\AA, but the integrated COHP is about -1.62 eV, which is comparable to that of the V-Al bonds.
Such features can be clearly observed from the DOS (cf. Fig.~S2(a) in the Supplemantary), where the hybridization between the $d$-orbitals of V and Nb is obviously strong.    
Such splittings of the original $t_{2g}$ and $e_g$ shells are not resulting into the separation of the \{d$_{xy}$, $d_{yz}$, $d_{zx}$\} (due to local tetragonal crystal fields) or \{$d_{x^2-y^2}$, $d_{3z^2-r^2}$\} (due to Jahn-Teller like distortions) orbitals (cf. Fig.~S6 in the Supplemantary),
like the local tetragonal distortions on the $d$-orbitals in the octahedral environment.
In contrast, the \{$d_{xy}$, $d_{yz}$, $d_{zx}$\} orbitals are still degenerated in the subshells $e_g$ and $b_{1g}$, which is the same for
the \{$d_{x^2-y^2}$, $d_{3z^2-r^2}$\} orbitals in the $a_{1g}^*$ and $b_{1g}^*$ subshells (cf. Fig.~S6 in the Supplemantary).

Following such a splitting scheme, the resulting magnetic moments for compounds with two magnetic elements can be easily understood. 
For the N$_V$=21 cases such as FeVNbAl (cf. Fig.~S2 (a) and Fig.~S4 in the Supplementary), 
the magnetic moment of 2.0 $\mu_B$ on the V atoms is as a result of the occupied $e_g$ subshell (originated form t$_{2g}$ shell) in the majority spin channel (Fig.~\ref{fig:crysalfield}(b)).
Moreover, for the Fe atoms,  the $e_g$, $b_{1g}$, and $a_{1g}^*$ subshells in the majority spin channel are occupied, while 
only the $e_g$ and $a_{1g}^*$ subshells in the minority spin channel are occupied, resulting in a magnetic moments of 1.0 $\mu_B$.
It is noted that in this case the $b_{1g}$ subshells can be higher in energy than the $a_{1g}^*$ subshells.
The magnetization of other two magnetic ion compounds with N$_V$=21 can also be understood in the similar way (not shown).


The similar picture can also be applied to the N$_V$=26 cases with two magnetic ions, where the total magnetic moments of 2.0 $\mu_B$ can be
attributed to 1.0 $\mu_B$ atomic moments from two atoms.
Here we take CoFeTaGe (cf. Fig.~S2 (b) and Fig.~S5 in the Supplementary) as an example. 
The bond lengths of nearest neighbor Co-Ge and Fe-Ge almost have the same value as  2.57 \AA. However the integrated COHP of Co-Ge is -1.62 eV, which is larger than that of Fe-Ge (-1.03 eV). 
Moreover, the next nearest neighbor Co-Fe bonds have a comparable bond length to the Co-Ge and Fe-Ge bonds (about 2.96 \AA), but a much weaker bonding with integrated COHP as -0.50 eV.
The resulting crystal field splittings are very comparable to those in the cases with N$_V$=21 case 
(cf. Fig.~S5  in the Supplementary).
For the Co atoms, the only unoccupied state is the $b_{1g}^*$ sub-shell in the minority spin channel, whereas the majority spin channel is fully occupied,
resulting in 1.0 $\mu_B$ magnetic moment. 
For the Fe atoms, the $t_{2g}$ is not split in both spin channels and lies below Fermi level. The  $e_g$ state is weakly split into $a_{1g}^*$ and $b_{1g}^*$ subshells below Fermi level in the majority spin channel. On the other hand, in the minority spin channel the $e_g$ state is also split into a wide spread $a_{1g}^*$ subshell below Fermi level and a localized  $b_{1g}^*$ above Fermi level. So the majority spin channel also has one more state than the minority spin channel, resulting in one $\mu_B$ magnetic moment (cf. Fig.~S2(b) and Fig.~S5 in the Supplementary). 
 
 In short, it is observed that the magnetization of quaternary Heusler compounds with  two magnetic ions can be understood based on
 the crystal splittings of the D$_{4h}$ like picture. Such splittings are originated from the anisotropic bonding between the ions, which breaks the local 
 O$_h$ (T$_d$) symmetry. 
 In this regard, the required band filling to achieve SGSs is more flexible for quaternary Heusler compounds than the ternary cases. 
 This explains also why we found more candidate SGSs in the quaternary Heusler systems, as mentioned above.
 On the other hand, as to the the only new SGS (NiFeMnAl) with 28 electrons (cf. Fig.~S3 in the Supplementary), the hybridization between $d$-orbitals from the Ni, Fe, and Mn atoms 
 is so strong that the atomic picture is not applicable.
This is also true for the other cases with three magnetic ions. 
  
 


\subsubsection{Properties of four types of SGSs}

\begin{center}
\begin{figure*}

\includegraphics[scale=0.52]{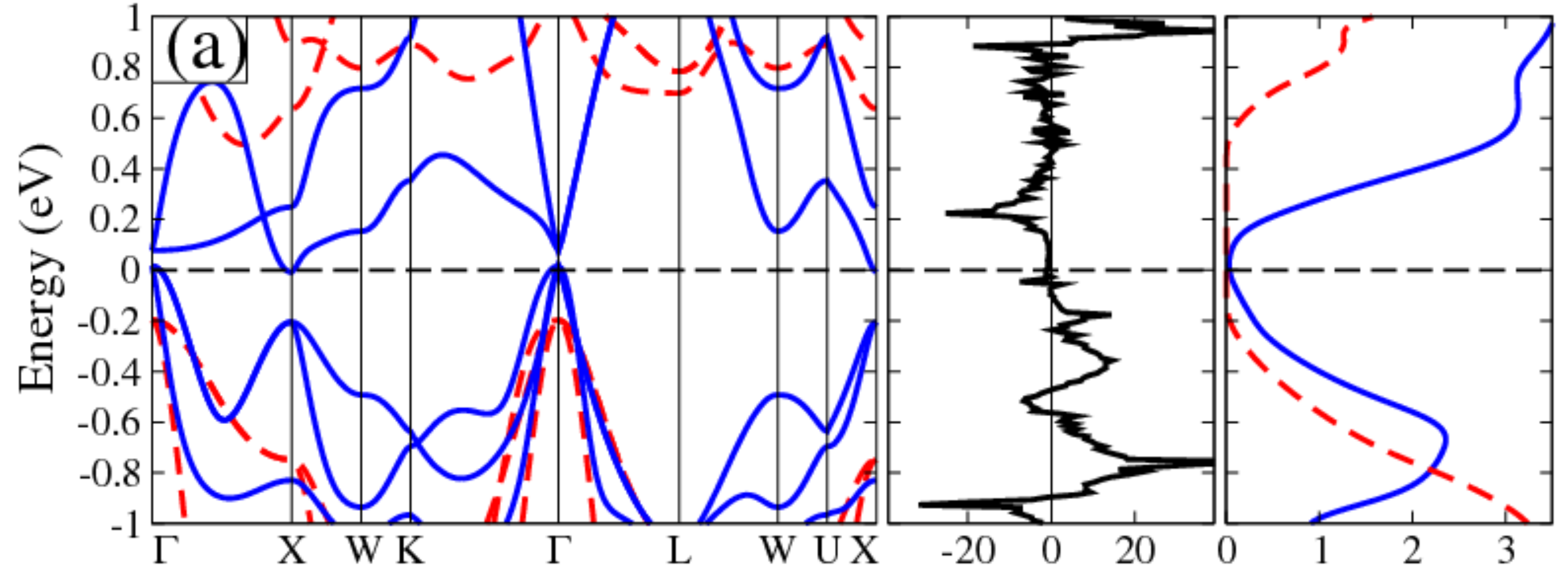}

\vspace{0.25cm}

\includegraphics[scale=0.52]{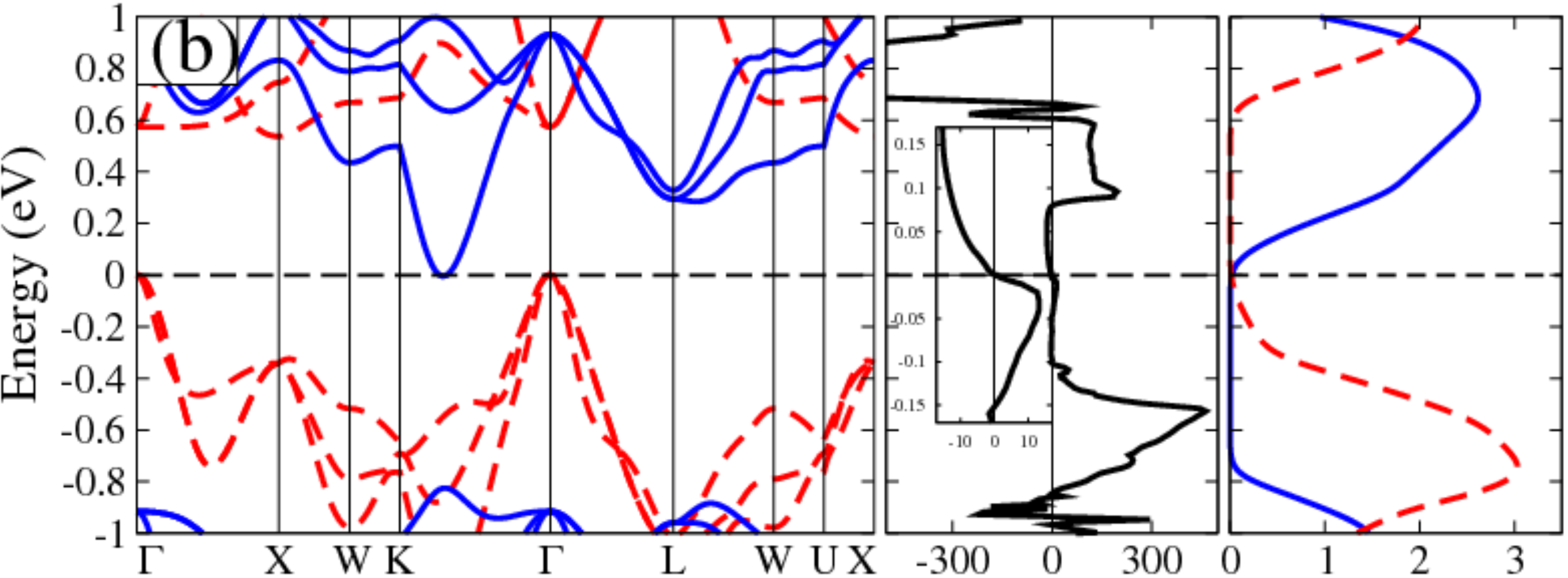}

\vspace{0.25cm}

\includegraphics[scale=0.52]{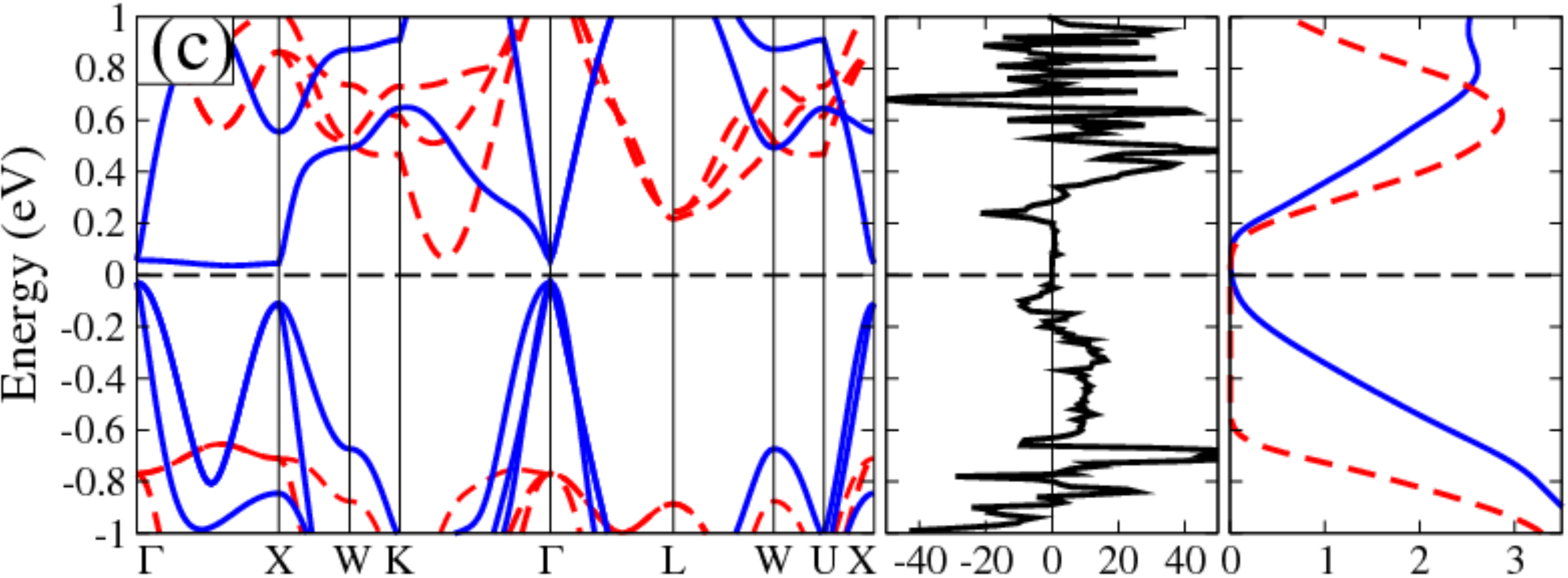}

\vspace{0.25cm}

\includegraphics[scale=0.52]{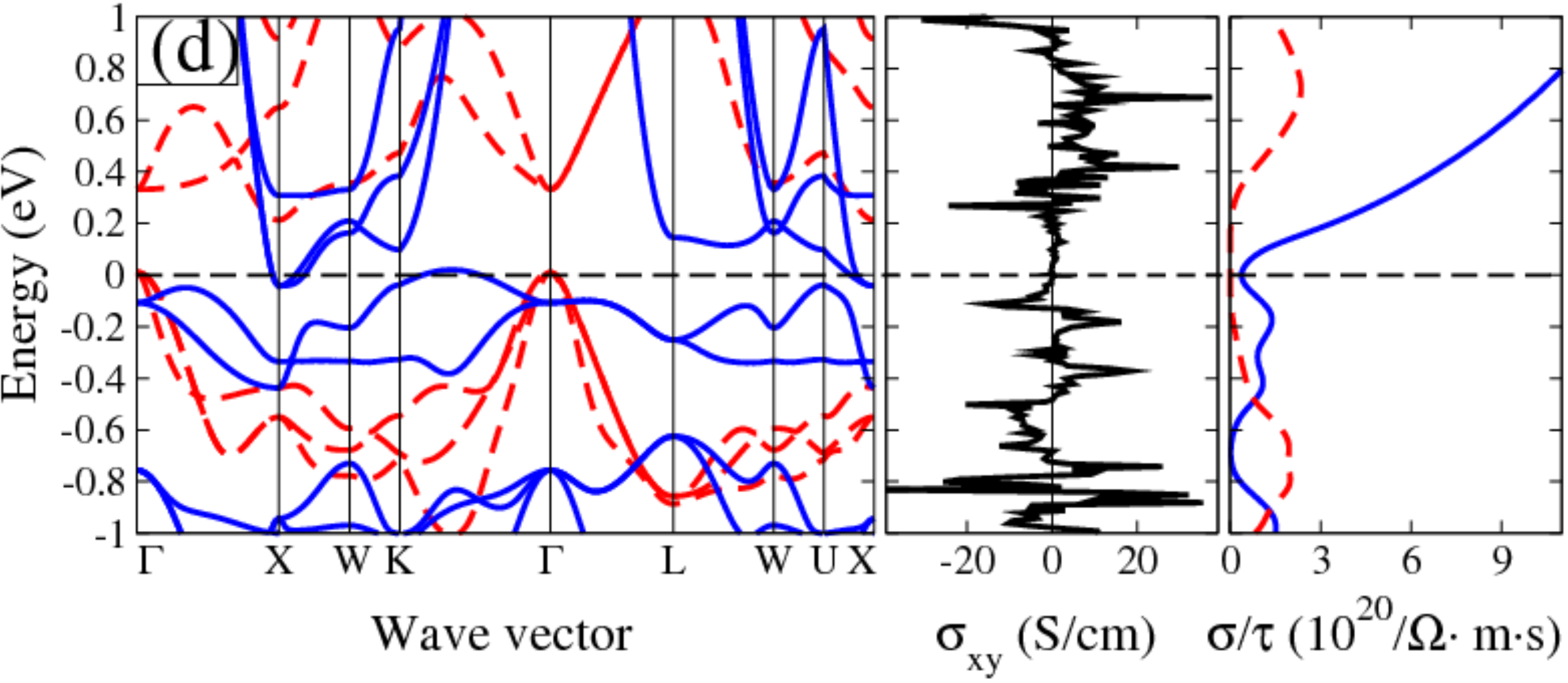}

\caption{(Color online) (a), (b), (c), and (d) are the band structure (left), anomalous hall conductivity (middle), and spin resolved longitudinal conductivity (right) of PtVYAl (type I), FeVHfSi (type II), RuCrZrAl (type III), and NiFeMnAl (type IV), respectively.
The insert in the middle panel of (b) displays the zoom-in of AHC $\pm$0.17 eV around the Fermi energy. 
The blue and red lines denote the majority and minority spin channels, respectively. The horizontal dashed lines indicate the Fermi energy.}
\label{fig:normalsgs}
\end{figure*}
\end{center}

For all four types of SGSs, the electronic structures for one representative case in each class are shown in Fig.~\ref{fig:normalsgs}, 
together with the AHCs and spin-resolved longitudinal conductivities.
For PtVYAl, which is a type-I SGS, VBM (at the $\Gamma$ point) and CBM (at the X point) touch with each other indirectly
in the majority spin channel, while there is a gap about 0.6 eV in the minority spin channel. 
Thus, the system is expected to show typical behavior of half metals, {\it i.e.}, 100\% spin polarized transport properties.
For type-II SGS as exemplified by FeVHfSi, the VBM and CBM have the opposite spin characters and touch with each other indirectly at the Fermi level (Fig.~\ref{fig:normalsgs}(b)). 
In this case, the spin polarization of the resulting current can be tuned by tailoring the Fermi energy.
For RuCrZrAl which represents type-III SGSs, the valence bands near the Fermi energy are mostly of the majority spin character, 
while the conduction bands constitute both majority and minority spin character carriers.
This is in contrast to the case of NiFeMnAl (a type-IV SGS), where the conduction bands are originated from one spin channel while the valence bands have both majority and minority spin characters.

Such specific electronic structures for four types of SGSs can be reflected in the transport properties in terms of the 
AHC and longitudinal conductivity, shown as well in Fig.~\ref{fig:normalsgs}.
Due to vanishing DOS at the Fermi energy, a common phenomenon for the four representative SGSs 
is that the AHC vanishes at Fermi level.
That is, the indirect band gaps for such compounds are topologically trivial, {\it i.e.}, there exist no quantum anomalous Hall effect.
This is comparable with the experimental AHC of Mn$_2$CoAl (also a type II SGS).~\cite{c-felsher-2013-prl}
Moreover, for type-II SGSs such as HfVFeSi, there is a sign change for AHC around the Fermi energy, 
due to the fact that the spin character of the carriers changes when they are excited from VBM to CBM. 
The resulting derivative of AHC is as high as 1597 S/(cm$\cdot$eV), corresponding to a large anomalous Nernst effect (ANE).
In this sense, such type-II SGSs are likely promising candidates for engineering spintronic field-effect transistors.

The right panels of Fig.~\ref{fig:normalsgs} display the spin resolved longitudinal conductivities at 300 K for four types of SGSs. Like the AHC, the longitudinal conductivities of all four SGSs are quite low due to the vanishing DOS at the Fermi energy. 
For type-I SGSs as exemplified by PtVYAl, the conductivity is mostly originated from the majority spin channel, showing typical behavior of half-metals. 
For type-II SGSs (FeVHfSi), due to the VBM and CBM with opposite spin characters, the spin polarization of the longitudinal conductivity can be conveniently tuned by controlling the chemical potential.
Such compounds may be used to fabricate spin valves which are switchable via electrostatic gating.
In case of type-III SGSs (RuCrZrAl), above the Fermi energy the conductivity has finite values for both spin channels, while the conductivity is nonzero only for one spin channel 
below the Fermi energy (Fig.~\ref{fig:normalsgs}). Such transport property is opposite to that of the type-IV SGSs (Fig.~\ref{fig:normalsgs}(d)). 
It is an interesting question how such two types of SGSs can be utilized for future spintronic devices.

\subsubsection{Effects of spin-orbit coupling}

\begin{figure}[t]
\centering
\includegraphics[scale=0.52]{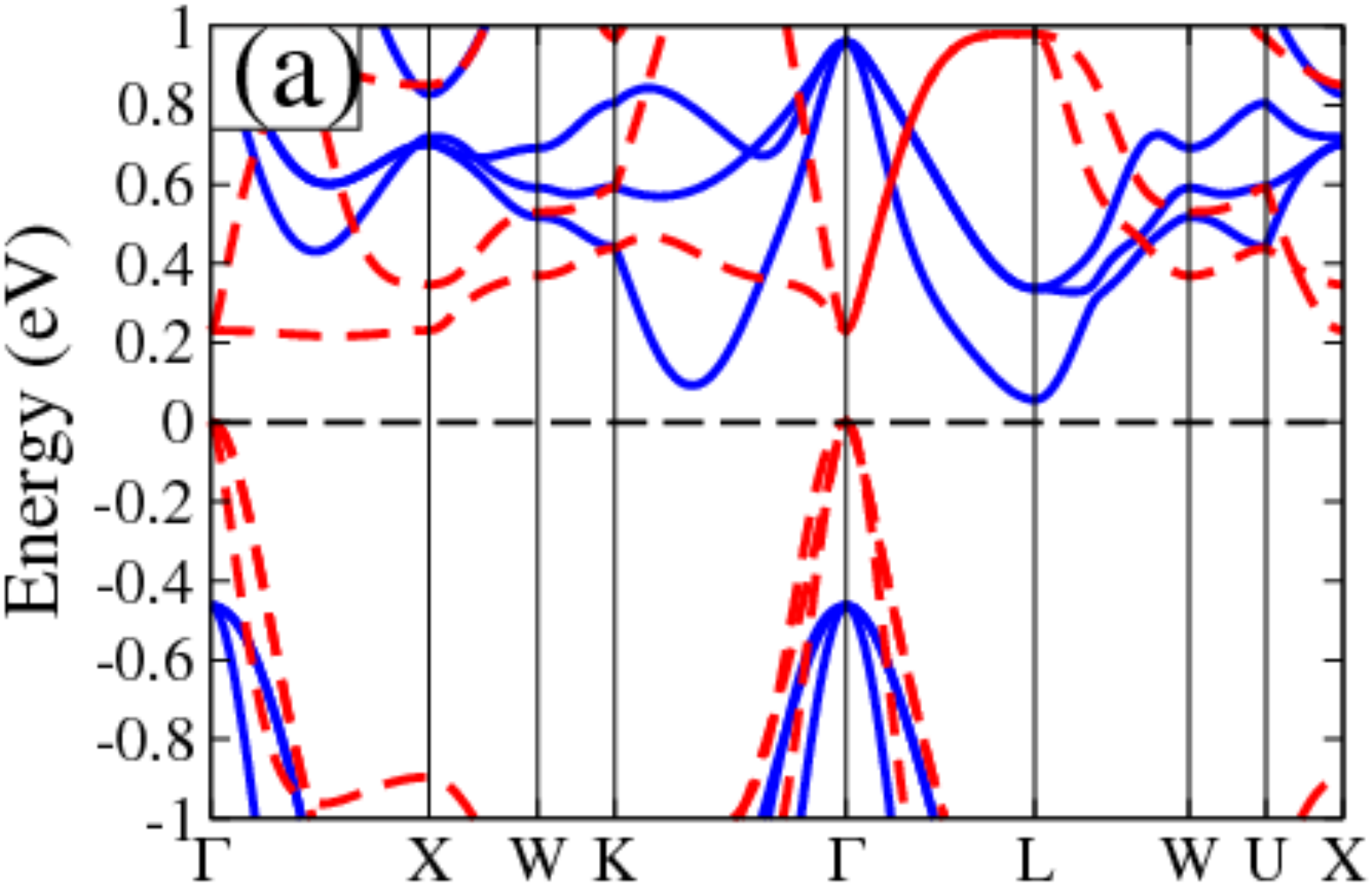}

\vspace{0.25cm}

\includegraphics[scale=0.52]{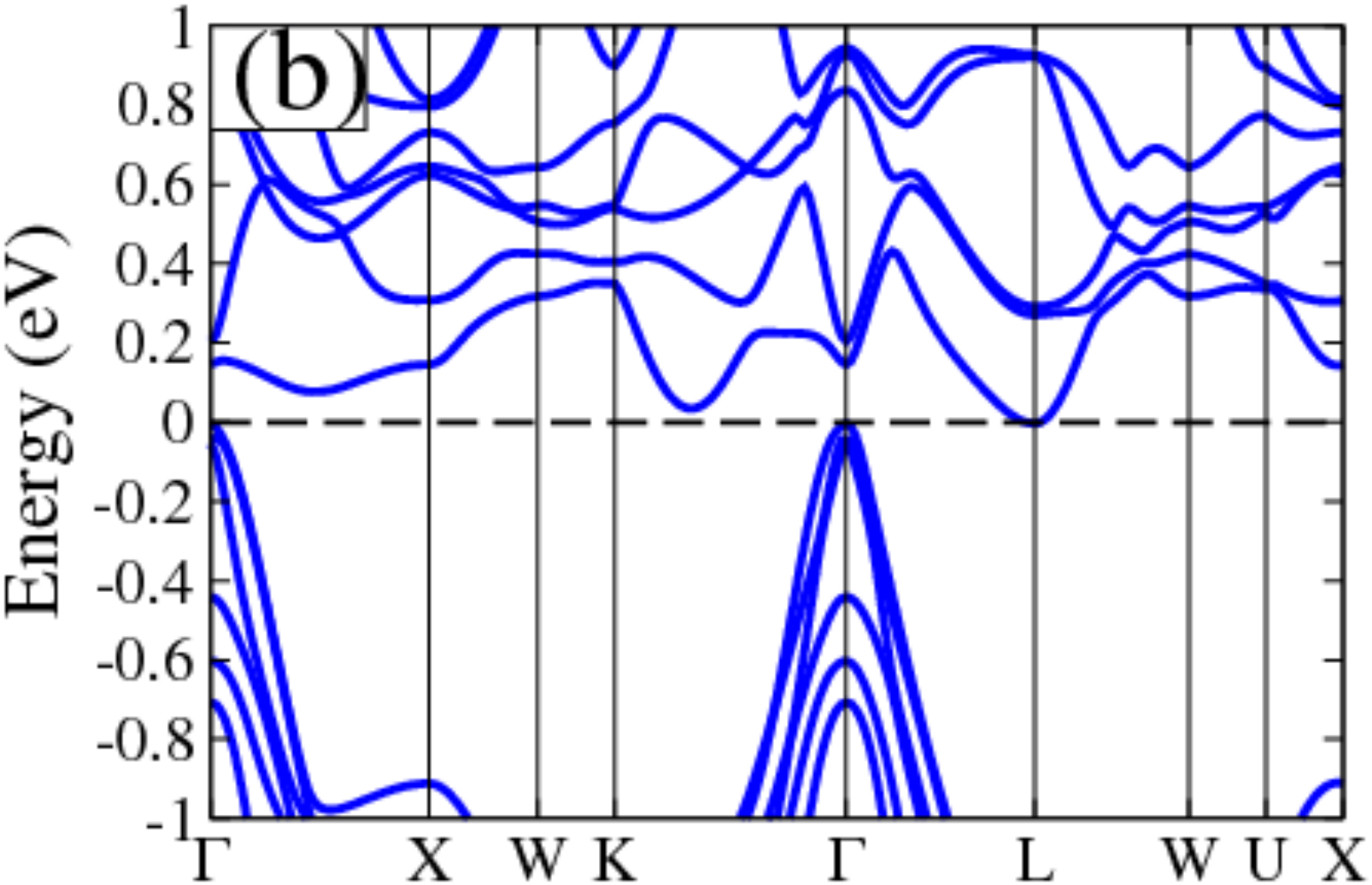}

\vspace{0.25cm}

\includegraphics[scale=0.52]{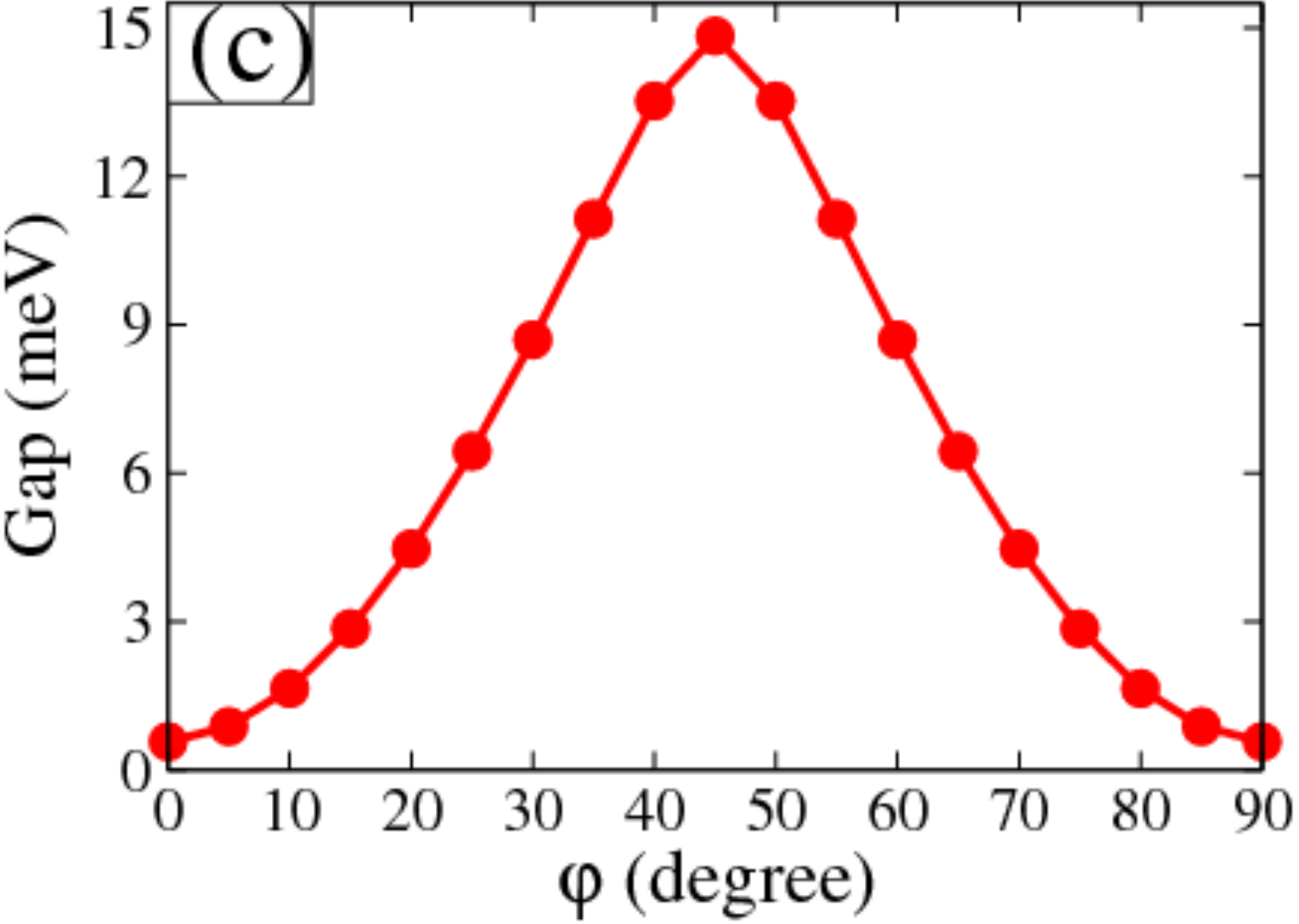}

\caption{(Color online) (a) and (b) are the band structures of IrVScSn without and with SOC. Without SOC, the blue and the red curves are majority and minority spin channels, respectively.  (c) The calculated gap as a function of azimuthal angle $\varphi$ (the angle
between the magnetization direction and the [100] axis) in the (001) plain. The horizontal dashed lines indicate the Fermi energy.}

\label{fig:socsgs}
\end{figure}

It is observed that SOC can induce significant changes in the electronic structure of SGSs, 
since the band gaps of SGSs are on average of small magnitude (cf. Section S4 in Supplemental Materials). 
For instance, for IrVScSn, the indirect band gap is about 58.4 meV without SOC (Fig.~\ref{fig:socsgs}(a)). 
When SOC is turned on with magnetization direction along the [001] direction, the band gap 
is reduced to only 0.6 meV. 
Such a large change in the magnitude of the band gap can be attributed to the fact that the CBM is 
mainly derived from the Ir-$d$ orbitals, where the atomic SOC strength is about 0.5 eV.  
Such SOC effect on electronic structure is particularly associated with compounds constituted of heavy elements such as Os, Ir, and Pt, due to the
strong atomic SOC strength.
Similarly, it is expected that SOC has significant influence on the electronic structure for compounds 
with heavy elements such as Os, Ir, and Pt.
This is indeed confirmed by our explicit DFT calculations for IrVScSn,  IrVScSi, IrVYSn,  PtVScAl, and OsCrZrAl   (cf.\textbf{ Section S4} in the Supplemental Materials),
where the band gap size can be fine tuned by about 15 meV on average.
 As to OsCrYSi, the gap is even closed and the CBM and VBM are overlapping. Here we select IrVScSn as an example for SOC effect on SGSs.


As the SGSs are magnetic, the combination of magnetic ordering with SOC lowers the symmetry of the systems,
leading to magnetization direction dependent physical properties. 
Fig.~\ref{fig:socsgs}(c) shows the magnetization direction dependence of
the band gap for IrVScSn, as the magnetization direction rotates in the (001) plane.
Obviously, the magnitude of the band gap shows a monotonous behavior of the sinusoidal type 
as a function of the azimuthal angle  $\varphi$ (the angle between the magnetization direction and the [100] axis).
A maximal band gap of 14.8 meV is achieved for $\varphi=\frac{\pi}{4}$.  
Such changes in the fine structure of electronic structure can be manifested by the 
anisotropic magnetoresistance (AMR) effect.
Using the constant relaxation time ($\tau$) approximation, we estimated the AMR ratio at 300K following the semiclassical transport theory, given by
  \begin{equation}
\frac{\rho(0)-\rho(\frac{\pi }{4})}{\rho(0)}=\frac{(\frac{1}{\sigma(0)}-\frac{1}{\sigma(\frac{\pi}{4})})}{\frac{1}{\sigma(0)}}=\frac{(\frac{1}{\sigma(0)/\tau}-\frac{1}{\sigma(\frac{\pi}{4})/\tau})}{\frac{1}{\sigma(0)/\tau}},
 \end{equation}
where $\sigma(\varphi)$ ($\rho(\varphi)$) is the longitudinal conductivity (resistivity) with the azimuthal angle $\varphi$  for the magnetization direction in the (001) plane. 
This results in an AMR ratio as large as 33\%. 
On the other hand, the magnetocrystalline anisotropy energy between such two cases with  azimuthal angle $\varphi$=0 and $\frac{\pi}{4}$ is only about $10^{-6}$ eV per formula unit, 
due to the underlying cubic symmetry.
Therefore, we suspect that such materials with large AMR ratio and easily tunable magnetization directions can be applied for future spintronic applications.



\begin{figure}[t]

\includegraphics[scale=0.52]{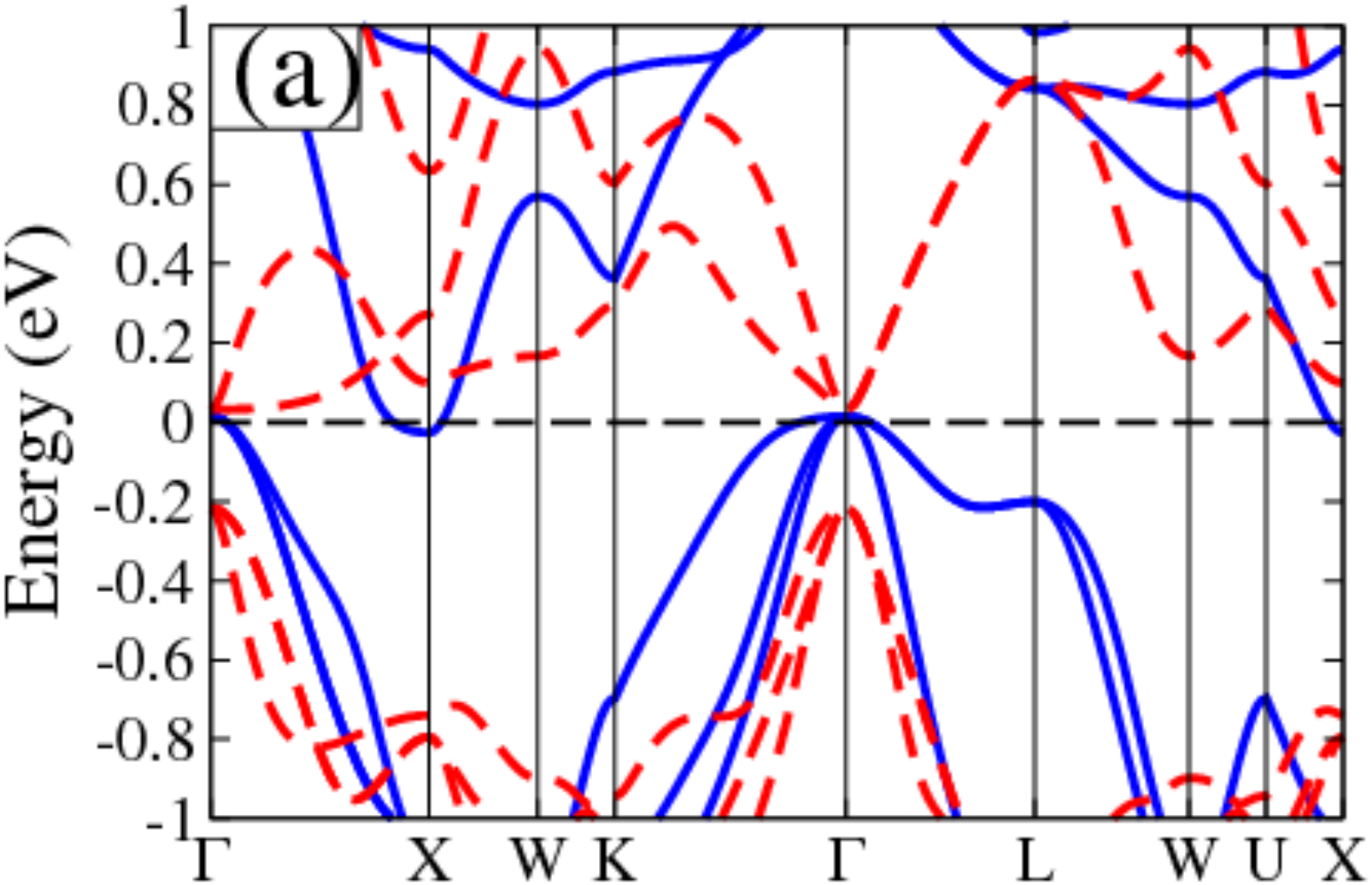}

\vspace{0.25cm}

\includegraphics[scale=0.52]{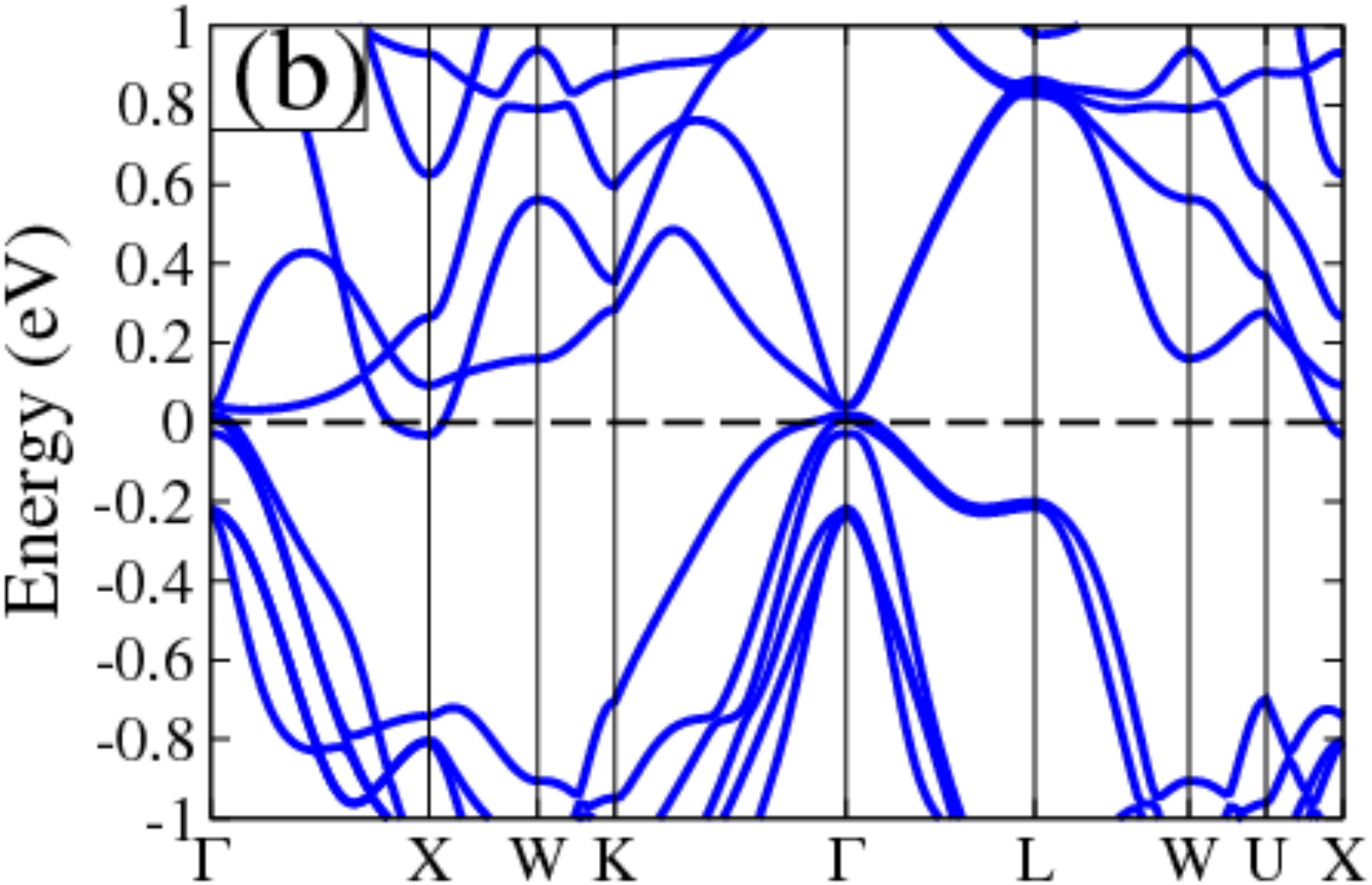}

\vspace{0.25cm}

\includegraphics[scale=0.52]{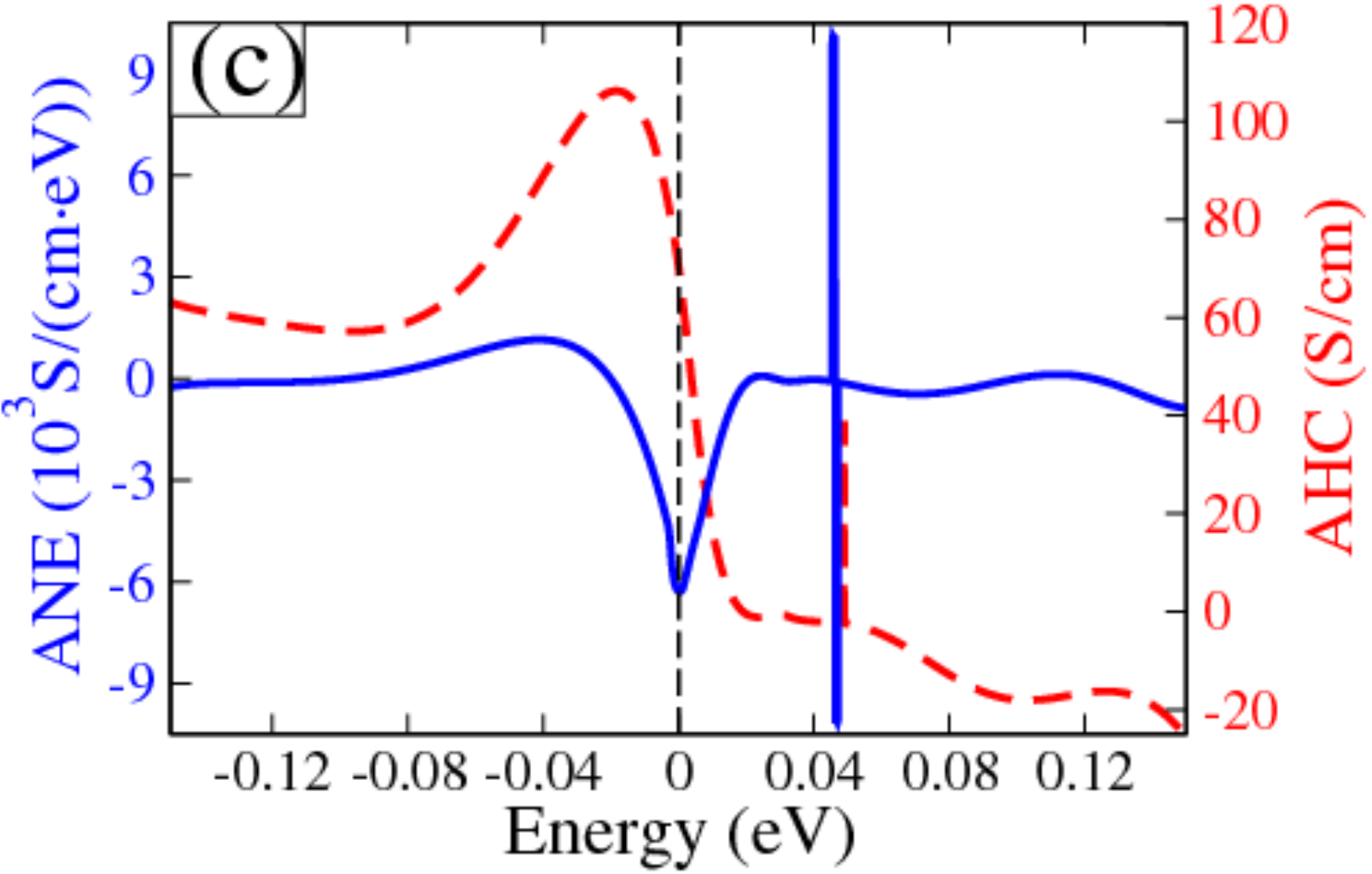}

\caption{(Color online)(a) and (b) are the band structures of NiCrMnAl without and with SOC. Without SOC, the blue and the red curves are majority and minority spin channels, respectively.  (c) The red and blue curves are AHC and ANE results $\pm$ 0.15 eV around Fermi level, respectively. The horizontal (a and b) and vertical (c) dashed lines indicate the Fermi energy.}

\label{fig:direct}
\end{figure}

\subsubsection{SGS with direct band touching}

As shown in Fig.~\ref{fig:direct}, we find NiCrMnAl is a special SGS, where a direct band touching occurs at the $\Gamma$ point.
Without considering SOC, the CBM from the minority spin channel touches the VBM with the opposite spin character.
That is, it is a type-II SGS following the classification discussed above. 
Unfortunately, due to the presence of a conduction band which goes slightly below the Fermi energy at the X point, the direct touching point
is hidden.
When SOC is turned on, a band gap of 24 meV is opened locally at the $\Gamma$ point. 
However, the resulting band gap is topologically trivial according to the AHC shown in Fig.~\ref{fig:direct}(c), 
since the magnitude of the AHC changes its sign around the Fermi energy, similar to the above discussions of FeVHfSi. 
Moreover, the AHC shows a singularity for an energy about 50 meV above the Fermi energy.  This indicates there is band anti-crossing in the electronic structure. 
Particularly, due to the drastic variation of the AHC with respect to the chemical potential around Fermi level, the resulting derivative of AHC is as large as -6,000 S/(cm$\cdot$eV) at the Fermi level.
That is, gigantic anomalous Nernst effect is expected in NiCrMnAl.  
Such a derivative of AHC is much larger than that of the recent experimental realized large anomalous Nernst effect in Mn$_{3}$Sn with a value of -845 S/(cm$\cdot$eV).\cite{mn3ga1, mn3ga2, mn3ga3} 
In this sense, type-II SGSs may be promising materials for anomalous Nernst applications.

\section{Conclusion}
To summarize, we have carried out a systematic high-throughput screening for spin-gapless semiconductors (SGSs) in quaternary Heusler compounds with 21, 26, and 28 valence electrons. 
After validating our calculations with the previously reported cases, we predicted 80 new stable (based on the formation energy) compounds as promising candidates of spin-gapless semiconductors,
where 70 cases are stable based on further evaluation of the mechanical and dynamical stabilities.
The magnetization of SGSs obeys the Slater-Pauling rule, which can be interpreted based on a new scheme of crystal field splitting of the D$_{4h}$ type.
Interestingly, all four types of SGSs have been identified among our candidate systems, where both the longitudinal conductivity and transversal anomalous Hall conductivity are calculated.
We find the type-II SGSs are particularly interesting for spintronic applications as the spin polarization of the longitudinal conductivity is very sensitive
to the chemical potential, while the anomalous hall conductivity changes its sign across the Fermi level, leading to possible significant anomalous Nernst effect. 
This is also true for the SGS candidate NiCrMnAl with direct touching.
Additionally, it is also demonstrated that spin orbit coupling can have significant effect on the electronic structure of SGSs with heavy elements, where 
the band gap can be tuned by the magnetization direction, resulting in large anisotropic magnetoresistance in cubic crystals. 
Therefore, we suspect that SGSs are promising materials for future spintronic applications, awaiting further experimental and theoretical explorations.

\section*{ACKNOWLEDGMENTS}

Qiang Gao thanks the financial support from the China Scholarship Council. The authors gratefully acknowledge computational time on the Lichtenberg High Performance Supercomputer.

{\footnotesize

\begingroup
\renewcommand{\section}[2]{}

\end{document}


\title{Supplemental Material}
\author{Qiang~Gao}
\author{Ingo~Opahle}
\author{Hongbin~Zhang}
\affil{Institute of Materials Science, Technische Universit$\ddot{a}$t Darmstadt, 64287 Darmstadt, Germany}

\date{\today}
\maketitle

\section{Detailed data}

 The detailed data is shown in the                                                                     following table, including lattice constants, total and partial magnetic moments, formation energies, elastic constants, mechanical and dynamical stabilities, magnetic orders, and SGS types. $a_{opt}$, $\bigtriangleup$H are lattice constant and formation energy.
 For mechanical and dynamical stabilities, 1 and 0 are stable and unstable. In magnetic order, FM and FIM are ferromagnetic and ferrimagnetic. In SGS type, 'SOC-' means spin-orbit coupling (SOC) has significant influence on such SGS. The compounds are arranged according to similar chemical component.

\begin{landscape}
\includepdfmerge[angle=90]{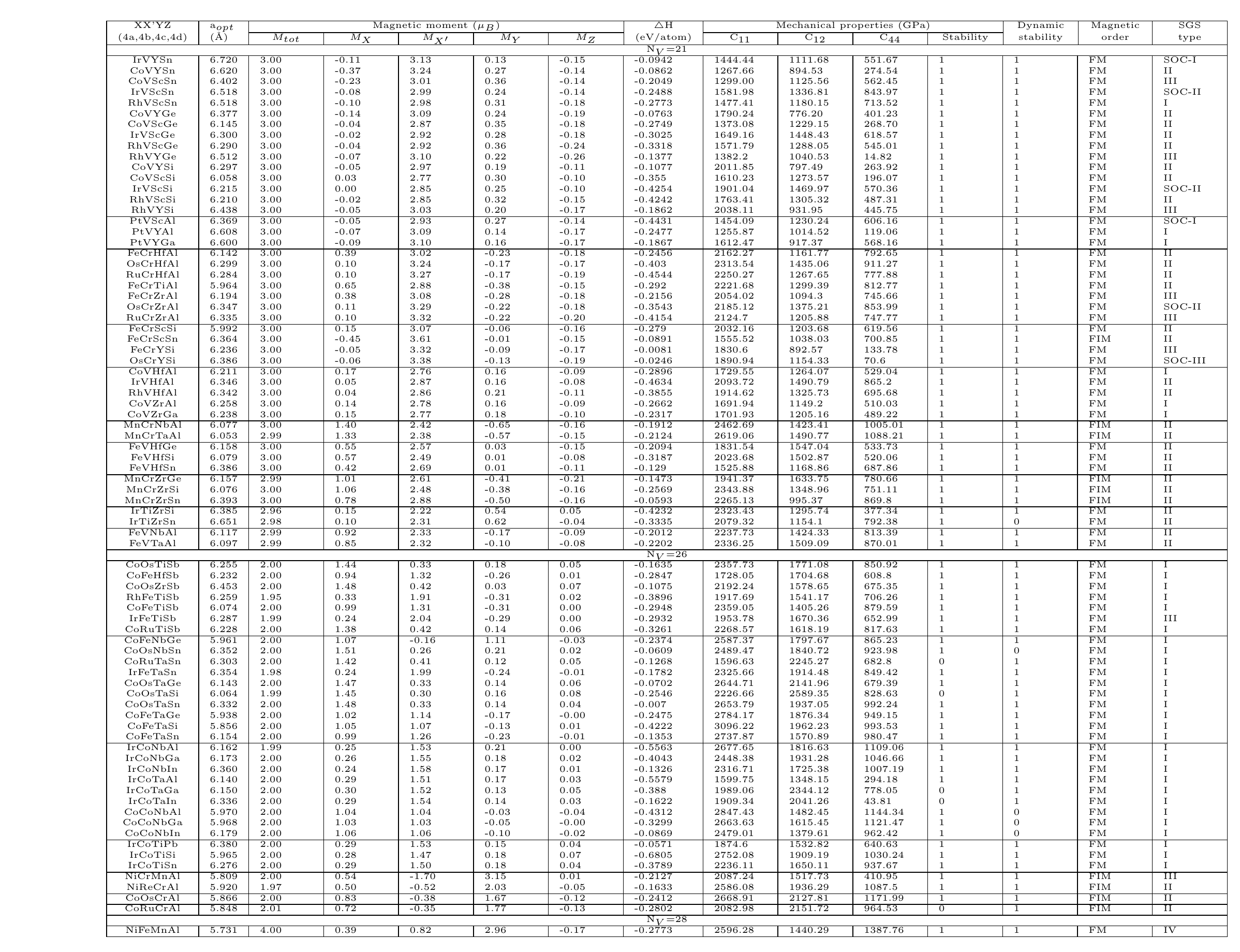}
\end{landscape}

\section{Density of sates for magnetism discussions}
We have discussed the magnetism of Heusler SGSs in the main text. For the understanding of exchange and crystal field splitting, we have shown the representative total and partial density of sates (DOS) of the considered 1 and 2 main magnetic compounds with N$_V$=21 or 26. In case of N$_V$=21, PtVYAl and FeVNbAl are the representatives of 1 and 2 main magnetic ions Heusler SGS. In case of N$_V$=26, IrFeTiSb and CoFeTaGe are the representatives of 1 and 2 main magnetic ions Heusler SGS.

\clearpage

\begin{figure}[h]
\centering

(a)\includegraphics[scale=0.48]{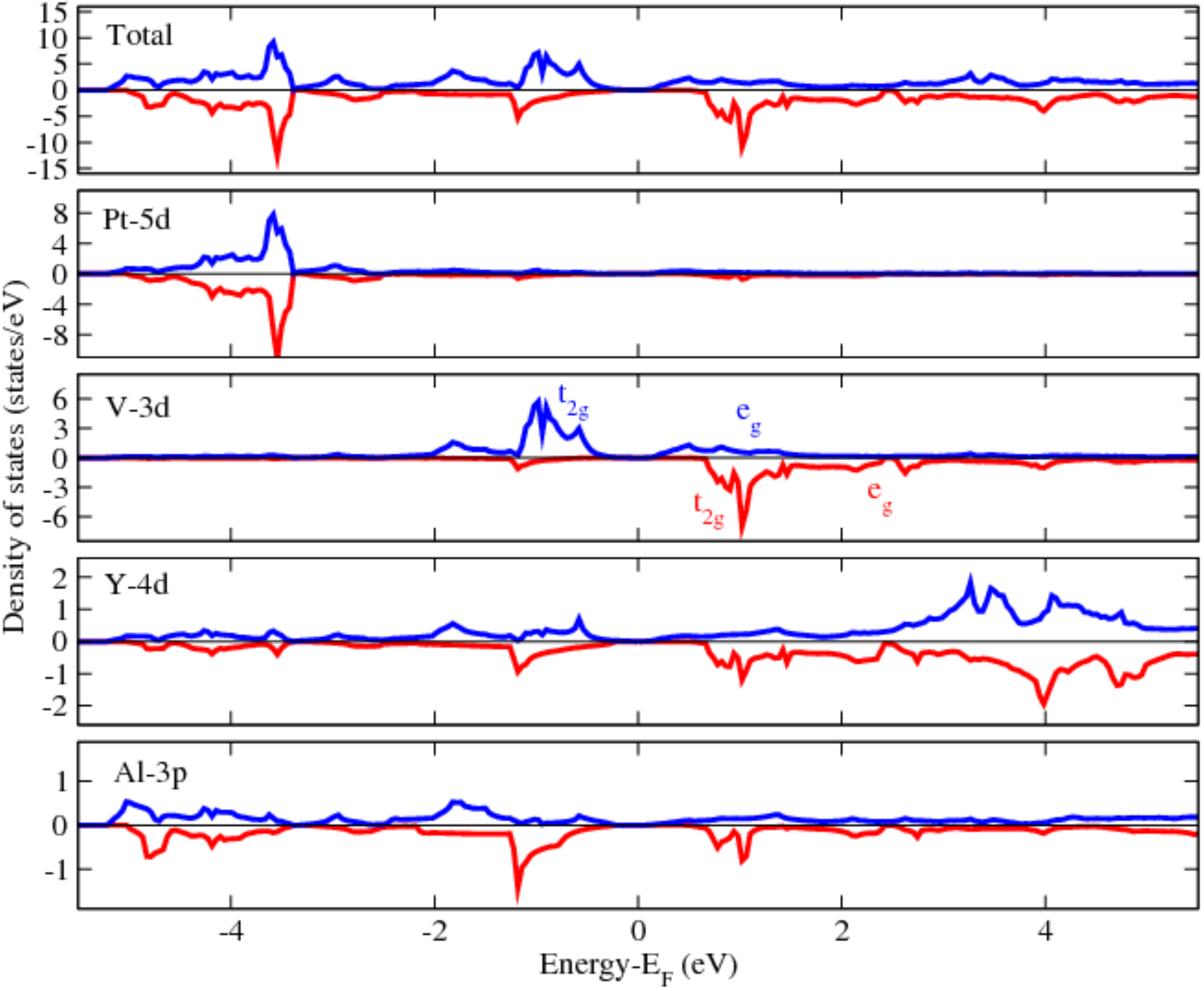}
(b)\includegraphics[scale=0.48]{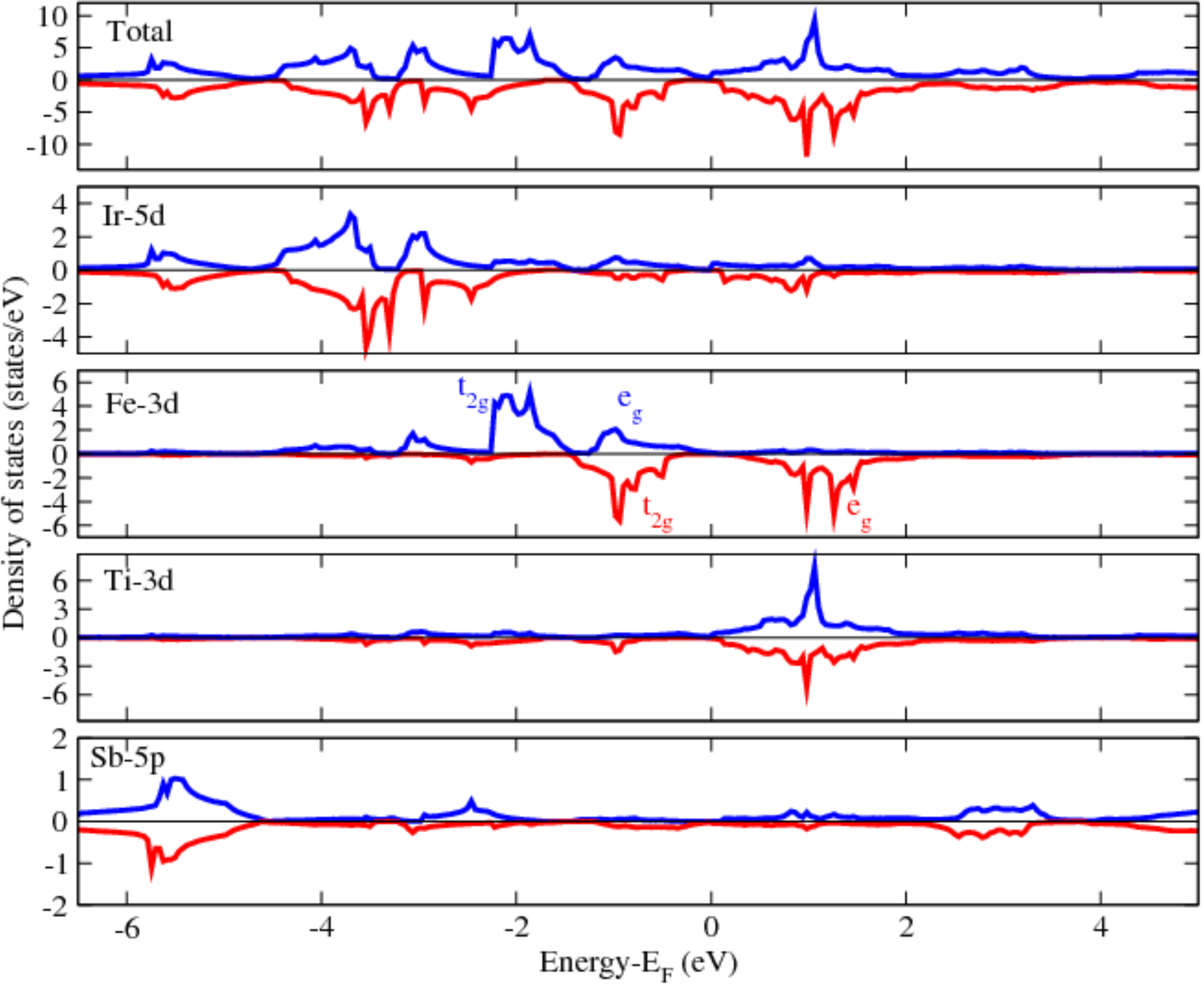}

\caption{  {\footnotesize     DOS for one magnetic ion Heusler SGS. (a)  N$_V$=21 Heusler SGS PtVYAl.  V makes main contribution to the magnetic moment.
 (b) N$_V$=26 Heusler SGS IrFeTiSb.  Fe makes main contribution to the magnetic moment.  Here the  blue and red curves are the majority and minority spin channels, respectively. (a) PtVYAl is one main magnetic ion SGS. V makes main contribution to the magnetic moment.  (b) }}
\label{}
\end{figure}

\begin{figure}[h]
\centering

(a)\includegraphics[scale=0.48]{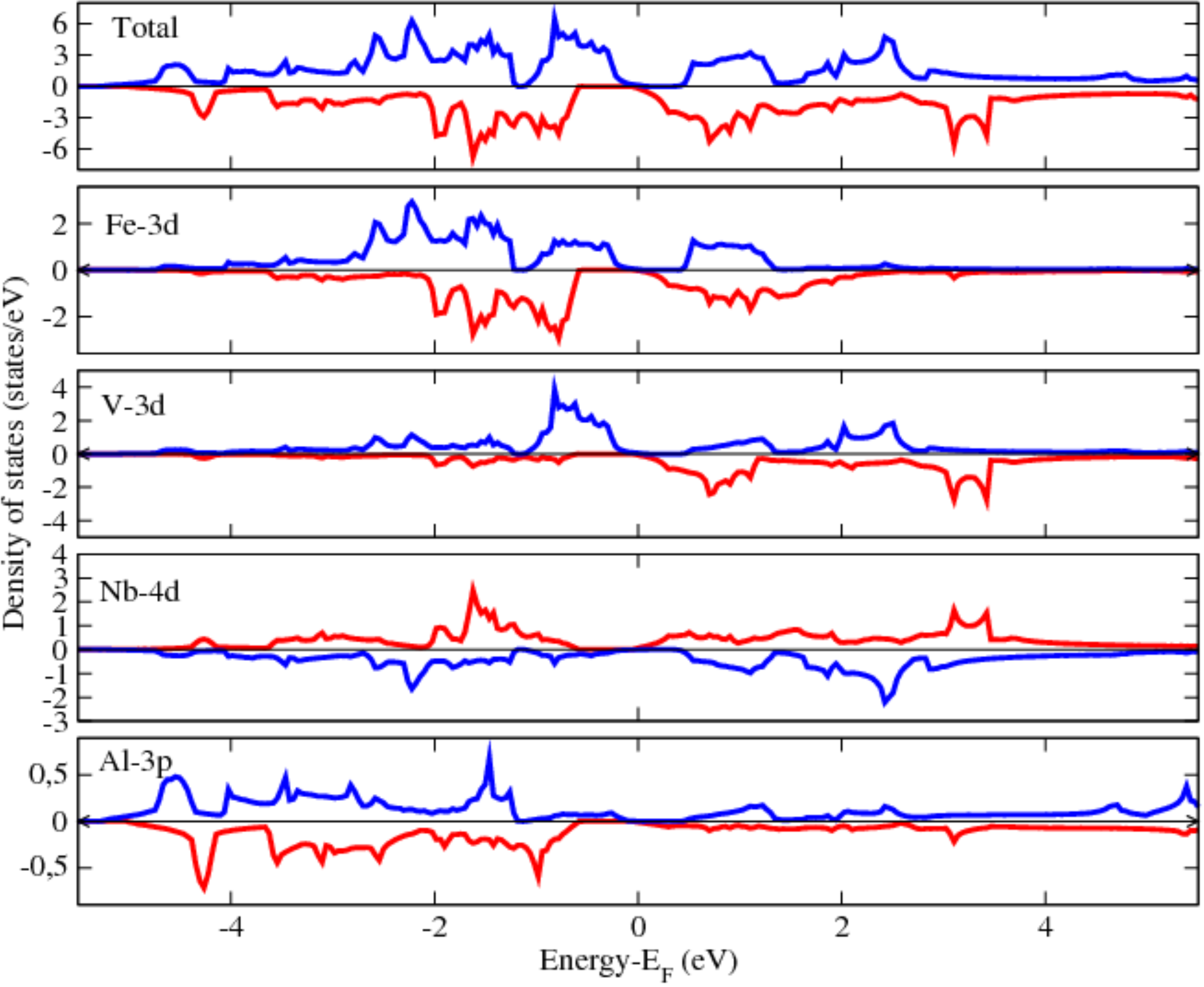}
(b)\includegraphics[scale=0.48]{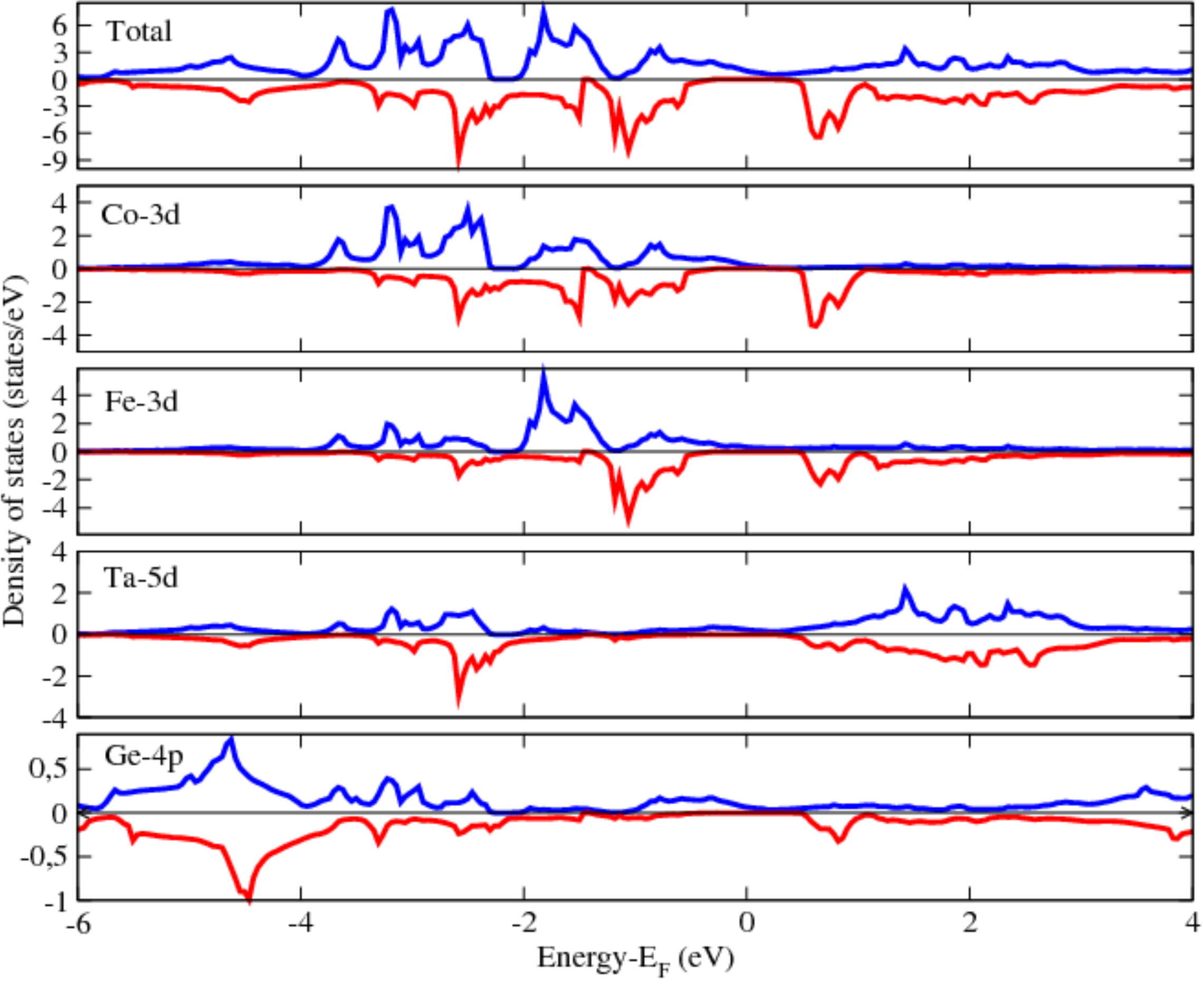}

\caption{{\footnotesize DOS for two magnetic ions Heusler SGS. (a)   N$_V$=21 Heusler SGSs FeVNbAl. It has two main magnetic ions with about 2 $\mu_B$ magnetic moment in V and 1 $\mu_B$ in Fe. (b) N$_V$=26 Heusler SGSs CoFeTaGe. It has two main magnetic ions with about 1 $\mu_B$ magnetic moment in both Co and  Fe. Here the blue and red curves are the majority and minority spin channels, respectively.  Here we do not show the simple atom model for the two magnetic ions SGSs.}}
\label{}
\end{figure}

\clearpage

\begin{figure}[htp]
\centering
\includegraphics[scale=0.60]{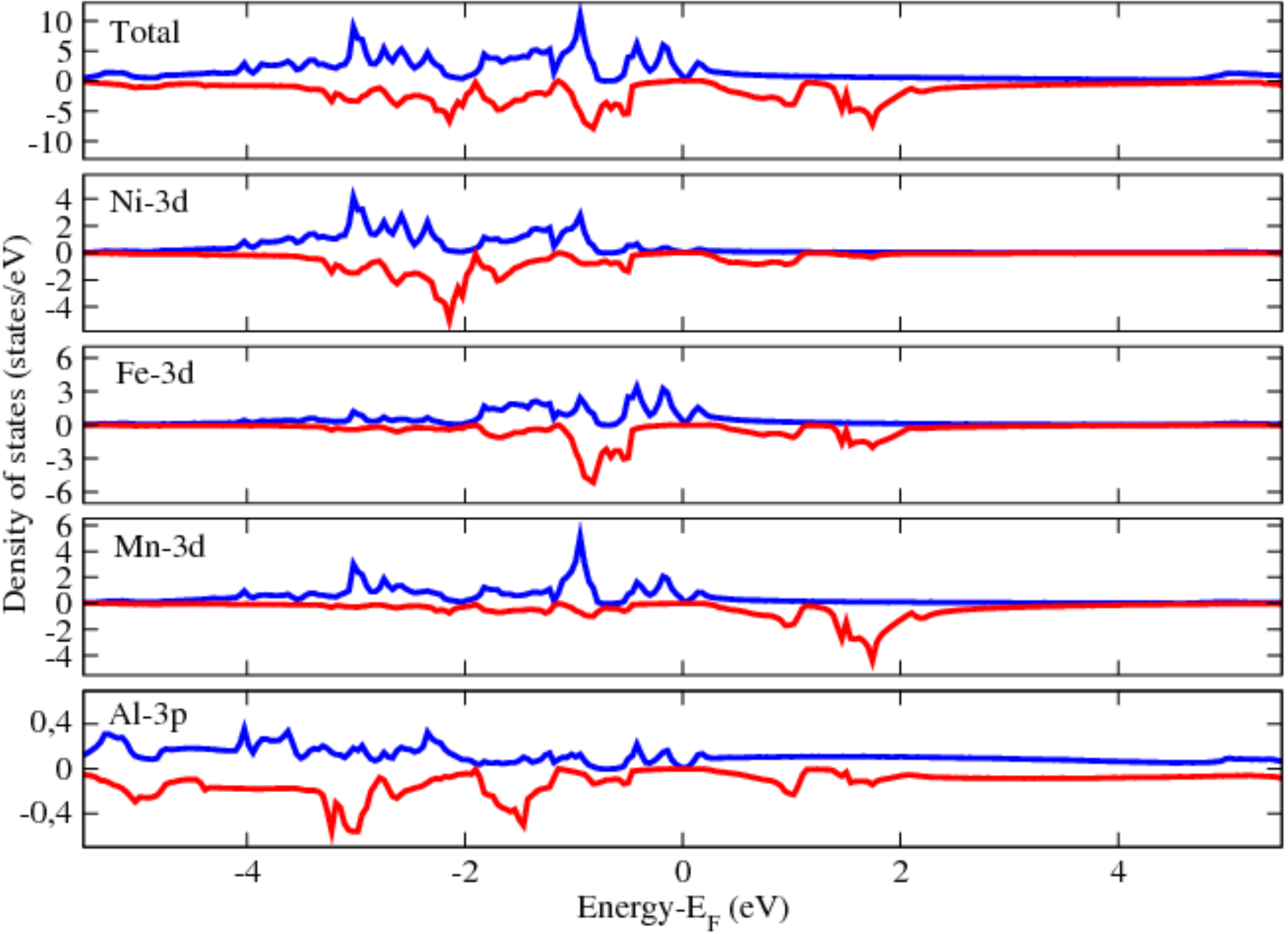}

\caption{DOS for 28 valence Heusler SGS NiFeMnAl. It can be seen that there is strong hybridization between Ni, Fe, and Mn.   }
\label{}
\end{figure}

\clearpage

In order to understand the spin splitting clearly for the two magnetic ion SGS clearly, we show the local density of states (LDOS) of the magnetic ions for FeVNbAl and CoFeTaGe  in the following.

\clearpage

\begin{figure}[htp]
\centering
\includegraphics[scale=0.60]{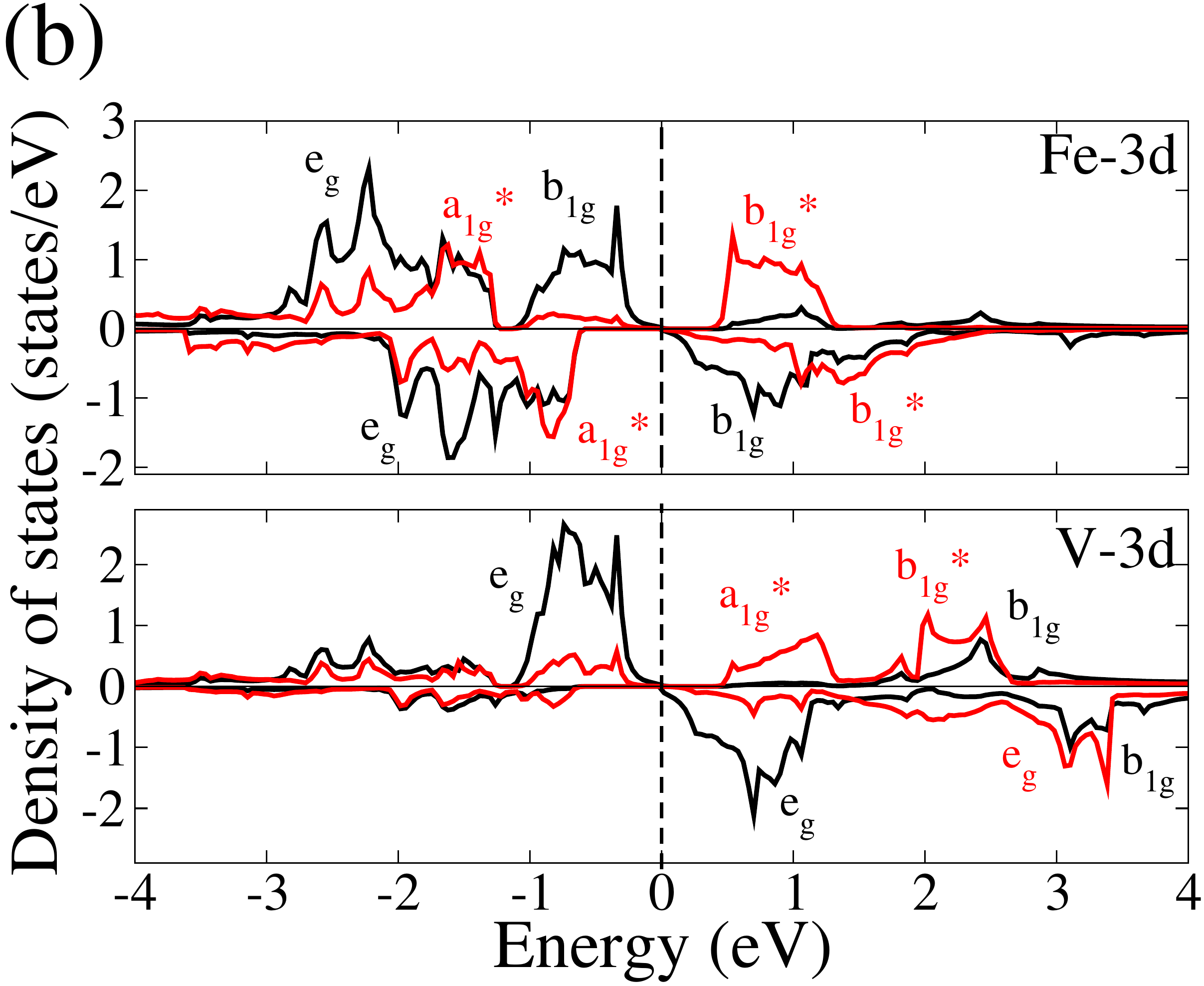}

\caption{LDOS for magnetic ions in FeVNbAl. Up: Fe 1 $\mu_B$ magnetic moments. Down: V with 2  $\mu_B$ magnetic moments. The black and red  curves are the original t$_{2g}$ and e$_g$ states, respectively. It should be noticed that for V in the minority spin channel the e$_g$ state is not splitted into a$_{1g}^*$ and b$_{1g}^*$ states, while the original t$_{2g}$ state splits into two fold e$_{g}$ and b$_{1g}$ states.}
\label{}
\end{figure}

\begin{figure}[htp]
\centering
\includegraphics[scale=0.60]{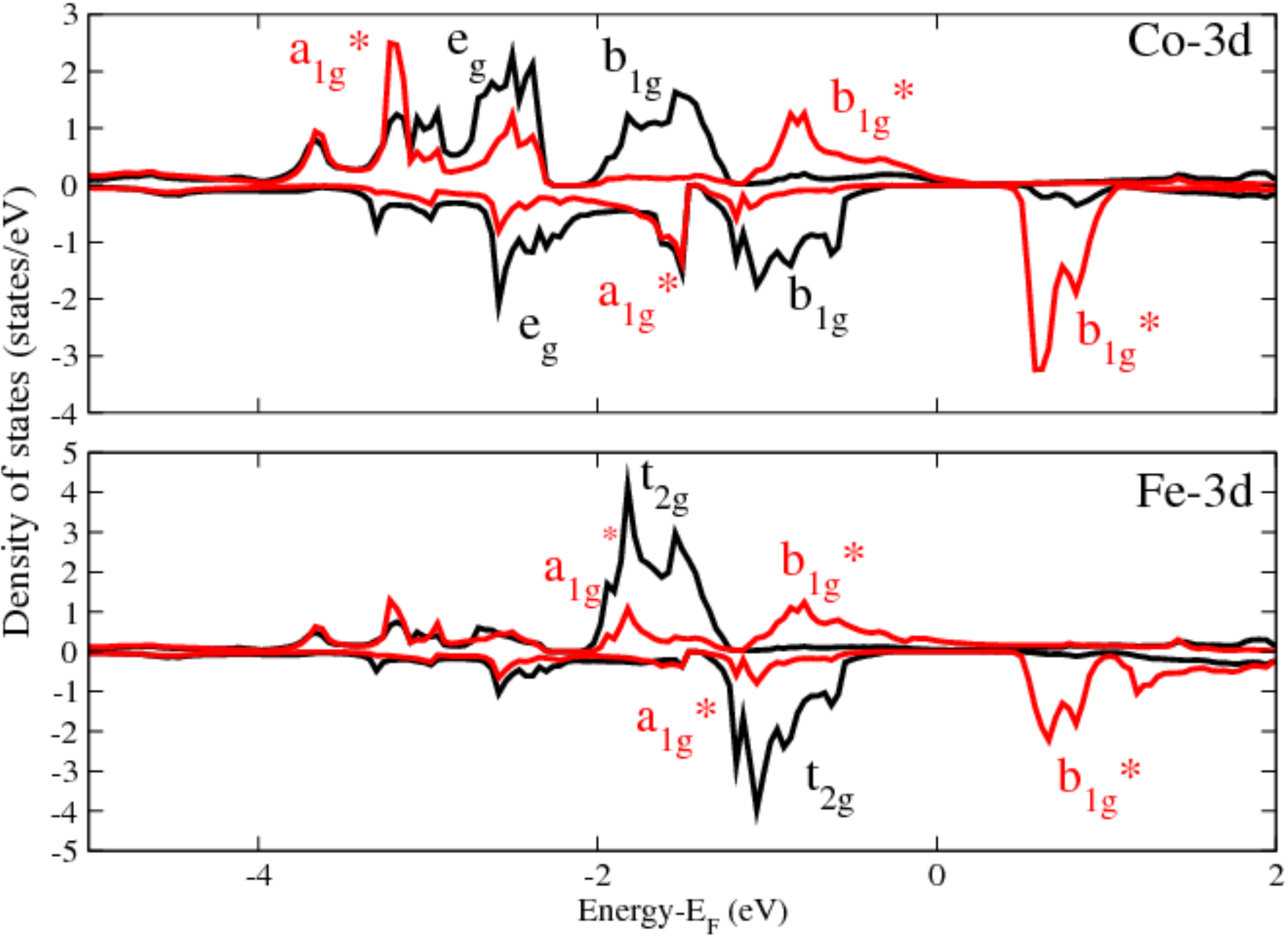}

\caption{LDOS for magnetic ions in CoFeTaGe. Up: Co  with 1 $\mu_B$ magnetic moments. Down: Fe with 1  $\mu_B$ magnetic moments. The black and red curves are the original t$_{2g}$ and e$_g$ states, respectively. }
\label{}
\end{figure}

\begin{figure}[htp]
\centering
\includegraphics[scale=0.50]{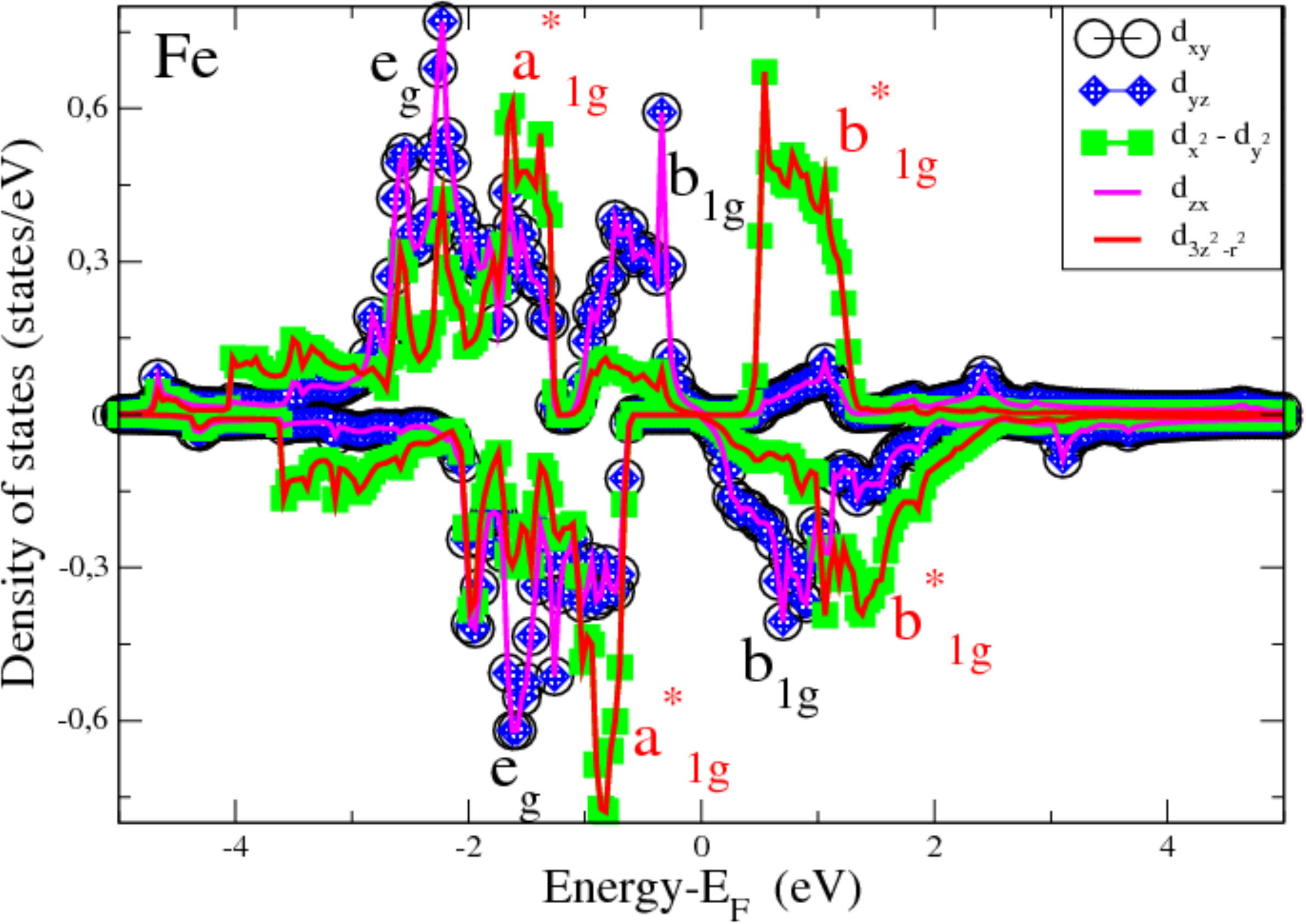}
\includegraphics[scale=0.5]{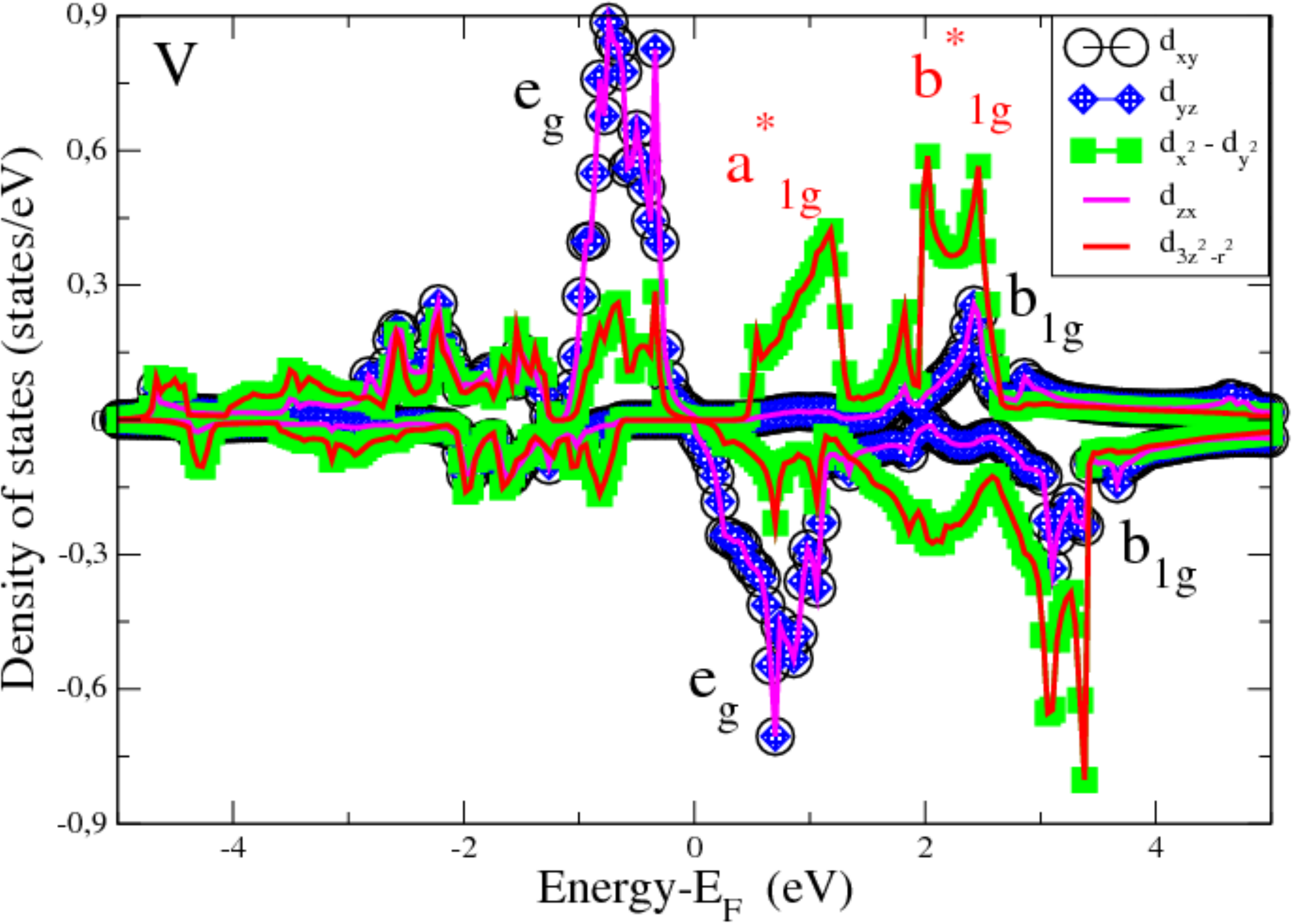}

\caption{ LDOS for decomposed orbitals of the two magnetic ions in FeVNbAl. Up and down are Fe and V.  We can see that the orbitals $d_{xy}$, $d_{yz}$, and $d_{zx}$ are always degenerated in the  $e_{g}$ and $b_{1g}$  subshells originated from the $t_{2g}$  shell, which is the same for the $d_{x^2-y^2}$, $d_{3z^2-r^2}$ orbitals in the $a_{1g}^*$ and $b_{1g}^*$ subshells originated from $e_{g}$ shell.}
\label{}
\end{figure}

\clearpage

\section{Band structures}
The band structures of newly predicted SGSs are shown in this part.

\subsection{Type I SGSs}

\centering

\includegraphics[scale=0.3]{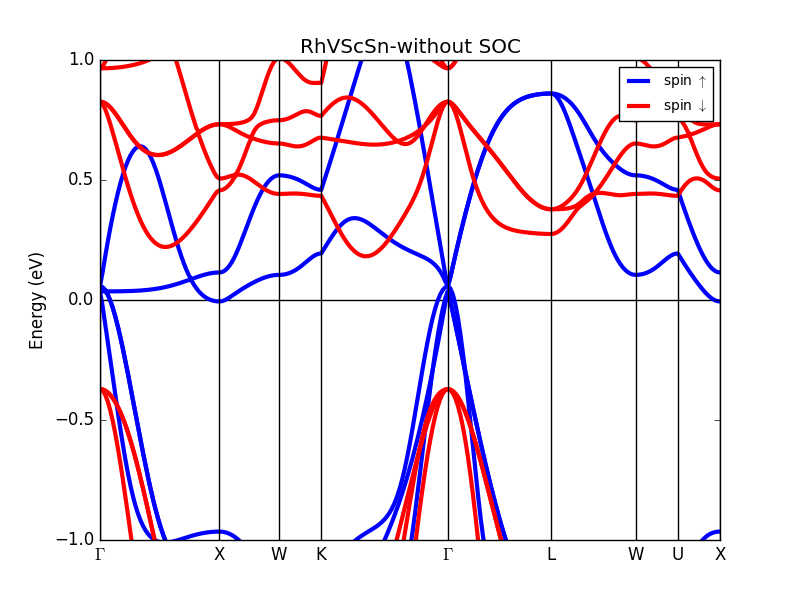}
\includegraphics[scale=0.3]{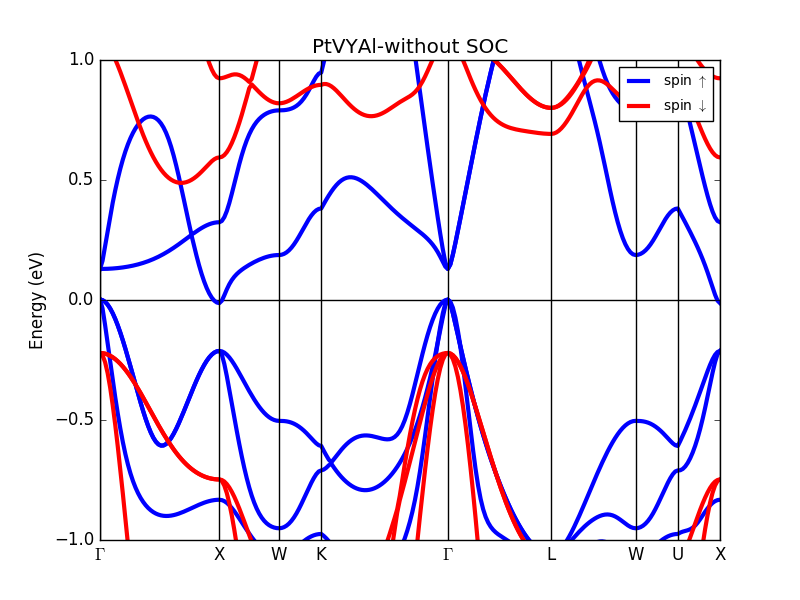}

\includegraphics[scale=0.3]{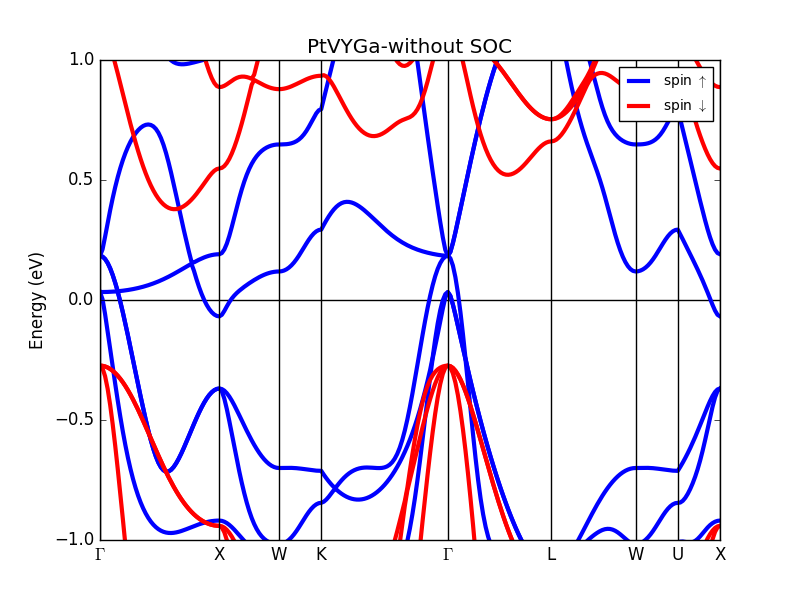}
\includegraphics[scale=0.3]{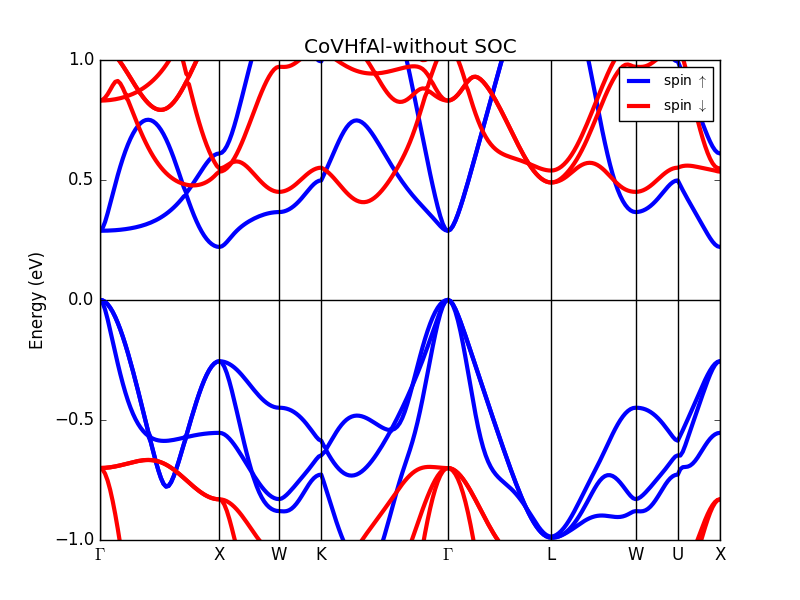}

\includegraphics[scale=0.3]{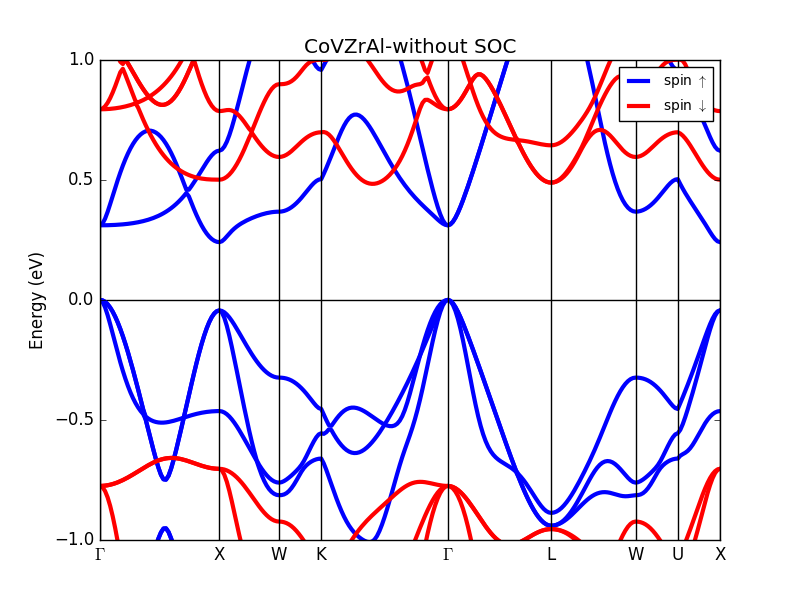}
\includegraphics[scale=0.3]{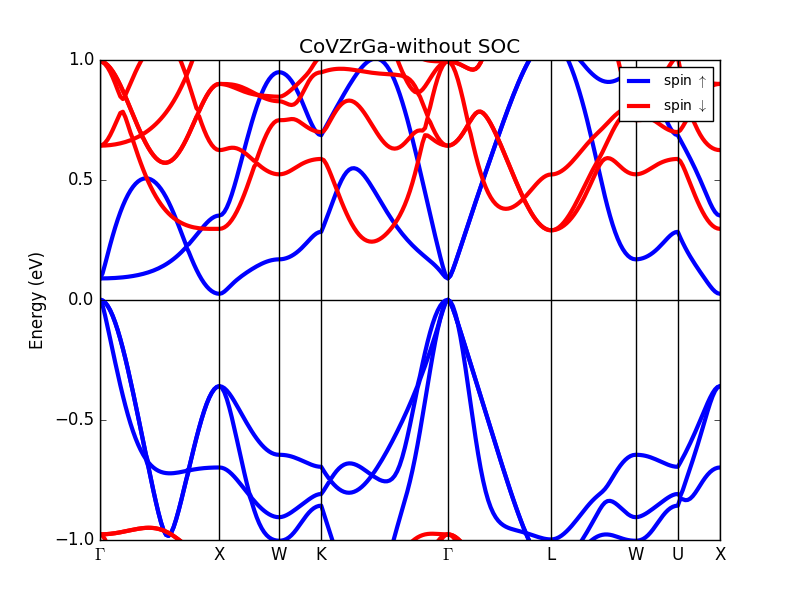}

\centering

\includegraphics[scale=0.3]{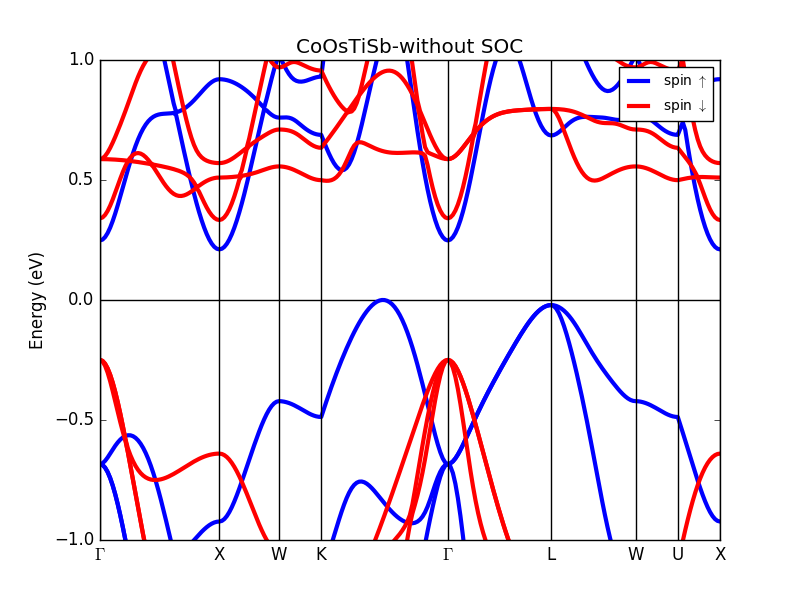}
\includegraphics[scale=0.3]{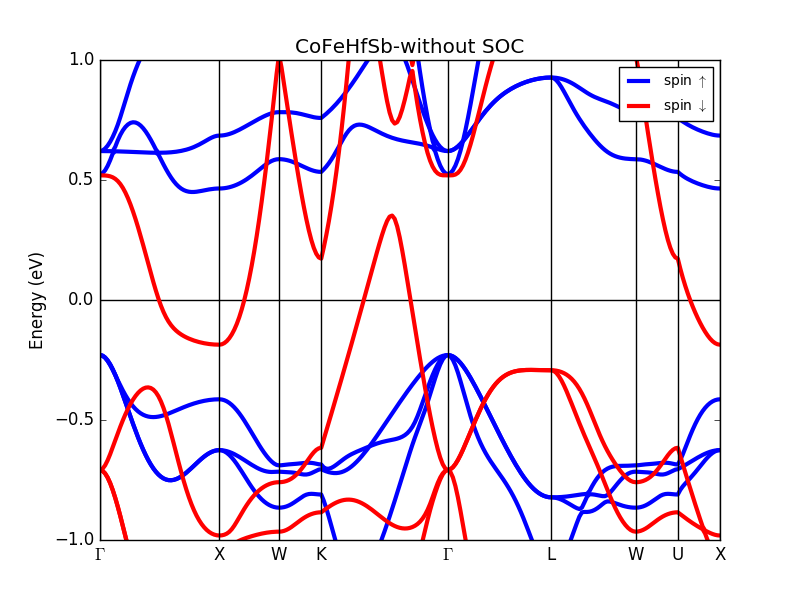}

\includegraphics[scale=0.3]{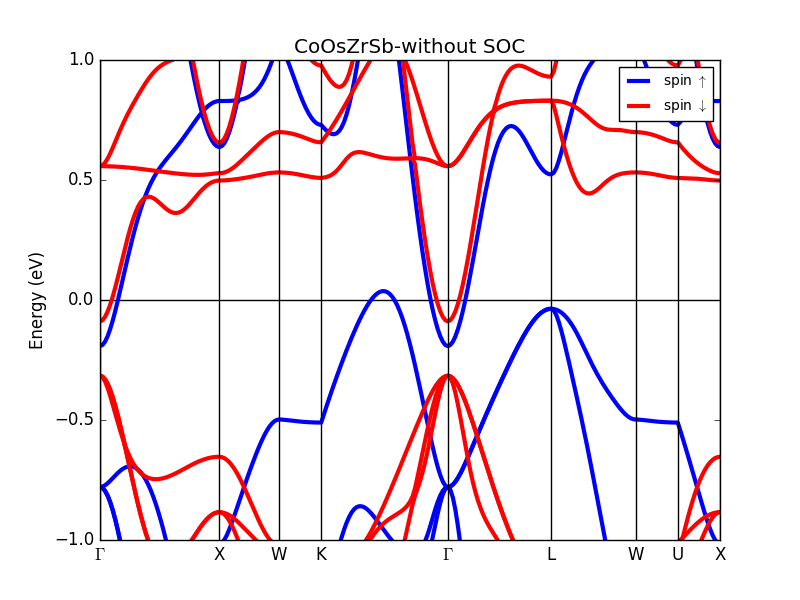}
\includegraphics[scale=0.3]{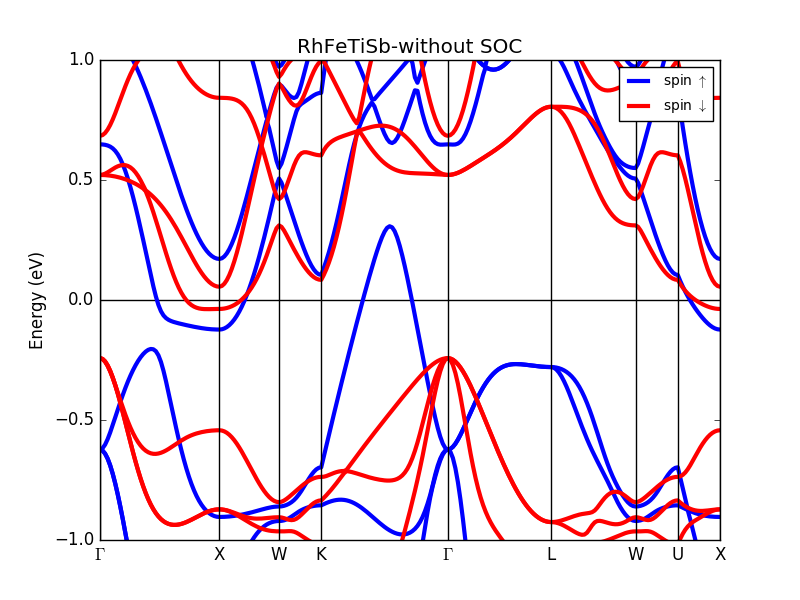}

\includegraphics[scale=0.3]{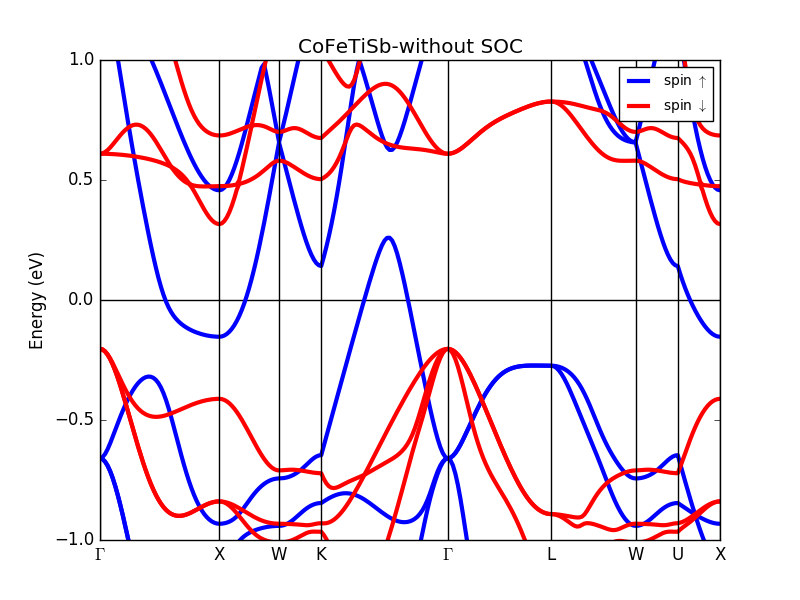}
\includegraphics[scale=0.3]{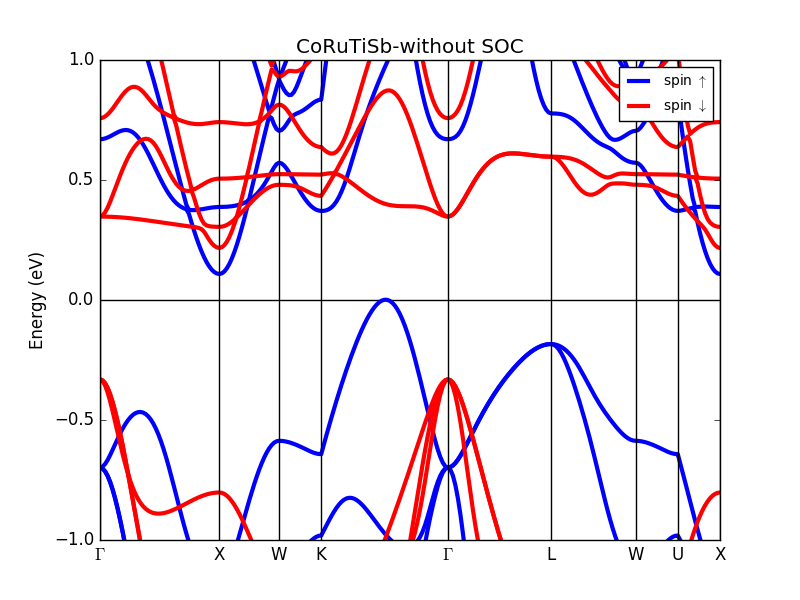}

\includegraphics[scale=0.3]{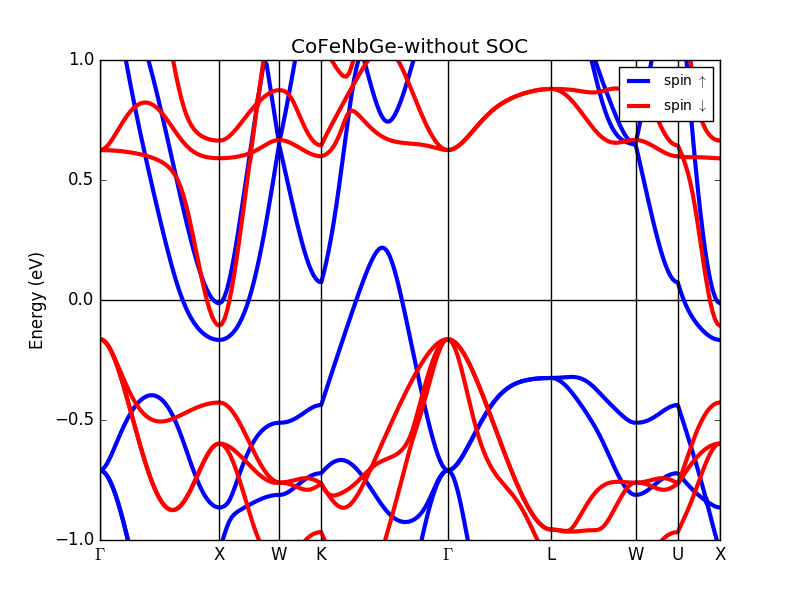}
\includegraphics[scale=0.3]{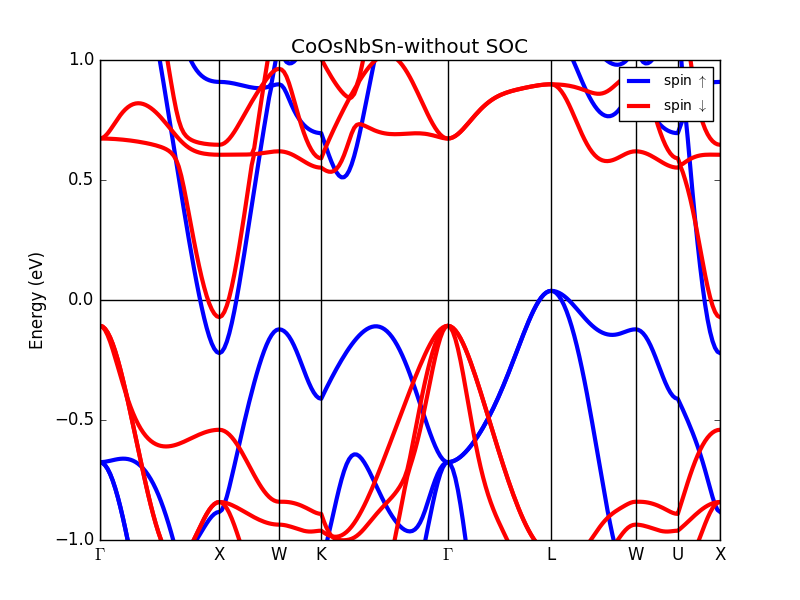}

\includegraphics[scale=0.3]{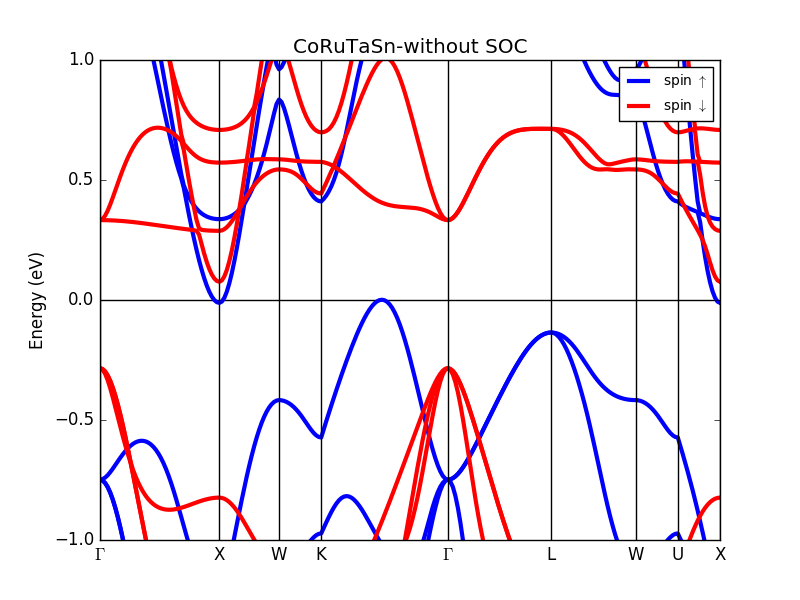}
\includegraphics[scale=0.3]{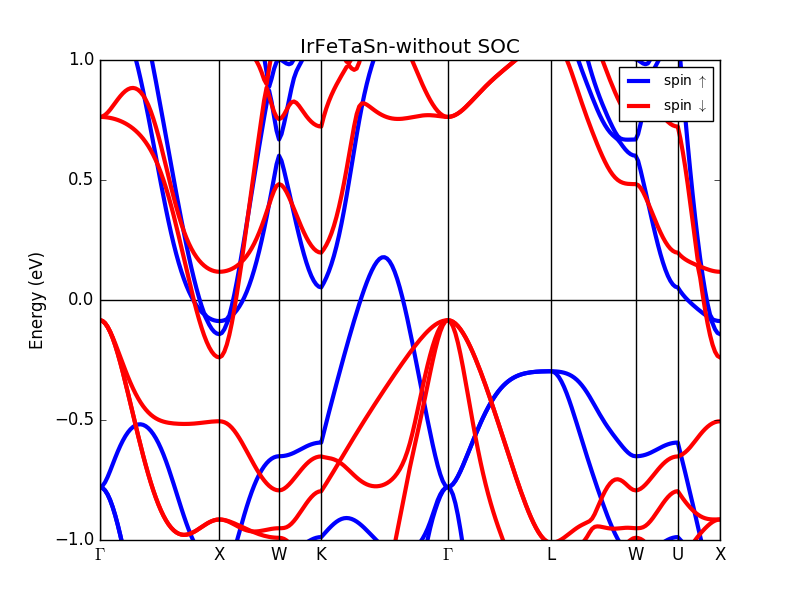}

\centering

\includegraphics[scale=0.3]{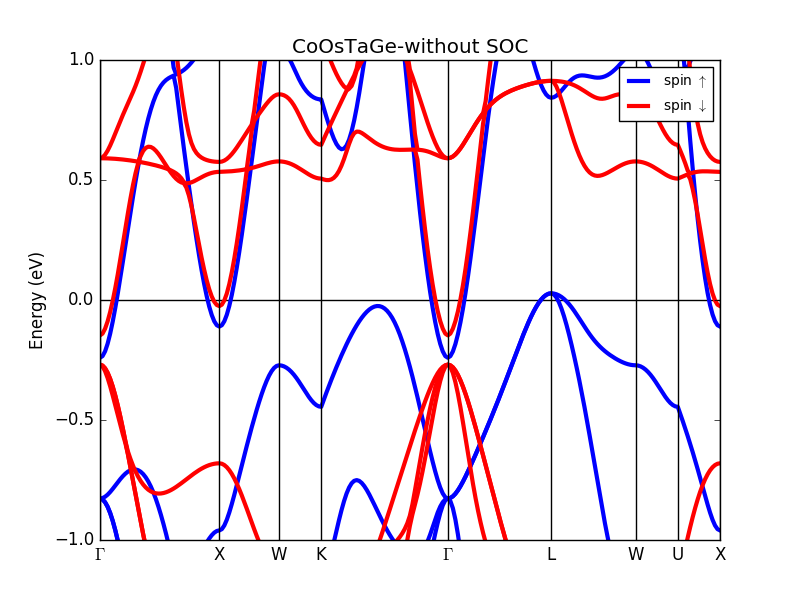}
\includegraphics[scale=0.3]{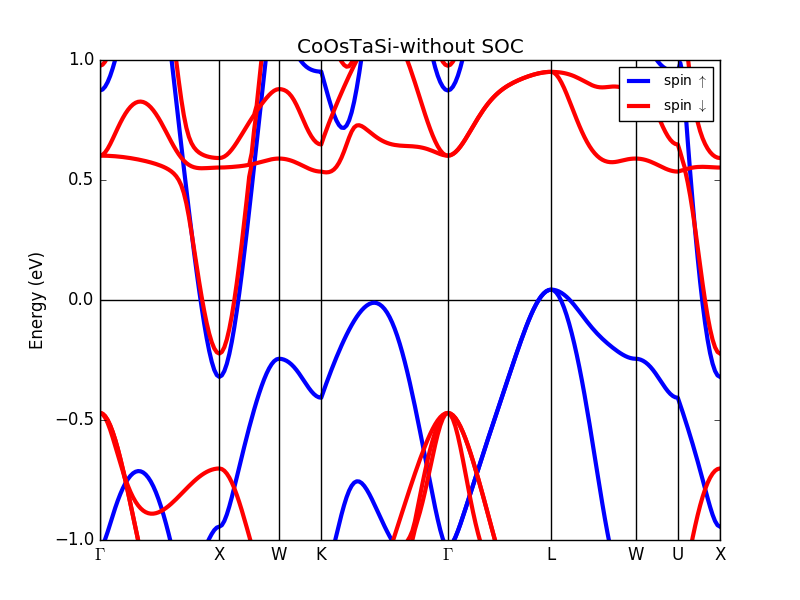}

\includegraphics[scale=0.3]{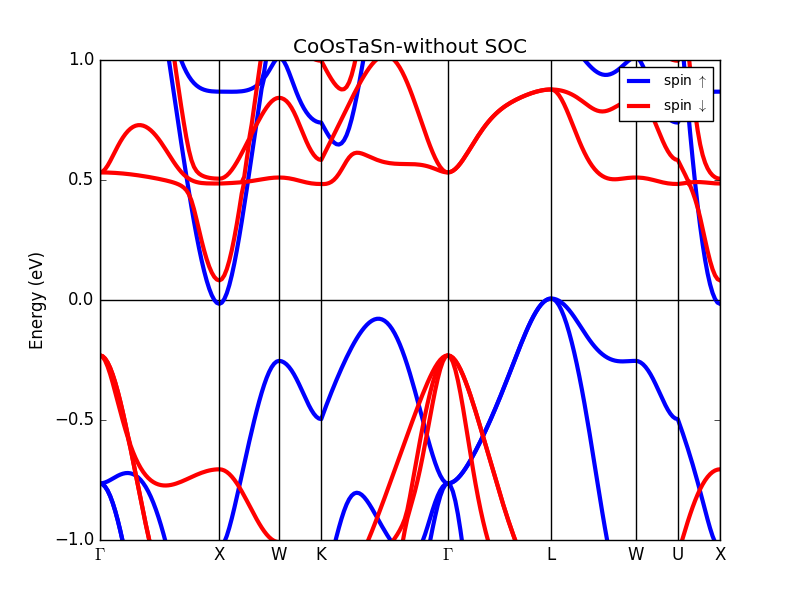}
\includegraphics[scale=0.3]{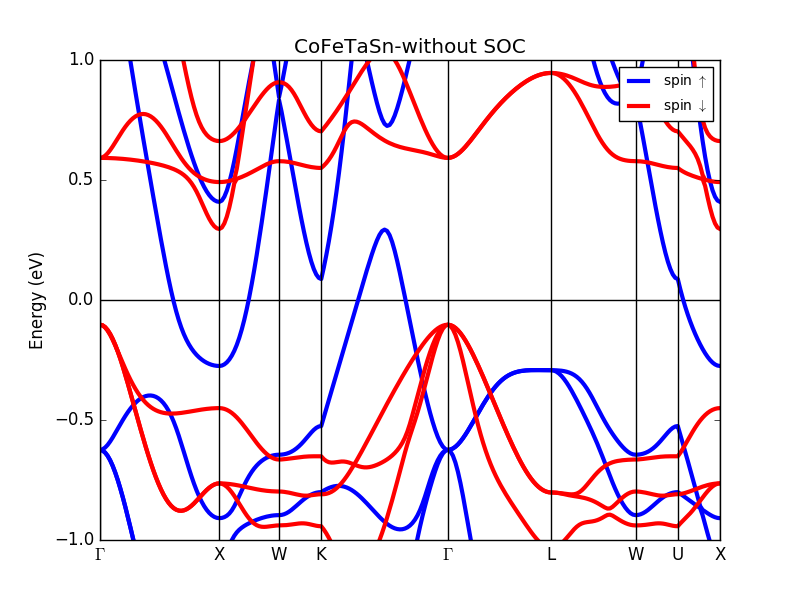}

\includegraphics[scale=0.3]{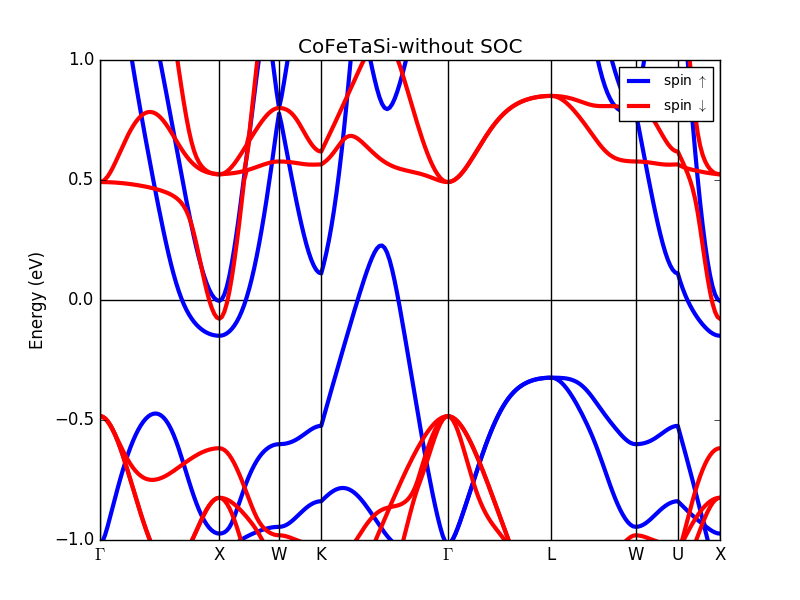}
\includegraphics[scale=0.3]{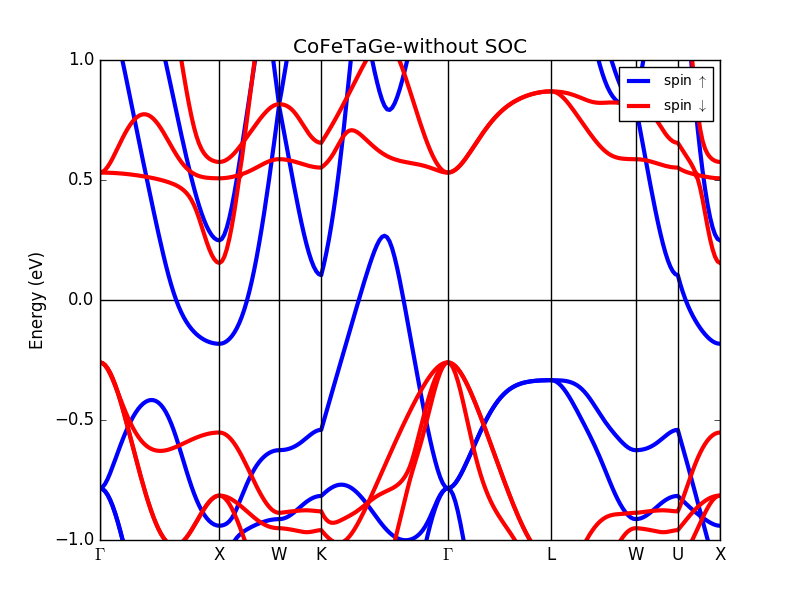}

\includegraphics[scale=0.3]{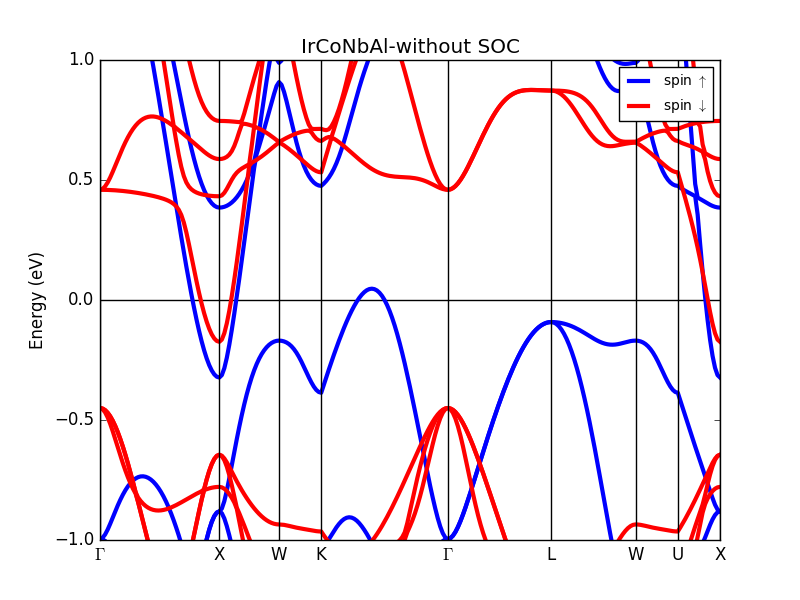}
\includegraphics[scale=0.3]{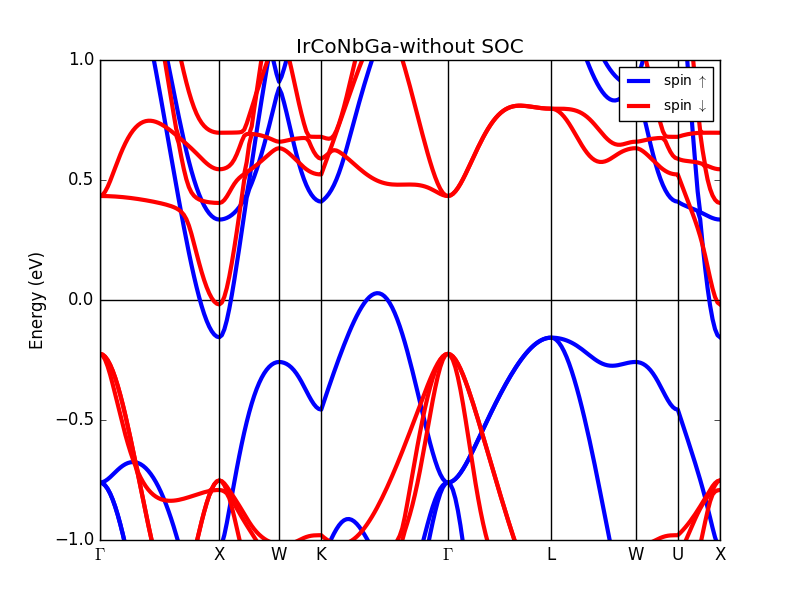}

\includegraphics[scale=0.3]{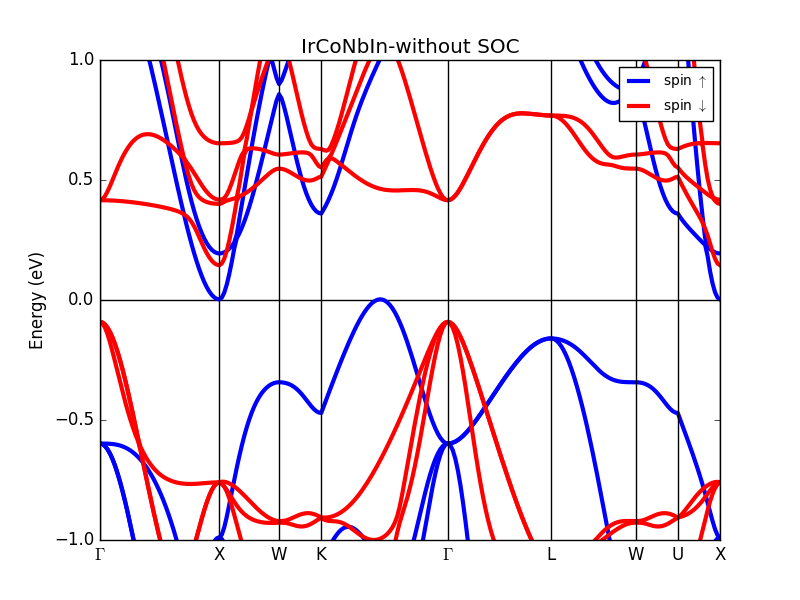}
\includegraphics[scale=0.3]{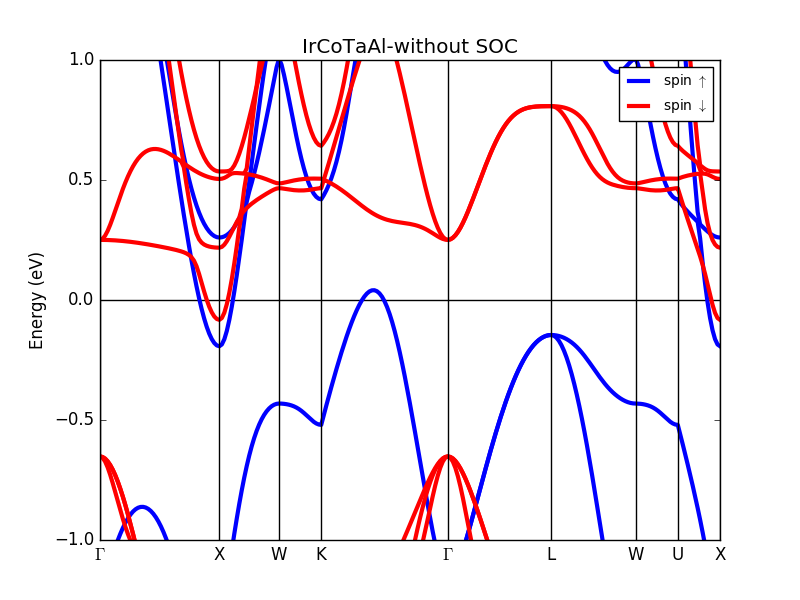}

\centering
\includegraphics[scale=0.3]{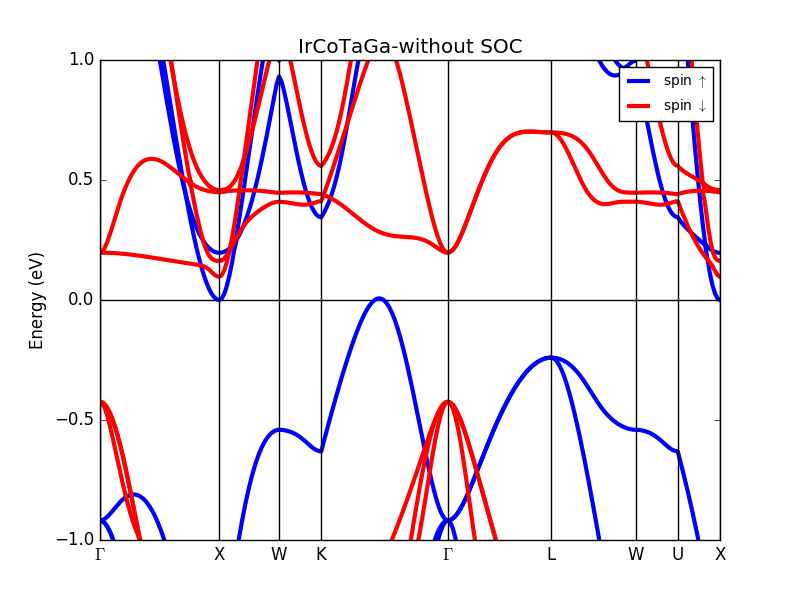}
\includegraphics[scale=0.3]{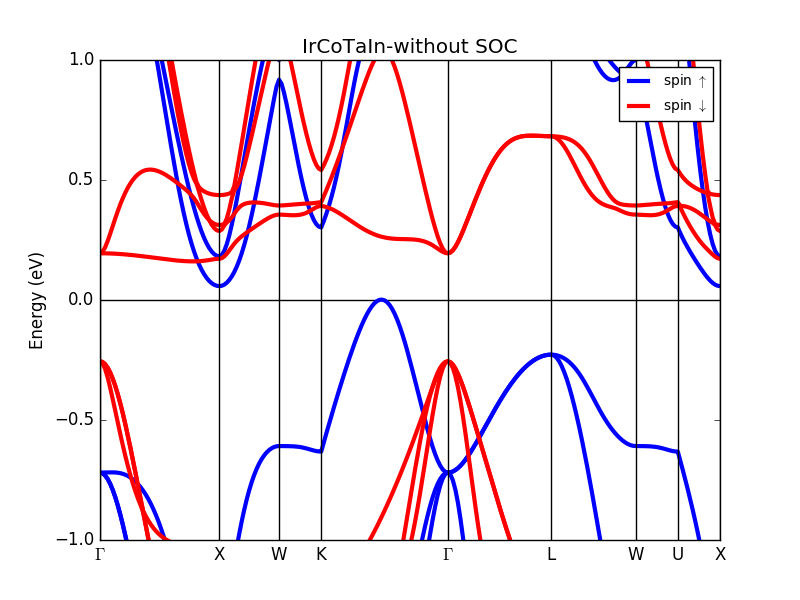}

\includegraphics[scale=0.3]{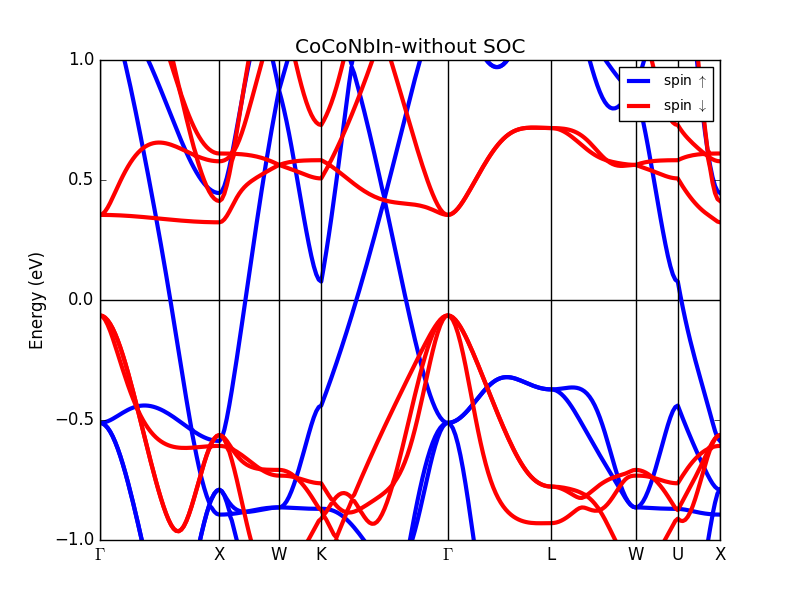}
\includegraphics[scale=0.3]{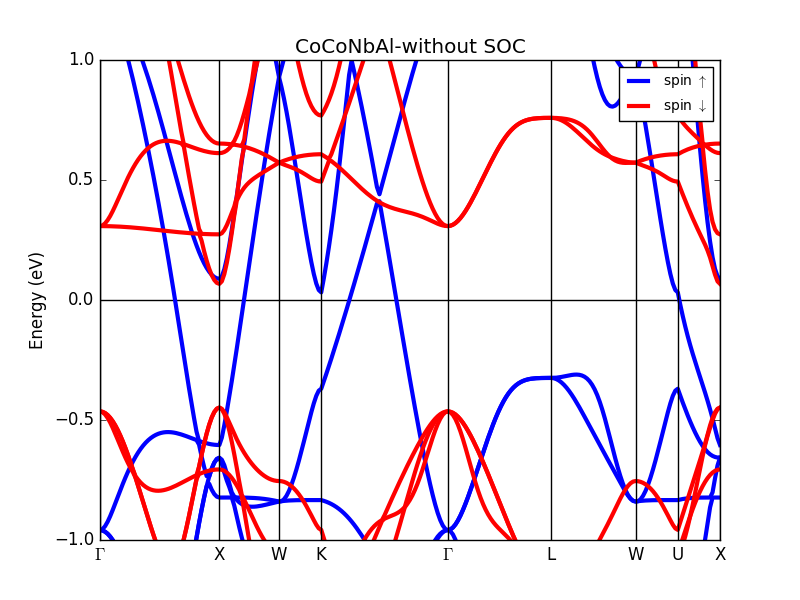}

\includegraphics[scale=0.3]{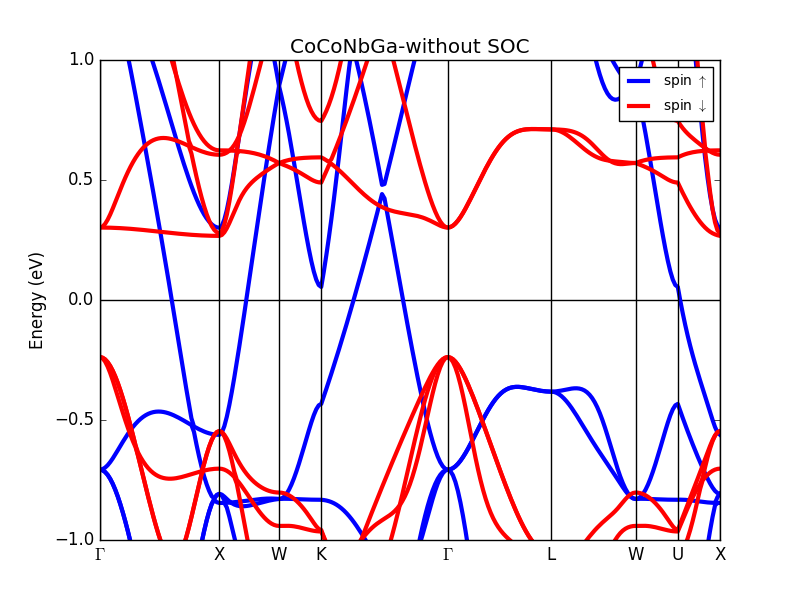}
\includegraphics[scale=0.3]{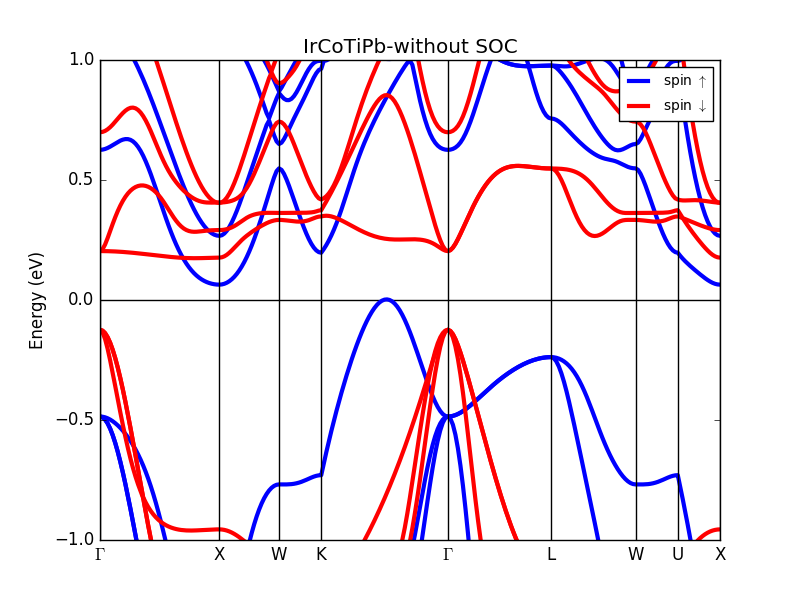}

\includegraphics[scale=0.3]{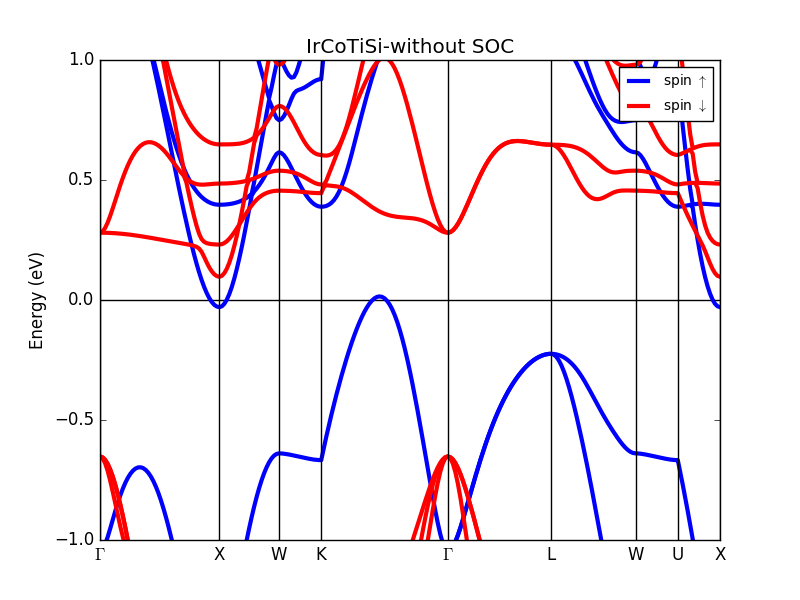}
\includegraphics[scale=0.3]{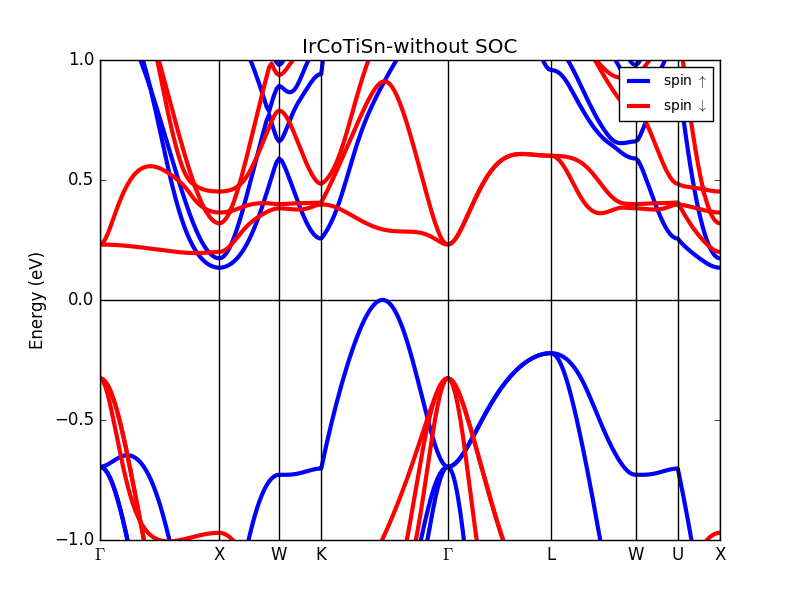}

\clearpage

\subsection{Type II SGSs}

\centering
\includegraphics[scale=0.3]{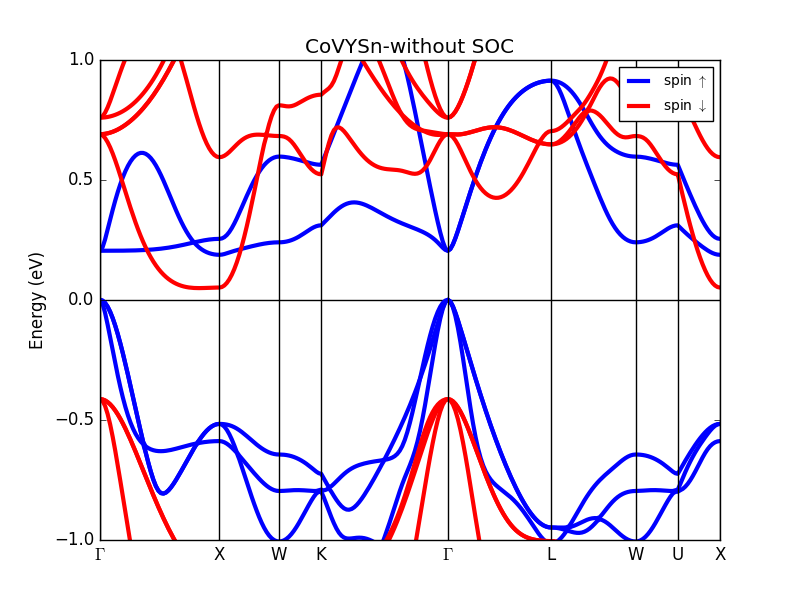}
\includegraphics[scale=0.3]{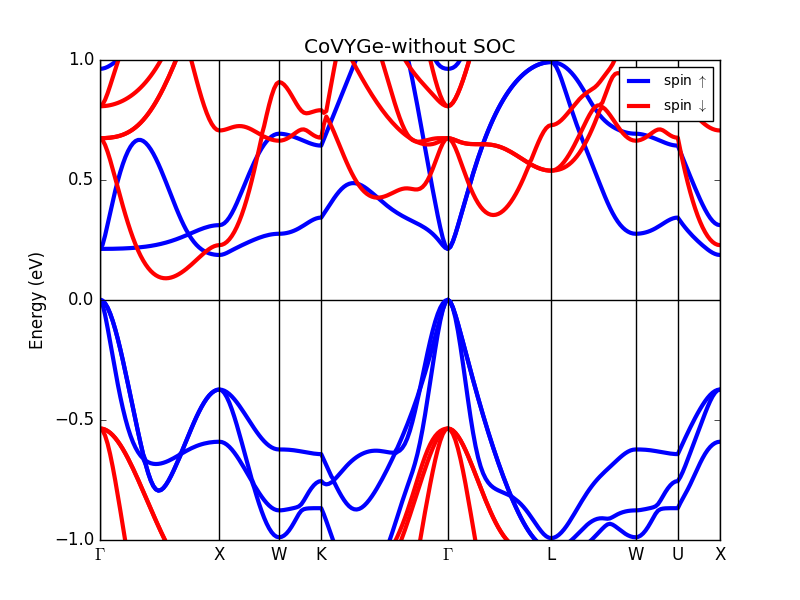}

\includegraphics[scale=0.3]{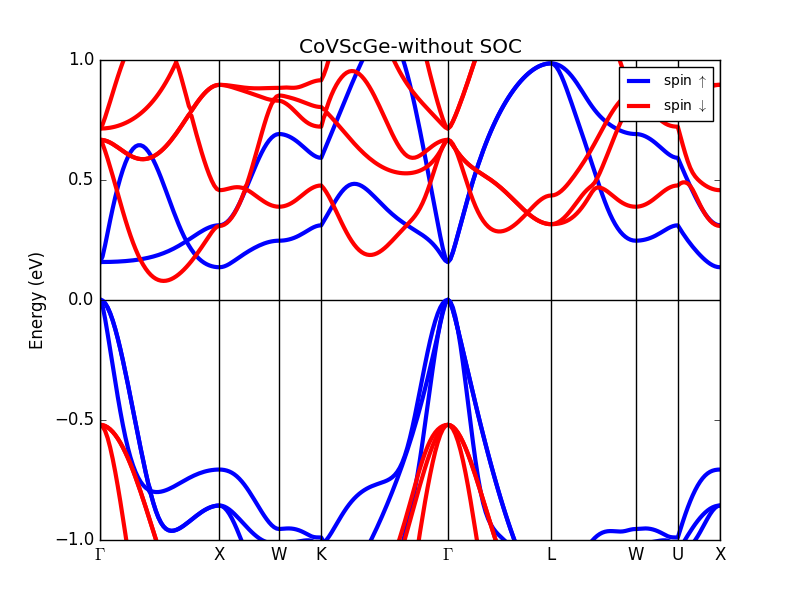}
\includegraphics[scale=0.3]{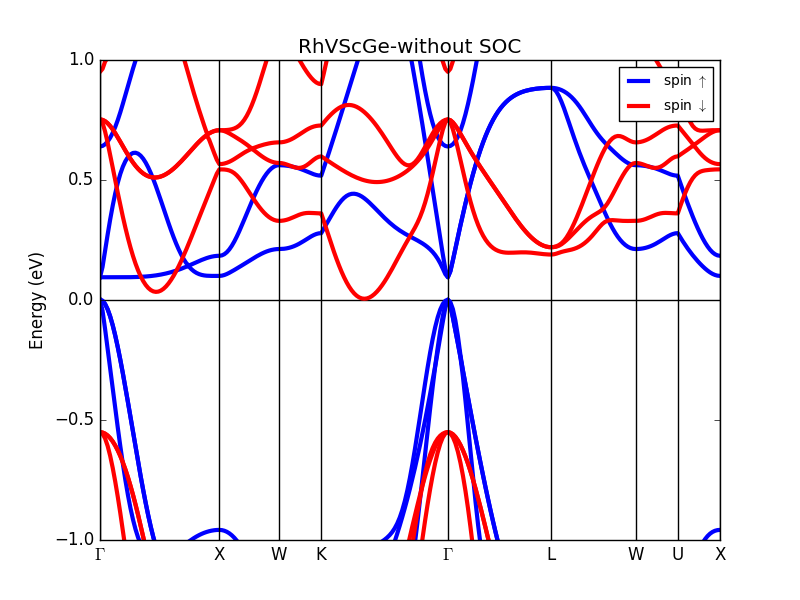}

\includegraphics[scale=0.3]{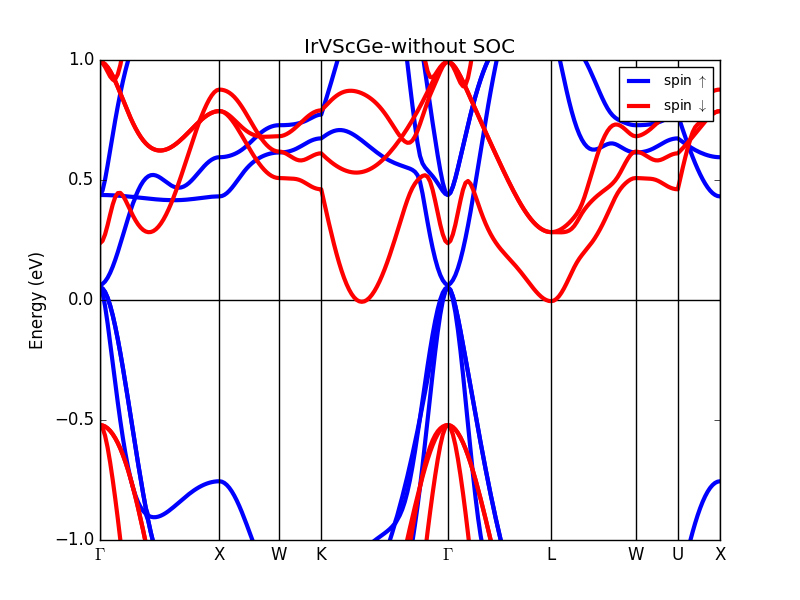}
\includegraphics[scale=0.3]{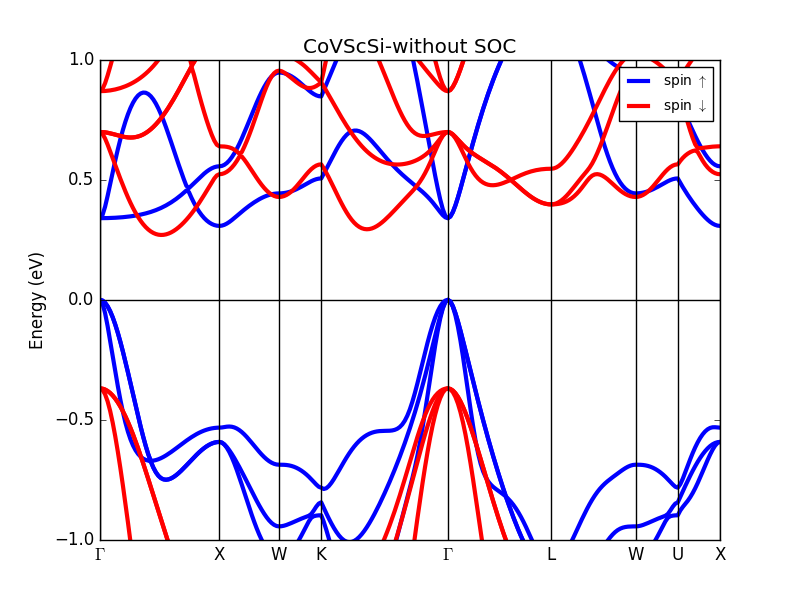}

\includegraphics[scale=0.3]{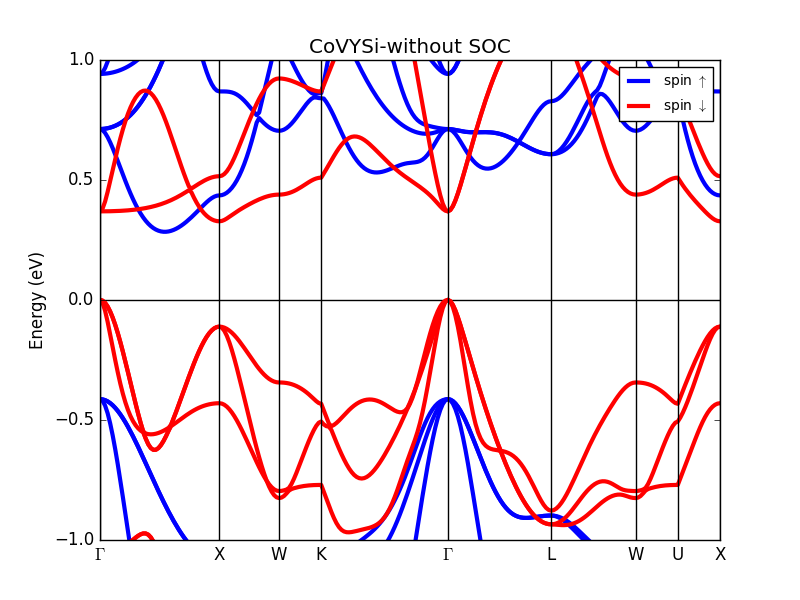}
\includegraphics[scale=0.3]{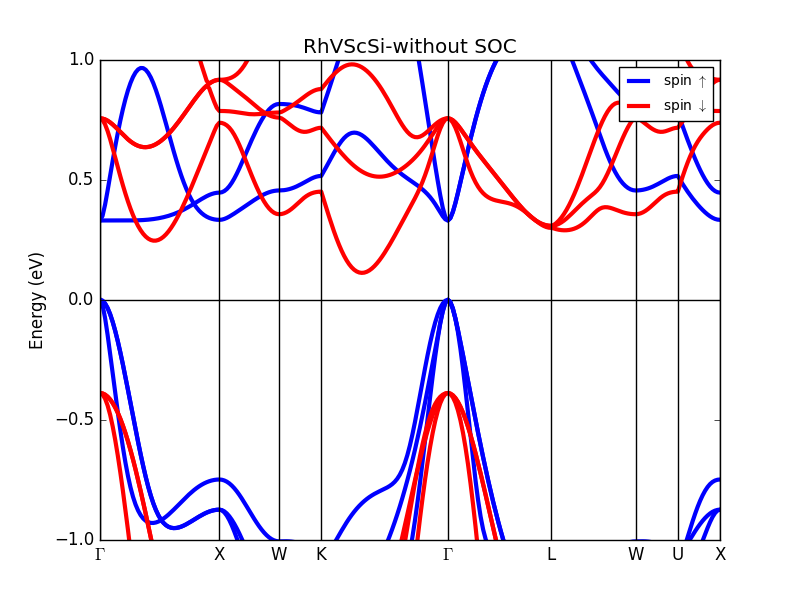}

\clearpage

\centering

\includegraphics[scale=0.3]{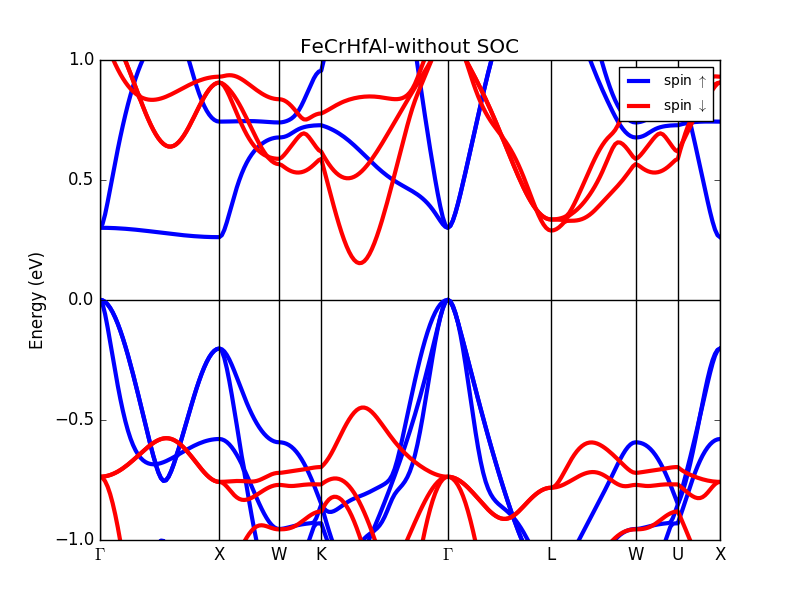}
\includegraphics[scale=0.3]{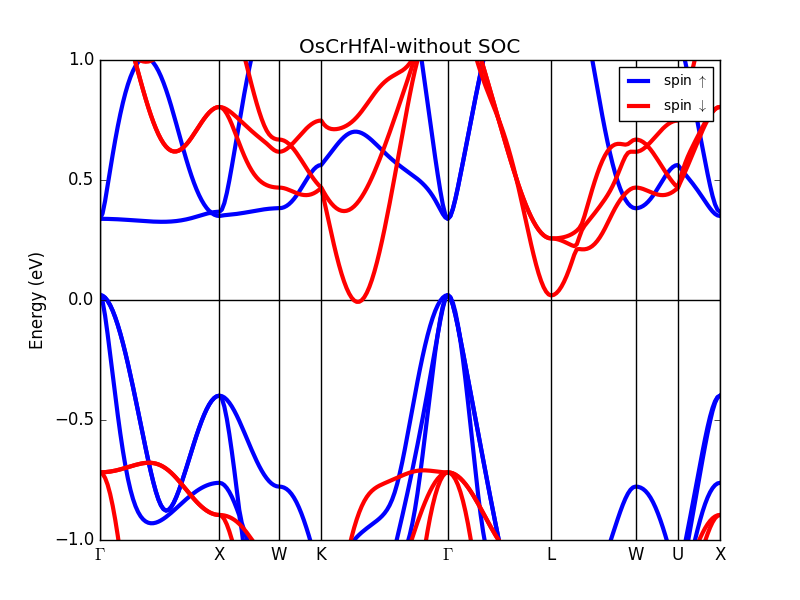}

\includegraphics[scale=0.3]{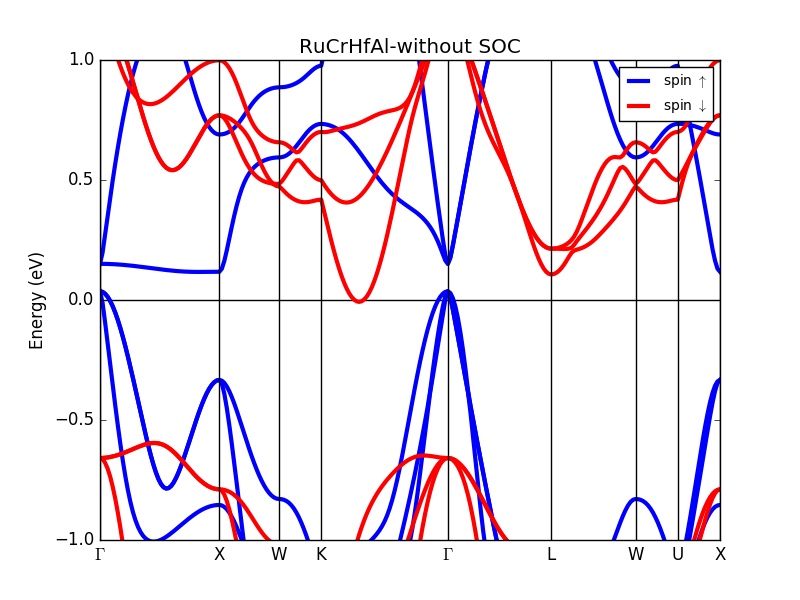}
\includegraphics[scale=0.3]{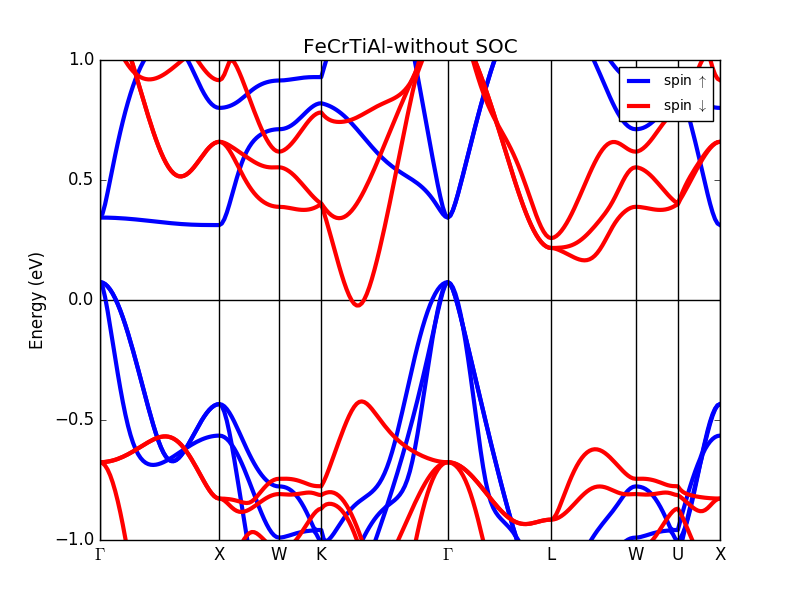}

\includegraphics[scale=0.3]{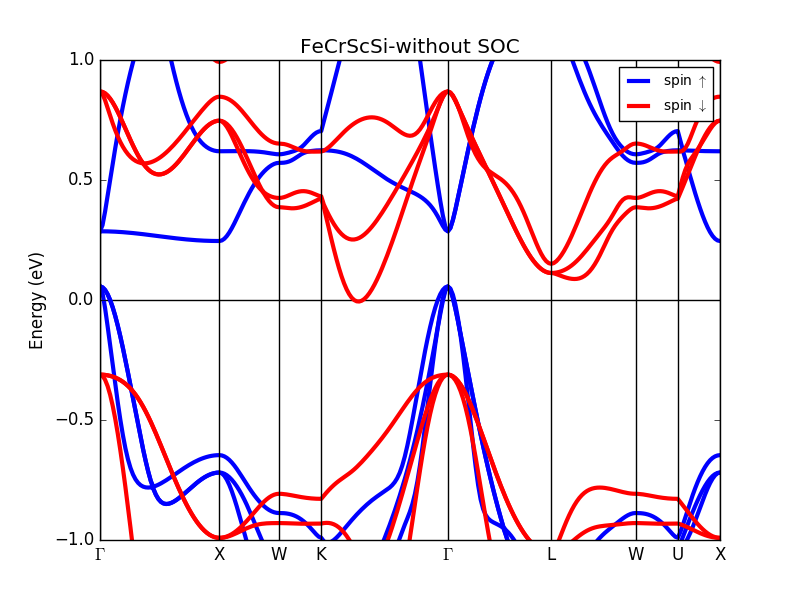}
\includegraphics[scale=0.3]{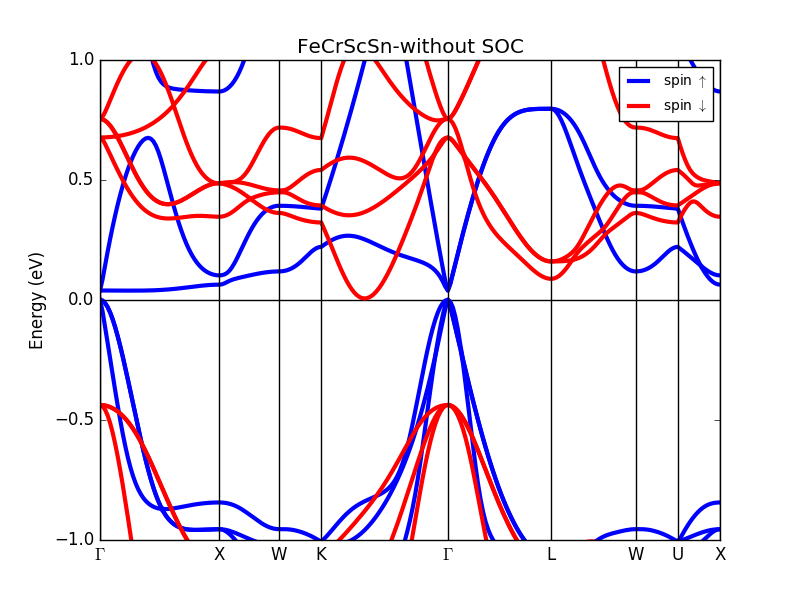}

\includegraphics[scale=0.3]{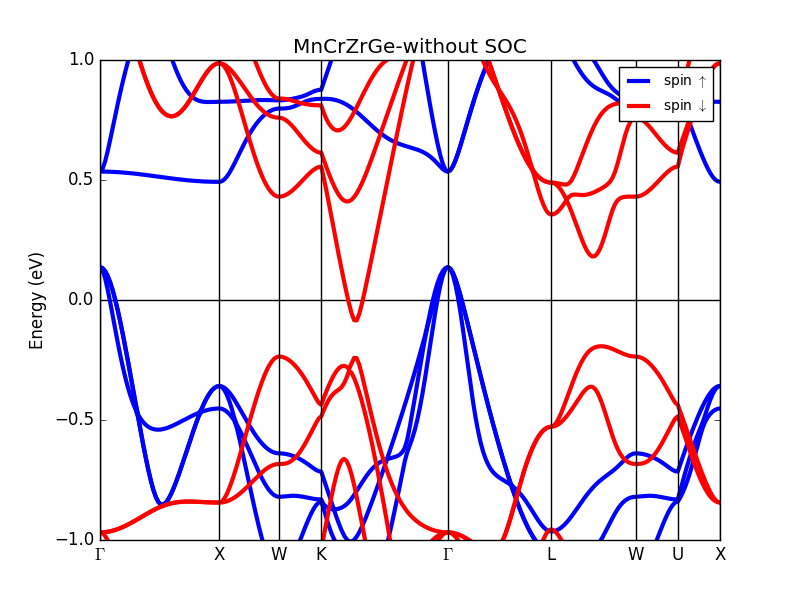}
\includegraphics[scale=0.3]{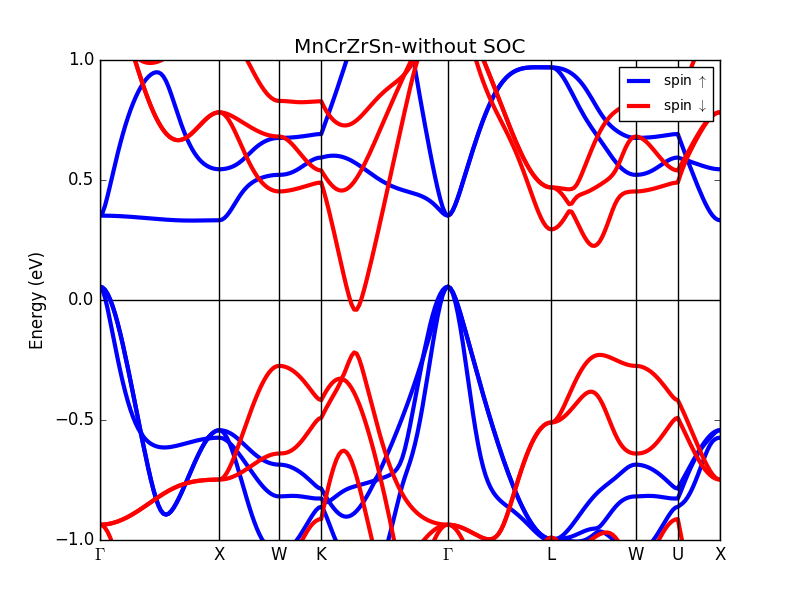}

\clearpage

\centering
\includegraphics[scale=0.3]{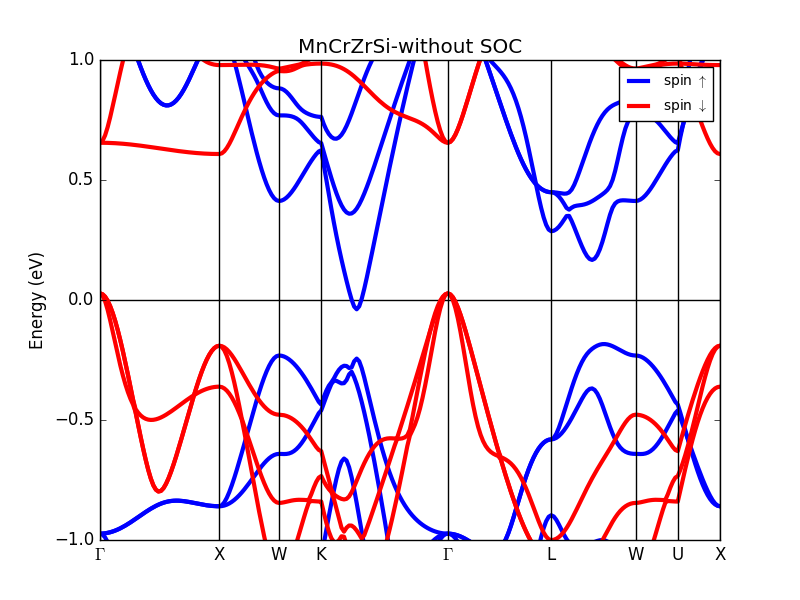}
\includegraphics[scale=0.3]{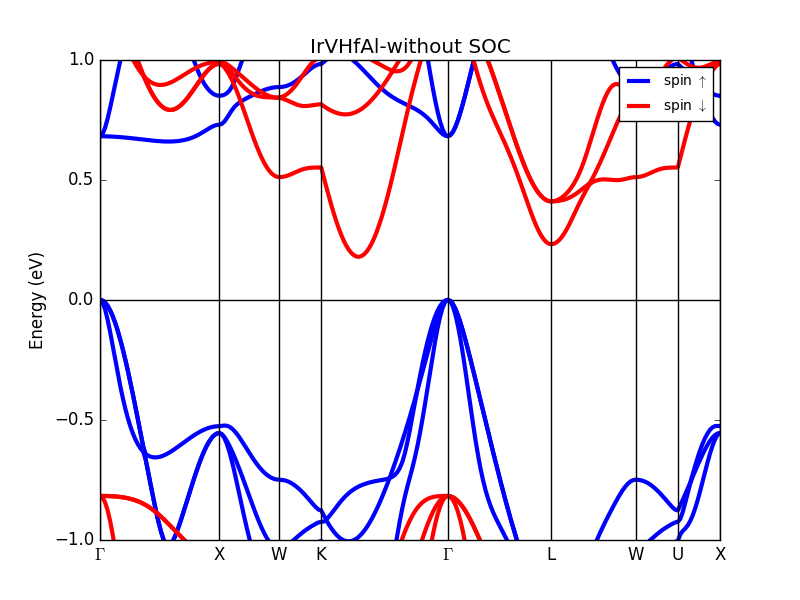}

\includegraphics[scale=0.3]{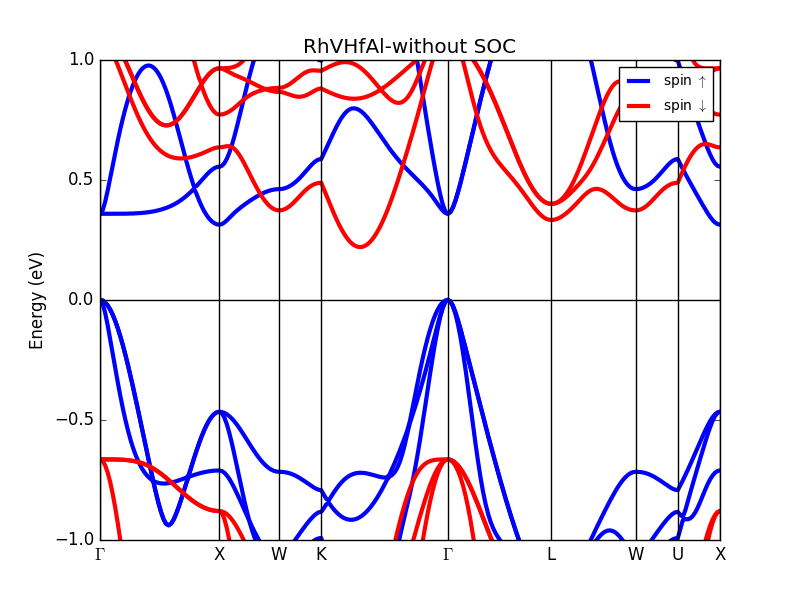}
\includegraphics[scale=0.3]{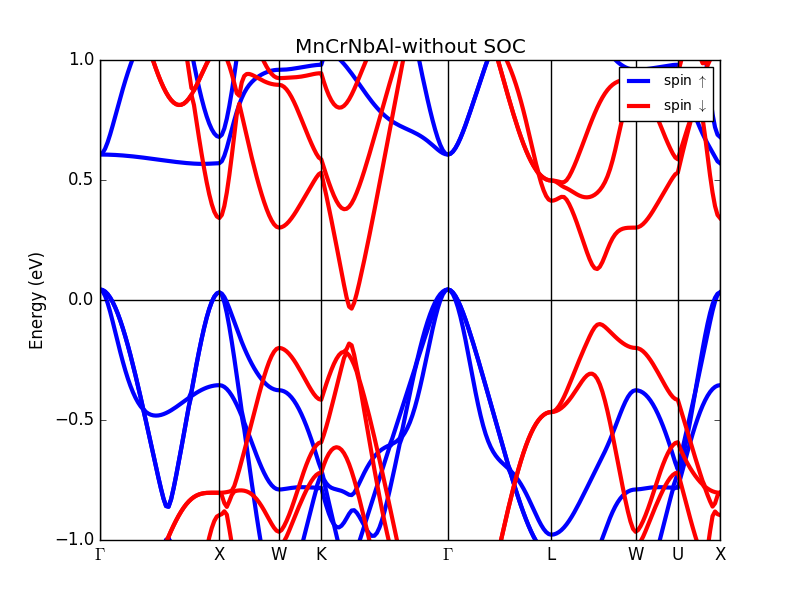}

\includegraphics[scale=0.3]{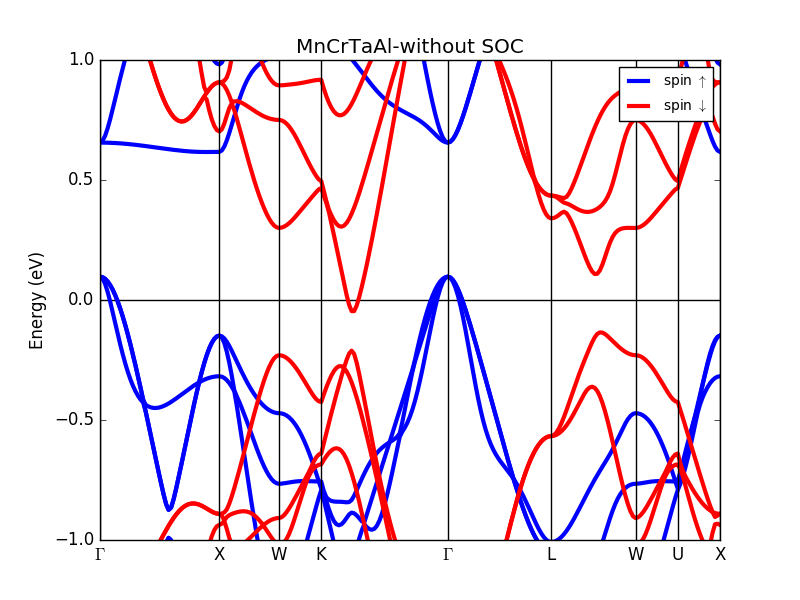}
\includegraphics[scale=0.3]{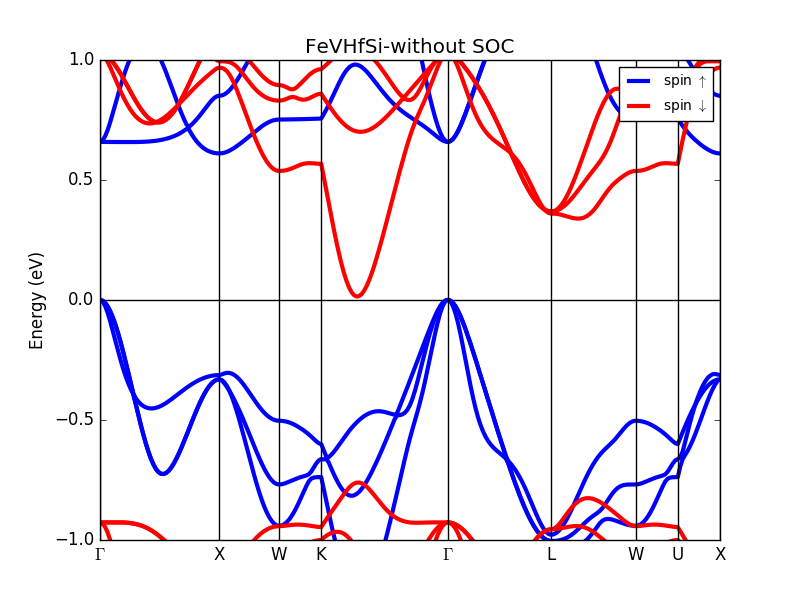}

\includegraphics[scale=0.3]{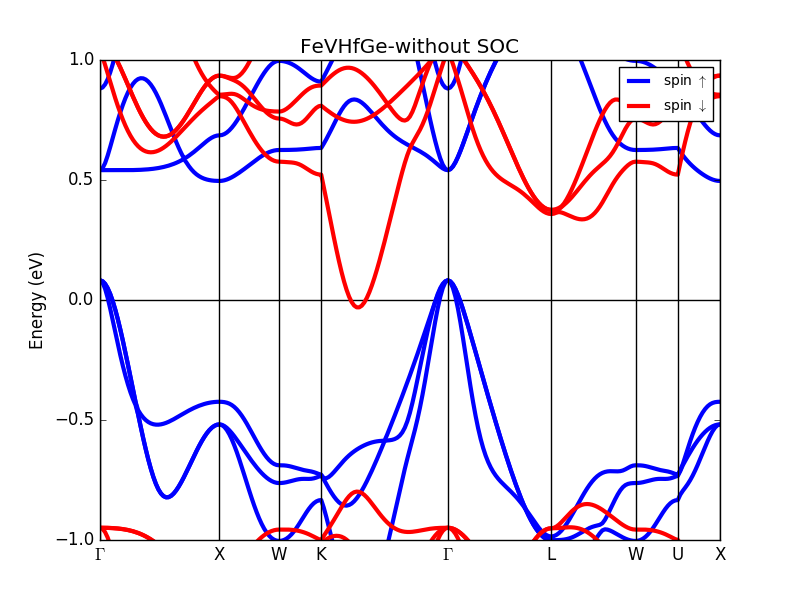}
\includegraphics[scale=0.3]{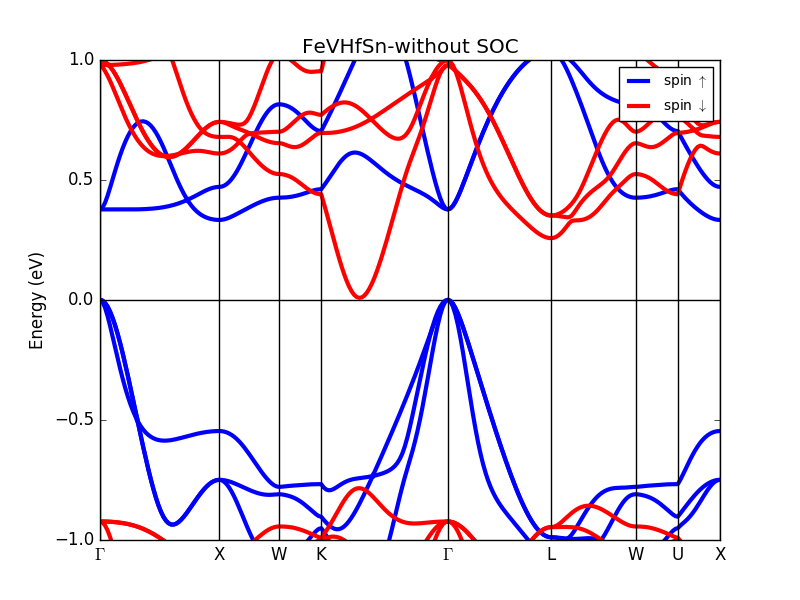}

\clearpage

\centering
\includegraphics[scale=0.3]{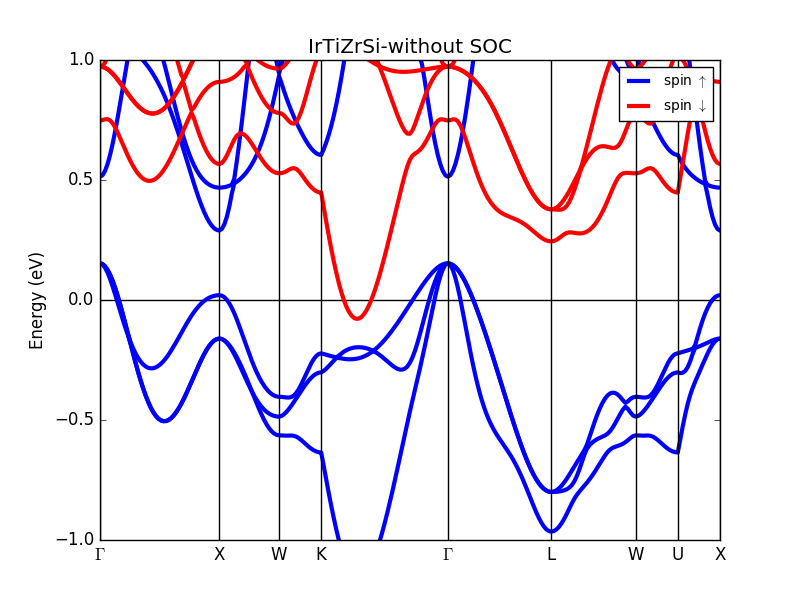}
\includegraphics[scale=0.3]{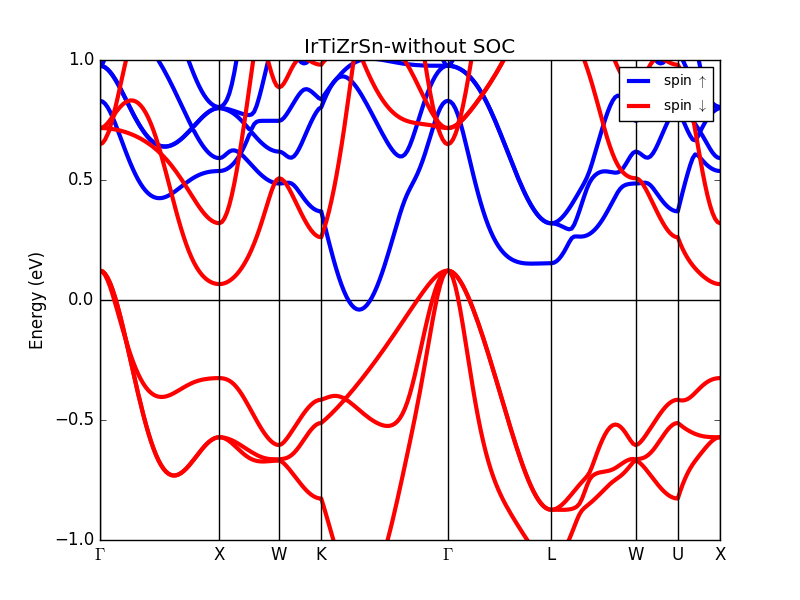}

\includegraphics[scale=0.3]{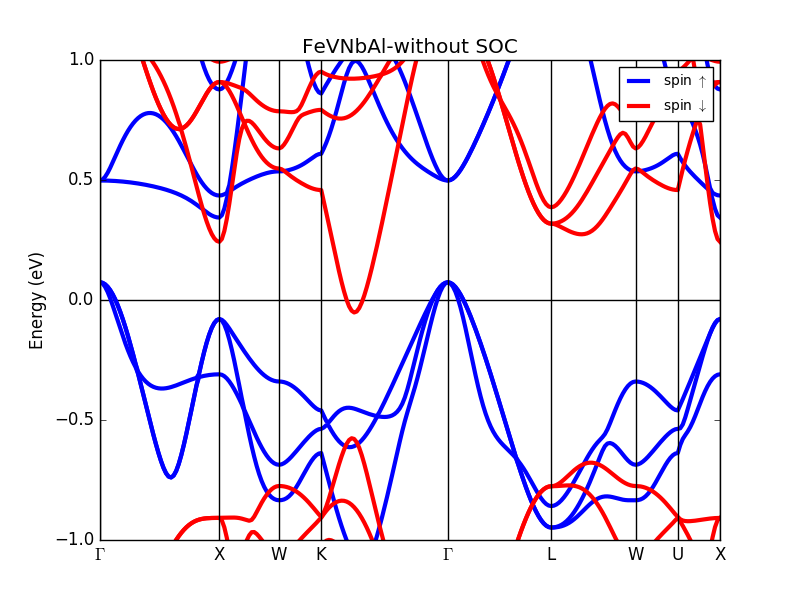}
\includegraphics[scale=0.3]{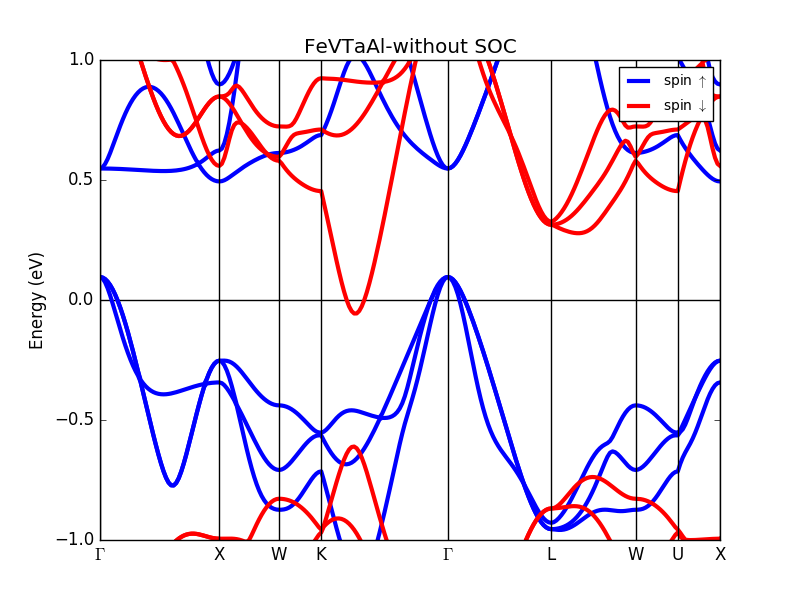}

\includegraphics[scale=0.3]{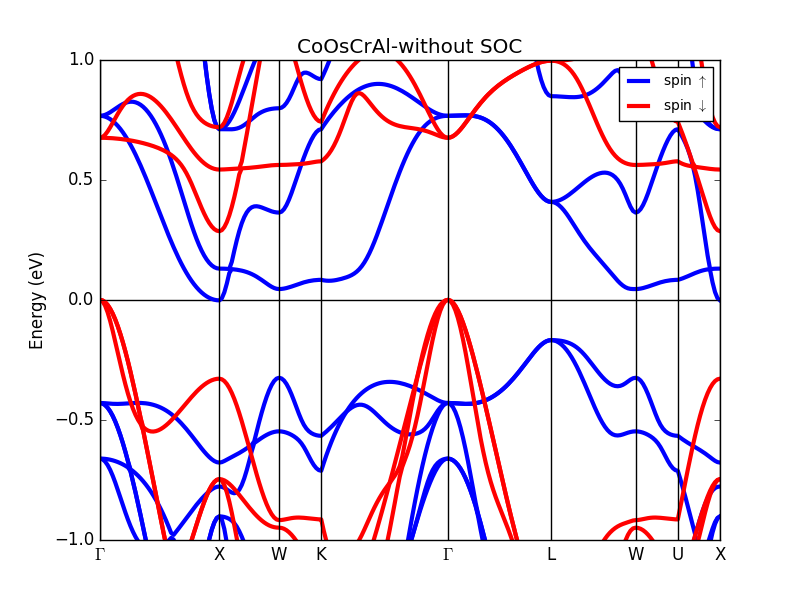}
\includegraphics[scale=0.3]{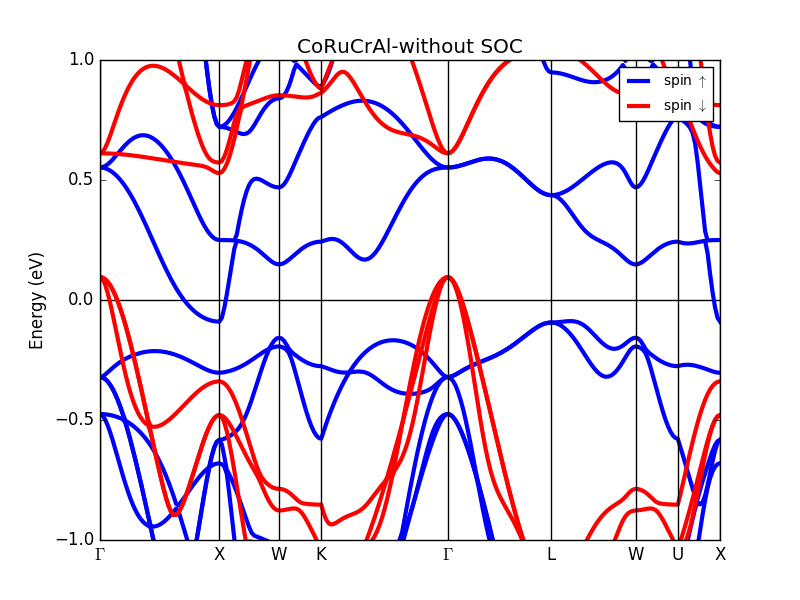}

\includegraphics[scale=0.3]{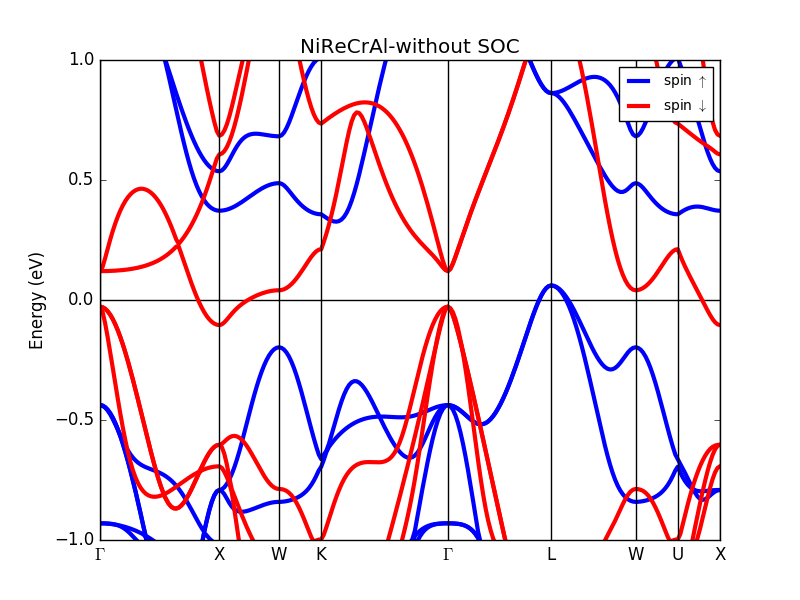}

\clearpage
\subsection{Type III SGSs}

\centering
\includegraphics[scale=0.3]{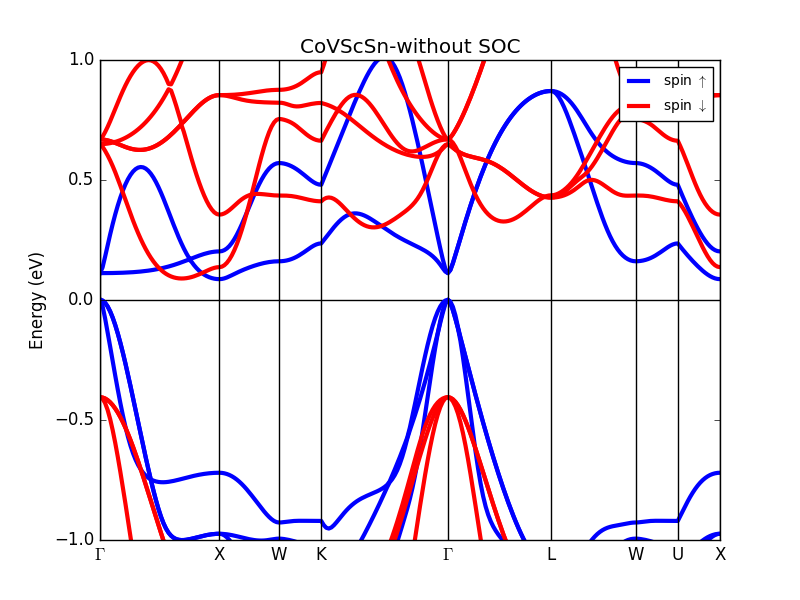}
\includegraphics[scale=0.3]{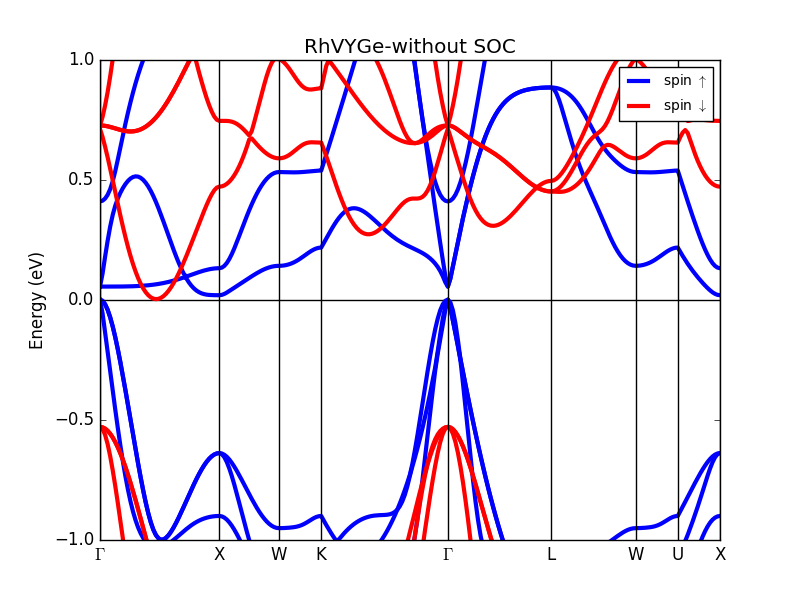}

\includegraphics[scale=0.3]{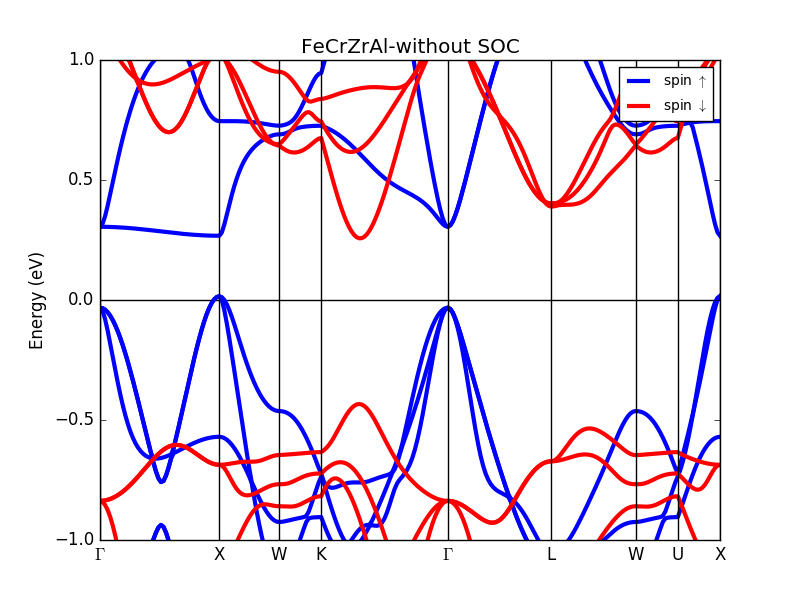}
\includegraphics[scale=0.3]{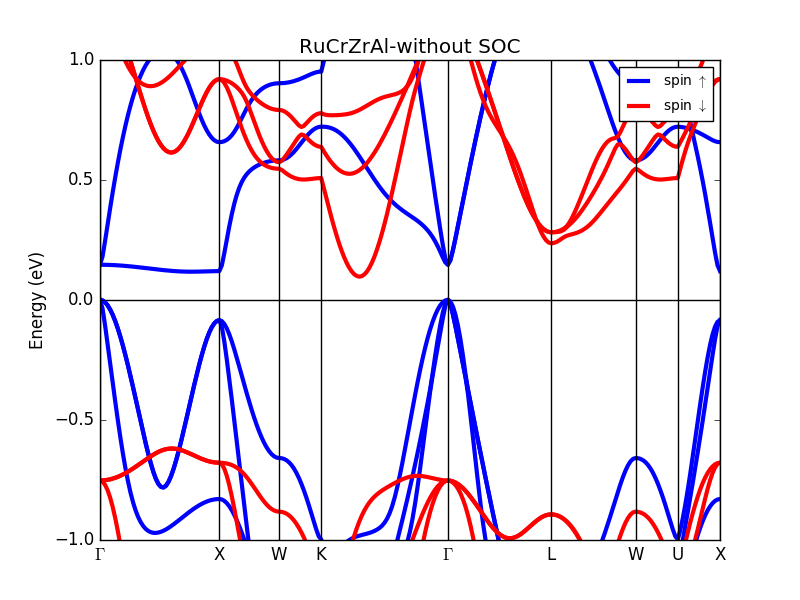}

\includegraphics[scale=0.3]{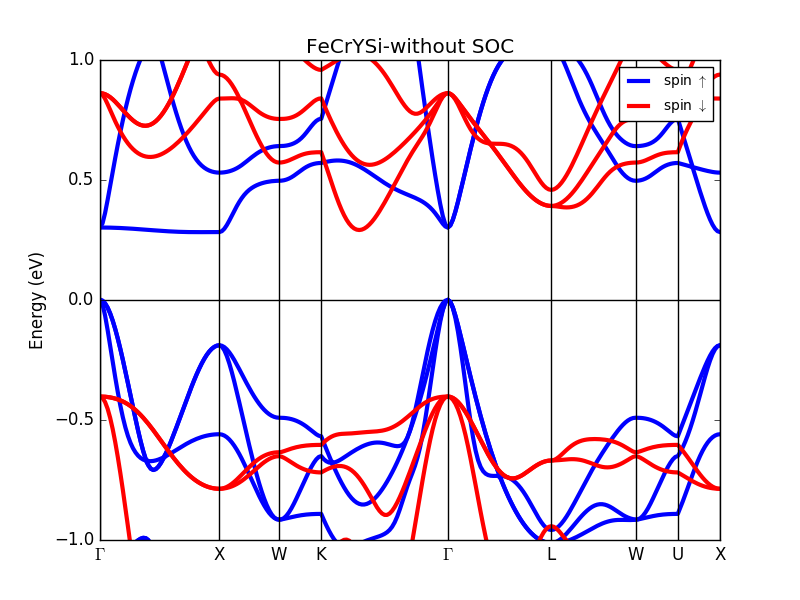}
\includegraphics[scale=0.3]{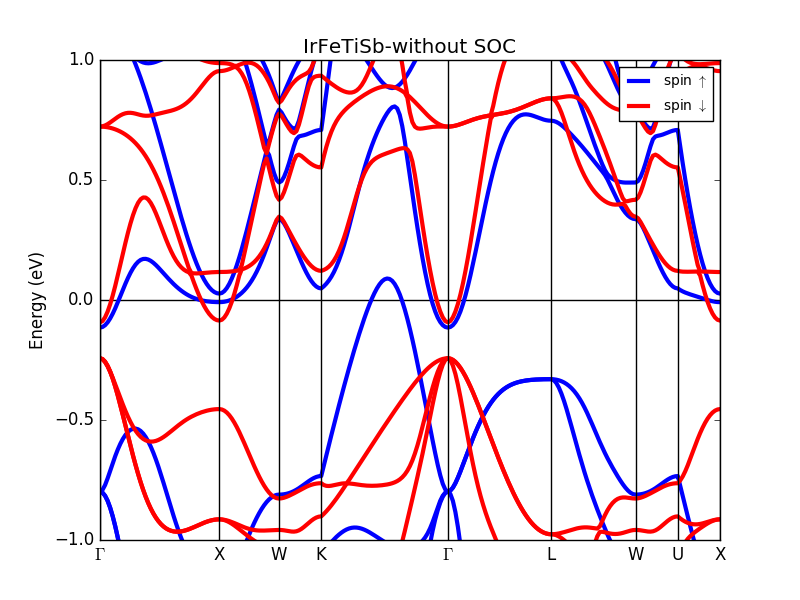}

\includegraphics[scale=0.3]{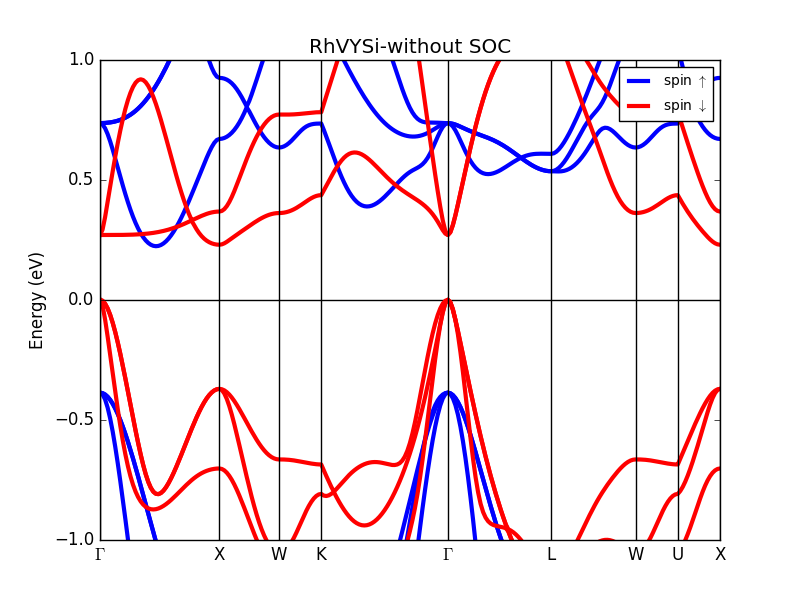}
\includegraphics[scale=0.3]{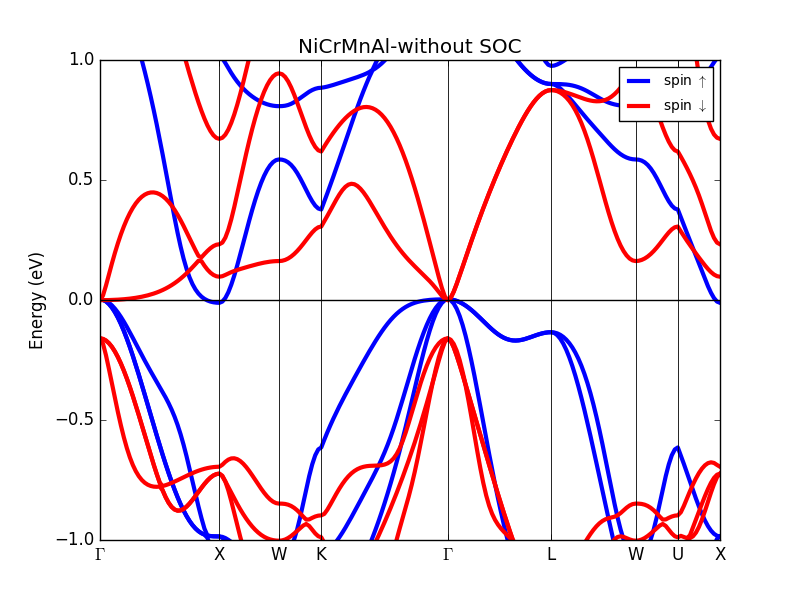}

\clearpage
\section{SOC effect for SGSs}
Spin orbit coupling (SOC) effect for SGS. Here SOC has significant effect on such SGSs.

\centering
\includegraphics[scale=0.3]{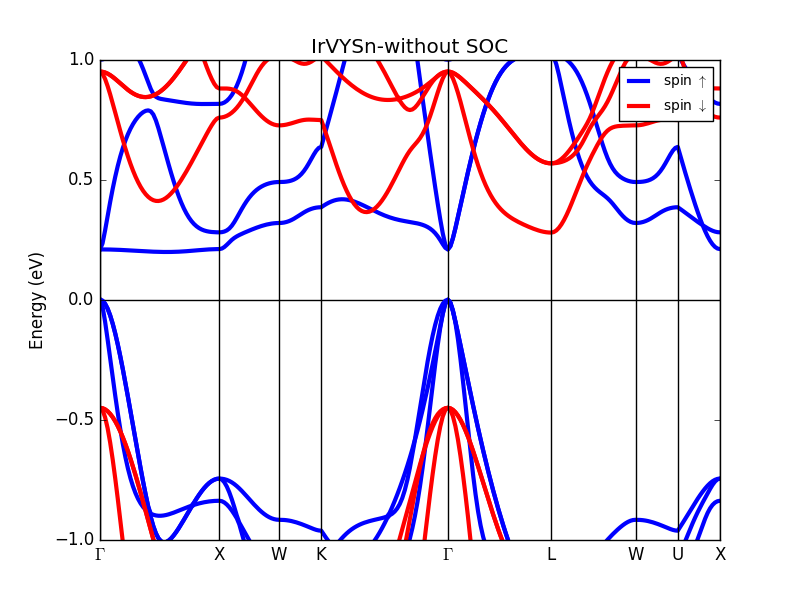}
\includegraphics[scale=0.3]{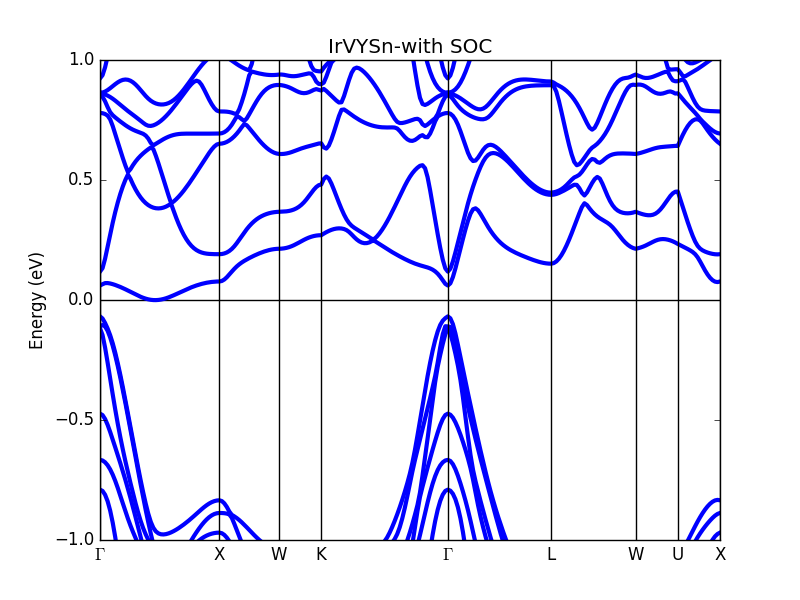}

\includegraphics[scale=0.3]{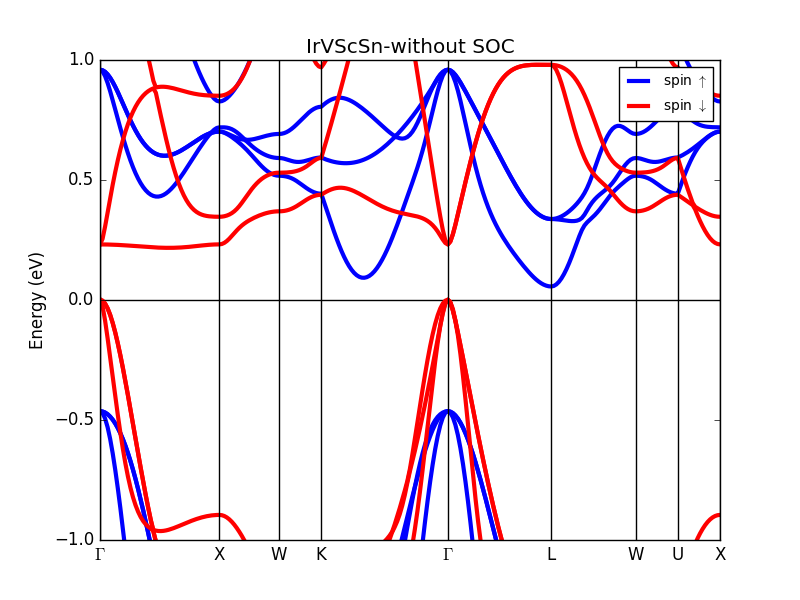}
\includegraphics[scale=0.3]{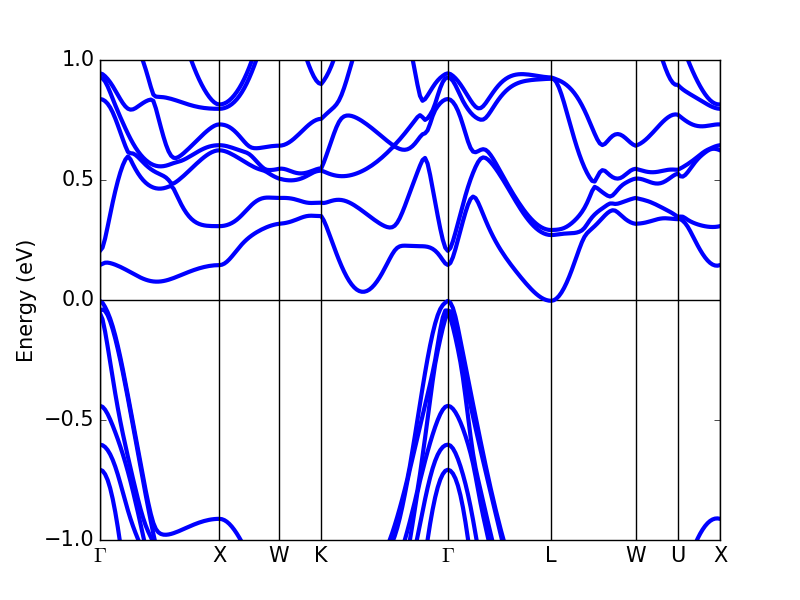}

\includegraphics[scale=0.3]{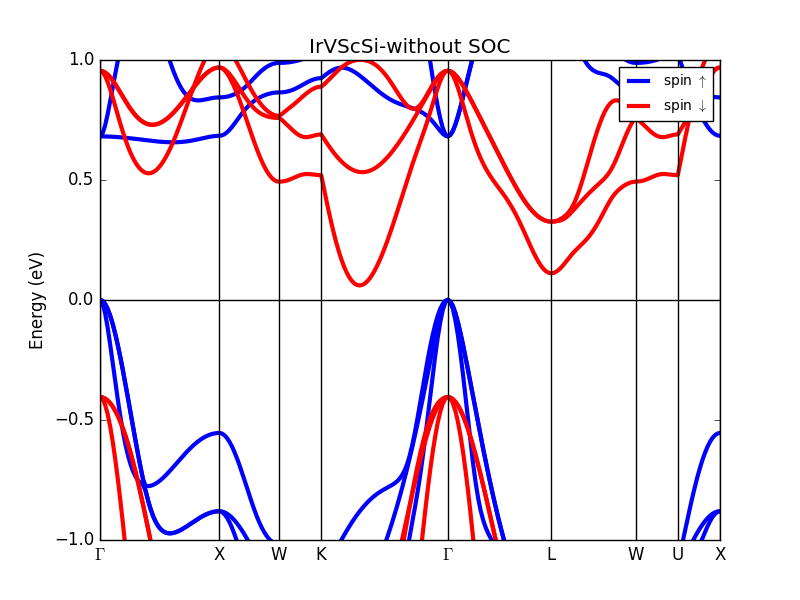}
\includegraphics[scale=0.3]{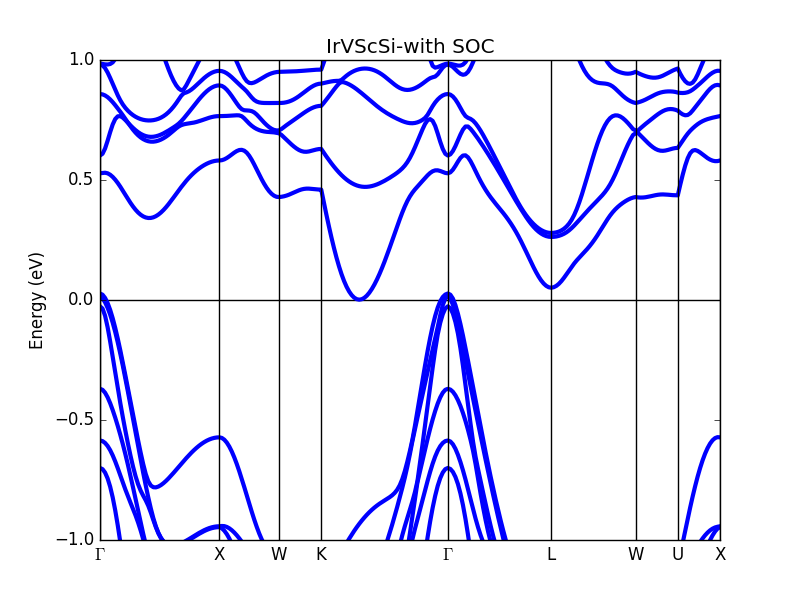}

\clearpage

\centering
\includegraphics[scale=0.3]{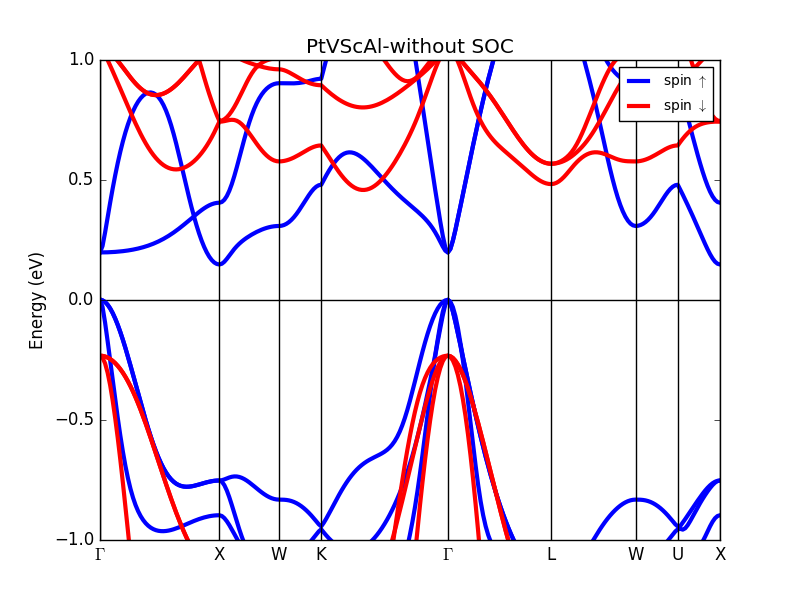}
\includegraphics[scale=0.3]{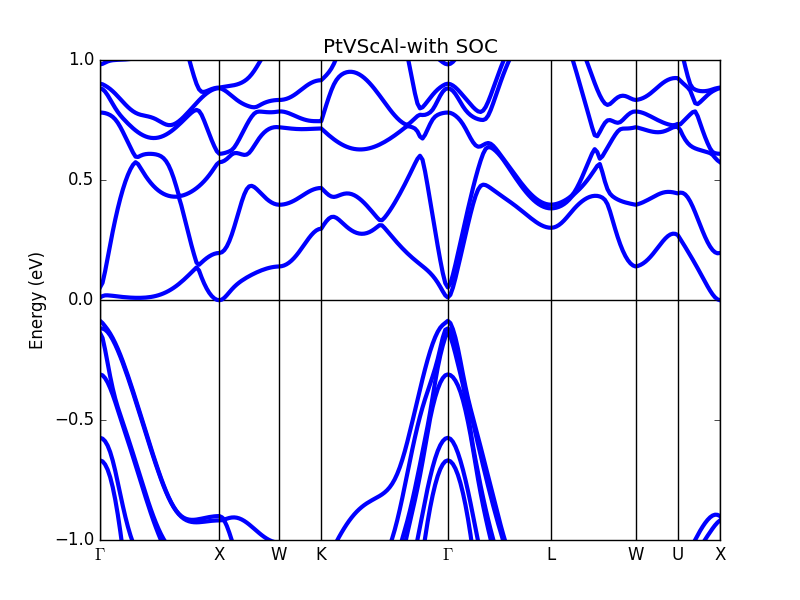}

\includegraphics[scale=0.3]{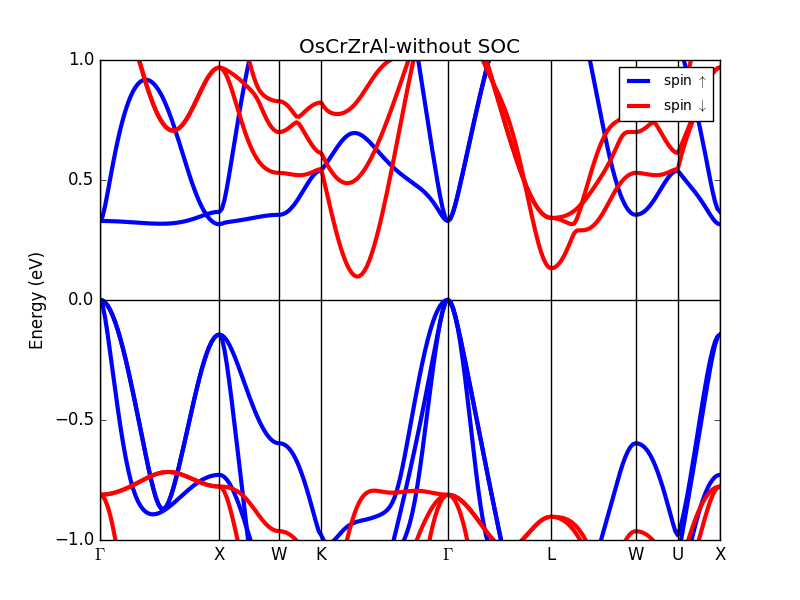}
\includegraphics[scale=0.3]{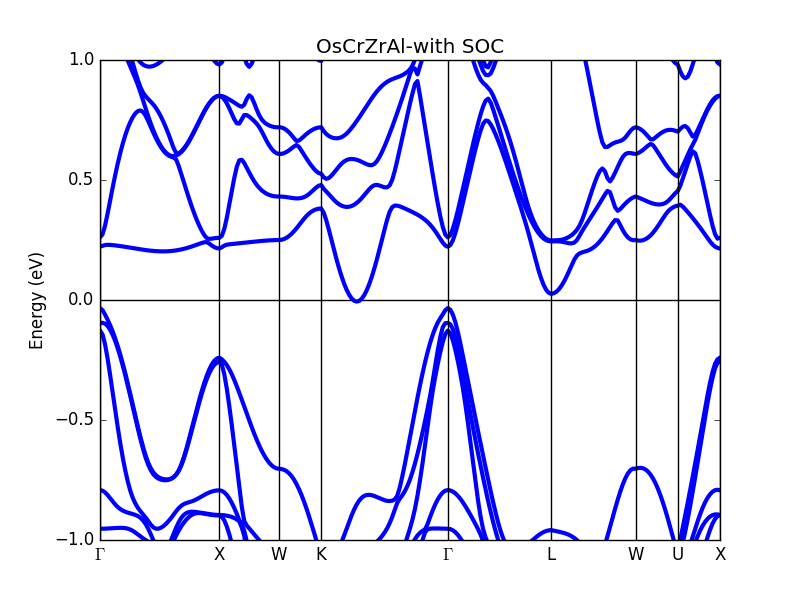}

\includegraphics[scale=0.3]{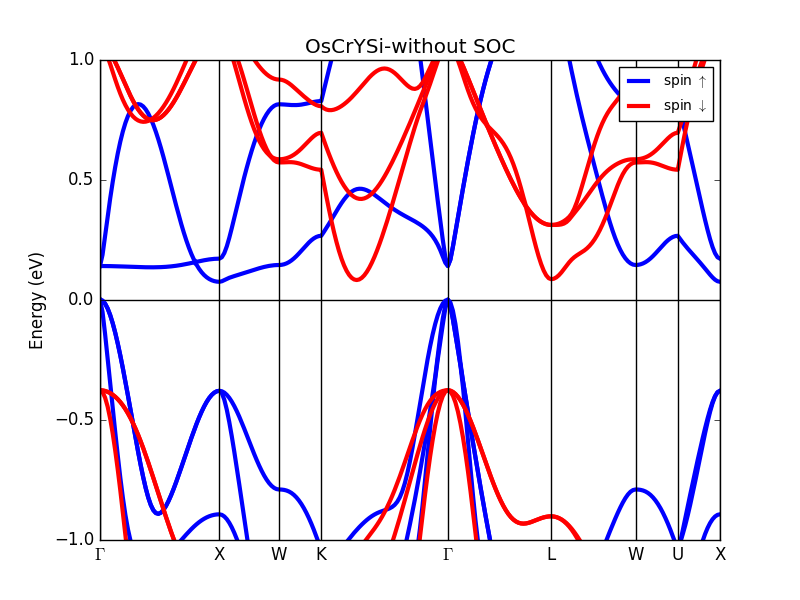}
\includegraphics[scale=0.3]{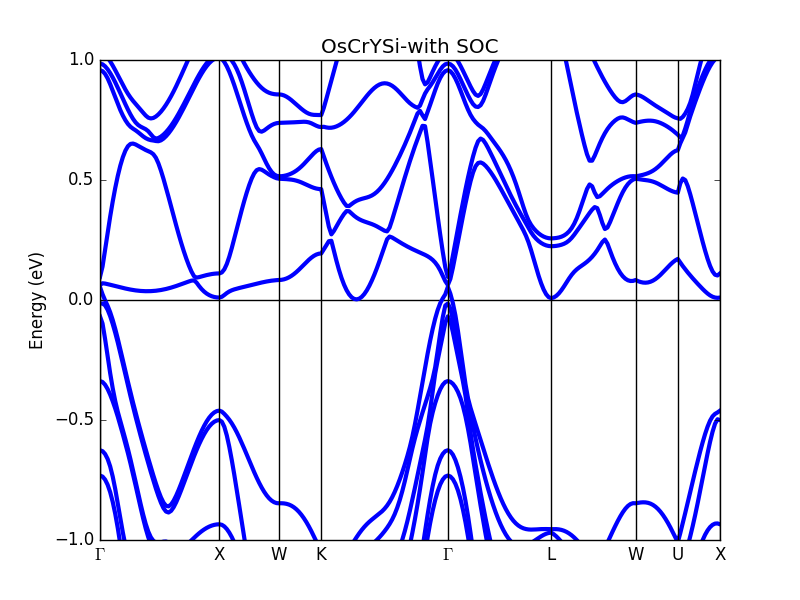}